\documentclass{article}
\pdfoutput=1
\usepackage{arxiv}
\usepackage{authblk}

\usepackage{dingbat}

\usepackage{hyperref}
\usepackage{graphicx}
\usepackage{amsmath}
\usepackage{url}
\usepackage{afterpage}

\usepackage{array}
\newcolumntype{R}[1]{>{\raggedleft\arraybackslash}p{#1}}
\newcolumntype{L}[1]{>{\raggedright\arraybackslash}p{#1}}
\newcolumntype{C}[1]{>{\centering\arraybackslash}p{#1}}

\newcommand{\repdate}{\today}
\newcommand{\ZeroIn}{\textbf{\textsc{ZeroIn}}}

\DeclareMathOperator\erf{erf}

\newlength\myfigwidth
\begin{document}

\title{{\ZeroIn}: Characterizing the Data Distributions of Commits in Software Repositories}

\author[1]{Kalyan Perumalla}
\author[1]{Aradhana Soni}
\author[1]{Rupam Dey}
\author[2]{Steven Rich}
\affil[1]{Department of Industrial and Systems Engineering, University of Tennessee, Knoxville\thanks{kperuma3@utk.edu, asoni5@vols.utk.edu, rdey1@vols.utk.edu}}
\affil[2]{Cisco Systems, Inc.\thanks{srich@cisco.com}}

\date{\repdate}
\maketitle

\begin{abstract}
Modern software development is based on a series of rapid incremental changes collaboratively made to large source code repositories by developers with varying experience and expertise levels. The {\ZeroIn} project is aimed at analyzing the metadata of these dynamic phenomena, including the data on repositories, commits, and developers, to rapidly and accurately mark the quality of commits as they arrive at the repositories.  In this context, the present article presents a characterization of the software development metadata in terms of distributions of data that best captures the trends in the datasets.  Multiple datasets are analyzed for this purpose, including Stack Overflow on developers' features and GitHub data on over 452 million repositories with 16 million commits.  This characterization is intended to make it possible to generate multiple synthetic datasets that can be used in training and testing novel machine learning-based solutions to improve the reliability of software even as it evolves.  It is also aimed at serving the development process to exploit the latent correlations among many key feature vectors across the aggregate space of repositories and developers. The data characterization of this article is designed to feed into the machine learning components of {\ZeroIn}, including the application of binary classifiers for early flagging of buggy software commits and the development of graph-based learning methods to exploit sparse connectivity among the sets of repositories, commits,  and developers.
\end{abstract}

\section{Introduction}

The {\ZeroIn} project is aimed at building a novel computational framework to enable the characterization and classification of buggy or vulnerable code changes at the very origin of source code, as early as the time at which they are committed to software repositories. We achieve this by novel use of machine learning for analyzing, classifying, visualizing, and modeling massive logs of version control in combination with key characterization of developers’ historical traits. The intrinsically temporal nature of software coder and repository interactions creates transitive and complex dependencies that could potentially reveal insights into the vulnerabilities \cite{10.5555/580808}. The research in this project is aimed at tapping the potential to broadly benefit modern software development platforms and increase software security globally.

In order to apply modern machine learning techniques to this difficult problem, large datasets are necessary for training, testing, and validating the techniques.  Most importantly, ground truth data at scale is needed not only to meet the accuracy measures but also to be relevant to the large sizes of modern software repositories and global developer team sizes \cite{alali2008s}.

Here, we present the data gathering and characterization of software development spanning institutions (coder team strengths, etc.), coder features (experience, expertise, etc.), software repository characteristics (commit volumes, frequencies, bugs, etc.).

To our knowledge, this is one of the first attempts at characterizing the metadata about modern software repositories at a large scale and across multiple dimensions including the software bases, commits, and coder features.

\subsection{Closed-form Distributions from Datasets}

The large sizes of the datasets are distilled into closed form distributions defined by probability density functions by applying multiple candidate distributions and sorting the distributions and their parameters by their goodness of fit and selecting the top candidates that provides the best fit. The cumulative distribution functions of these probability density functions are necessary to derive the inverse cumulative distribution functions, which are in turn necessary for accurately sampling the distributions in the generation of synthetic ground truth driven by the real data.  To illustrate, the probability density function of the Log-Normal distribution is given by
\begin{equation*}
\begin{array}{lcl}
f(x-\theta) & = & \frac{1}{(x-\theta)\sigma\sqrt{2\pi}}\exp\left(\frac{-(\ln(x-\theta)-\mu)^2}{2\sigma^2}\right),
\end{array}
\end{equation*}
where
$\theta$ is the shift parameter, and
$\sigma$ and $\mu$ are such that $\log(x-\theta)$ follows the Normal distribution $N(\mu,\sigma)$ with a mean $\mu$ and standard deviation $\sigma$.

The cumulative density function of $f(x-\theta)$ is given by
\begin{equation*}
\begin{array}{lcl}
F(x-\theta) & = & \int_\theta^x f(t-\theta)dt \\
 & = & \frac{1}{2} [1+\erf(\frac{ln(x-\theta)-\mu)^2}{\sigma\sqrt{2}})],
\end{array}
\end{equation*}
where, the error function is defined as:
\begin{equation*}
\begin{array}{lcl}
\erf(z) & = & \frac{2}{\pi} \int_0^z \exp{(-t^2)}dt.
\end{array}
\end{equation*}

The inverse, $F^{-1}(\cdot)$, of the cumulative distribution function $F(\cdot)$ is given by

\begin{equation*}
(x-\theta) = F^{-1}(u) = \exp\left(\mu+\sqrt{\sigma\sqrt2\erf^{-1} (2 u-1)}\right),
\end{equation*}
where, $u=F(x-\theta)$, and $\erf^{-1}$ is defined by the pair of relations $q=\erf(p)$ and $p=\erf^{-1}(q)$.

For the other distributions used in later sections (such as Exponential and Negative Binomial), a similar set of derivations provide the inverse CDF equations from their corresponding PDF equations.

The inverse CDF $F^{-1}(x)$ is directly useful in regenerating any probability density function $f(x)$ using random sampling techniques (see, for example, chapter 12 on reversible distribution sampling \cite{perumalla2013}).  We exploit this key feature of the relation between empirically-determined fit for $f(x)$ (and, by implication, $F(x)$) and its inverse CDF $F^{-1}$ to develop concise representations of all large datasets used in our study.

\subsection{Machine Learning Model}

\begin{figure}
    \centering
    \includegraphics[width=\myfigwidth]{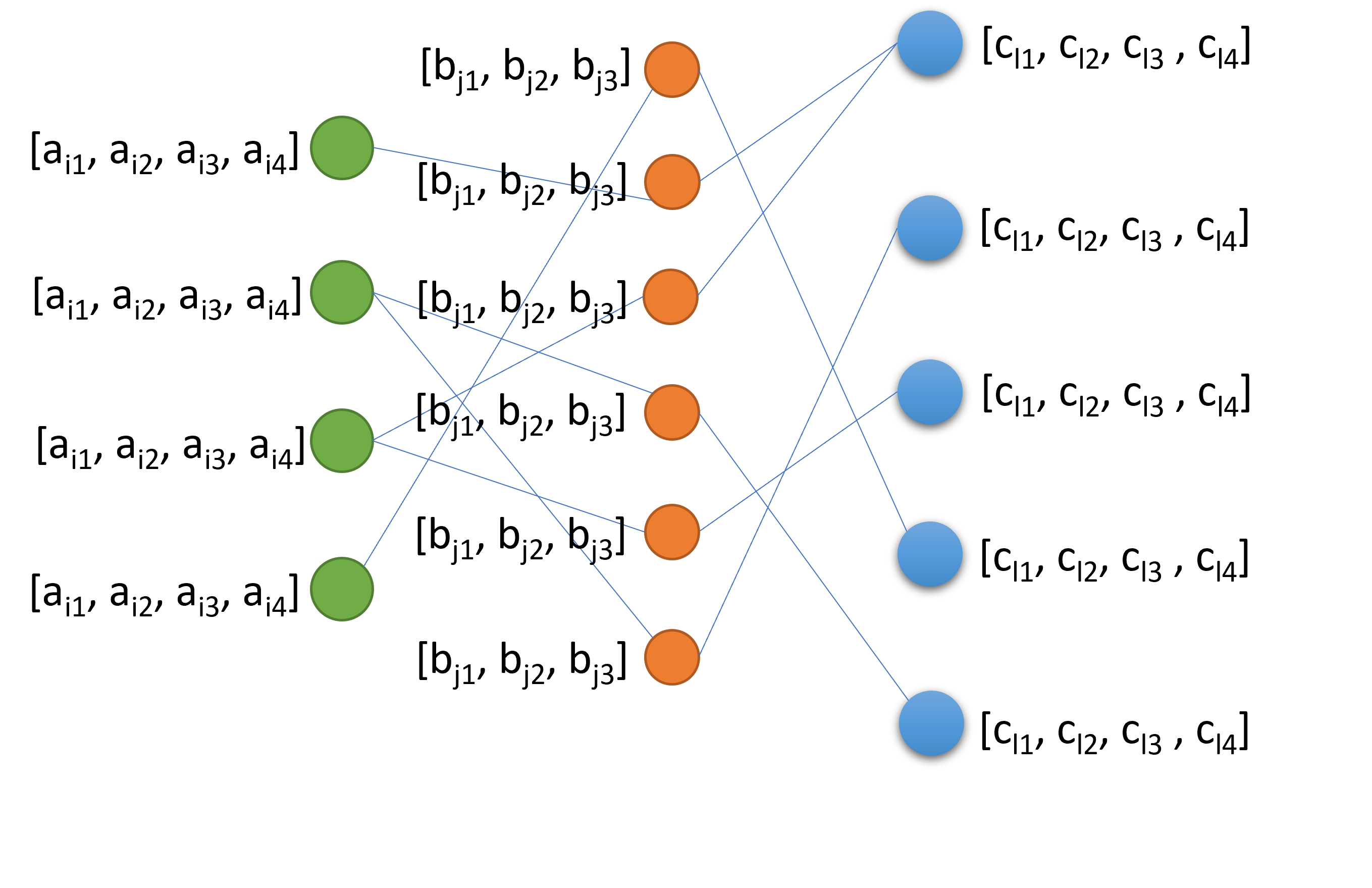}
    \caption{Our tripartite graph model of the software development network formulated to suit machine learning. The columns represent coders, commits, and repositories, and the vectors attached to the nodes represent the feature vectors corresponding to them}
    \label{fig:GNN-tripartite}
\end{figure}

Figure~\ref{fig:GNN-tripartite} depicts our graph model for modelling the software development process. The first column of nodes denotes coders, each with its own feature vector.  The second column of nodes (along with their own feature vectors) represents the commits performed by the coders, and the last column of nodes (with their feature vectors) represents the repositories to which the commits are made. The datasets are used to ultimately generate the distributions needed to populate this tripartite graph and their feature vectors for use in training, testing, and validating our machine learning algorithms.  In the later inferencing stage, the same graph model will be used to represent the production data on which the learner will be exercised to watch every commit and signal the prognoses of vulnerabilities \cite{hu2020open}.

\section{GitHub Dataset with 452 Million Commits}

To capture and characterize the volumes of commits to software repositories, we undertook a search for publicly available metadata on large software repository storehouses.  We found the \texttt{data.world} repository on 452 million commits to the popular GitHub software storehouse as a suitable dataset.

The dataset is described at the following website: \url{https://data.world/vmarkovtsev/452-m-commits-on-github}.  The direct URL to the statistics file containing the metadata in comma-separated value (CSV) format is
\url{https://query.data.world/s/7euzfiycvbfxikevuc2cg2p4pbuish}.
This dataset contains metadata from 16 million repositories on GitHub. The primary data in this dataset is organized into four main columns: repository name, number of commits, number of contributors and average length of the commit.

Using Python’s Scipy Library, we performed regression using multiple candidate distributions.  Among the different distributions evaluated, the Exponential and Log-Normal distributions fared the best in terms of goodness of fit.  The best fitting distributions are shown in Table~\ref{tbl:Fit}.  The distribution provides \textit{loc} and \textit{scale} parameters: the \textit{loc} parameter shifts the distribution by the appropriate amount and the \textit{scale} parameter stretches the distribution as required.  These values are indicated in the table as (\textit{loc}, \textit{shape}) values against the distribution.  Similar parameters are used to define the other distributions as well.

\begin{table}
\centering
\caption{Distributions of commits by number of repositories}
{\small
\begin{tabular}{|C{0.25\textwidth}|R{0.20\textwidth}|C{0.20\textwidth}|C{0.25\textwidth}|}
\hline
\textbf{Commits} & \textbf {Repositories} & \textbf {Best Fit} & \textbf {Second-Best Fit}\\
\hline\hline
\textless 20 & 13,156,036 & Exponential 
(-0.83,0.83) & Log-Normal (5.67,-0.832,0) \\
\hline
20 - 100 & 2,235,831 & Exponential 
(-1.07,1.07) & Log-Normal (1.01,-1.17,0.76) \\
\hline
100 - 1000 & 554,079 & Log-Normal (1.30,-0.83,0.41) & Weibull Min (0.81,-0.81,0.71) \\
\hline
1000 - 4000 & 28,549 & Exponential (-1.07,1.07) & Weibull Min (0.93,-1.07,1.11)\\
\hline
4000 - 10,000 & 4,766 & Exponential (-1.26,1.26) & Gamma (1.17,-1.26,1.07) \\
\hline
10,000 - 100,000 & 2,221 & Log-Normal (1.30,-0.81,0.40) & Inv. Gaussian (2.13,-0.851,0.40) \\
\hline
\textgreater 100,000 & 128 & Exponential (-0.94,0.94) & Log-Normal (1.33,-0.96,0.48) \\
\hline
\end{tabular}
}
\label{tbl:Fit}
\end{table}

Table~\ref{tbl:Fit} gives the parameters of the best fitting and second best fitting distributions over the different range of number of commits.

\subsection{Distributions of Commits to Repositories}

\begin{figure}
    \centering
    \includegraphics[width=\myfigwidth]{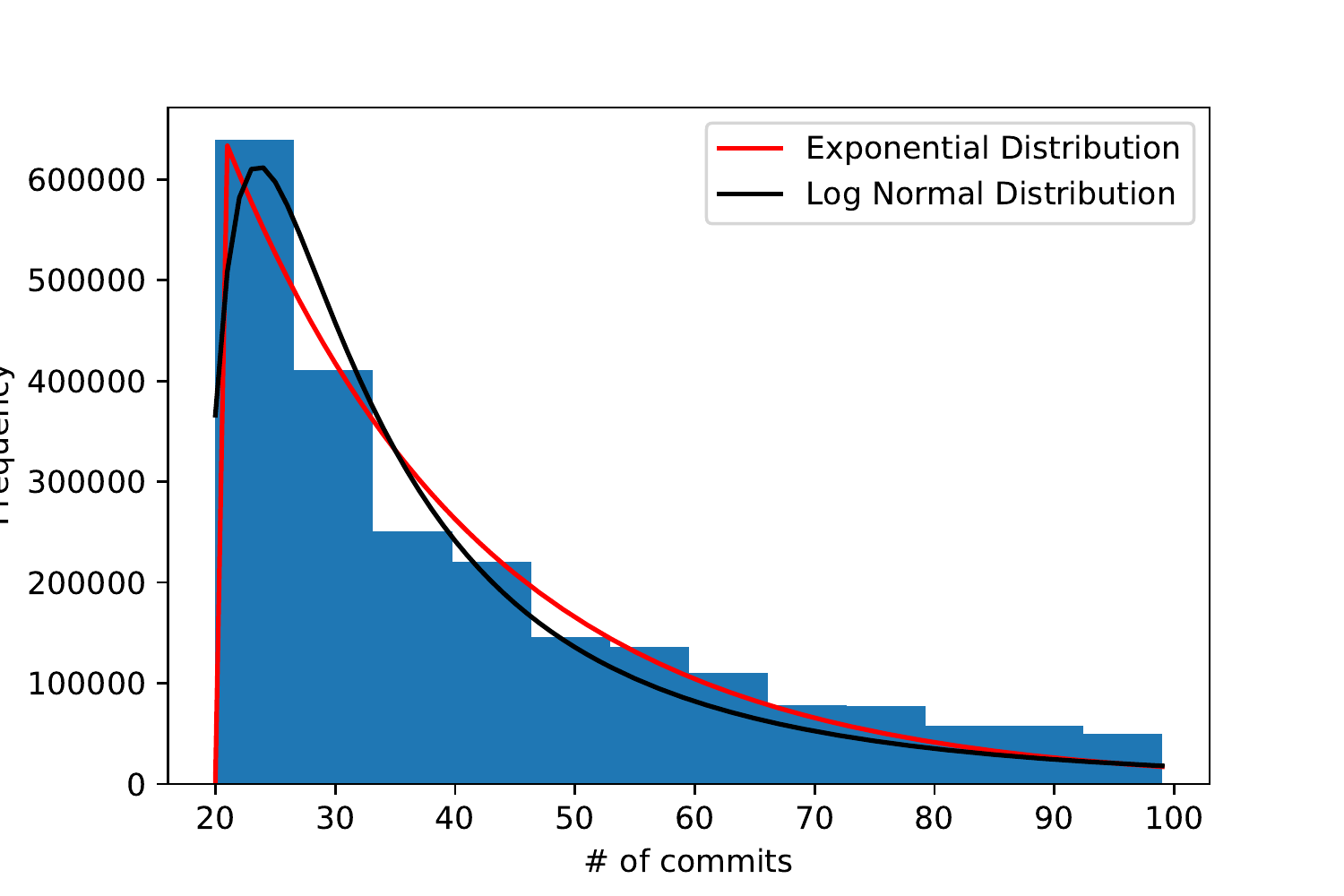}
    \caption{Histogram of number of commits between 20 and 100}
    \label{fig:452M_hist_bw 20 & 100}
\end{figure}

\begin{figure}
    \centering
    \includegraphics [width=\myfigwidth] {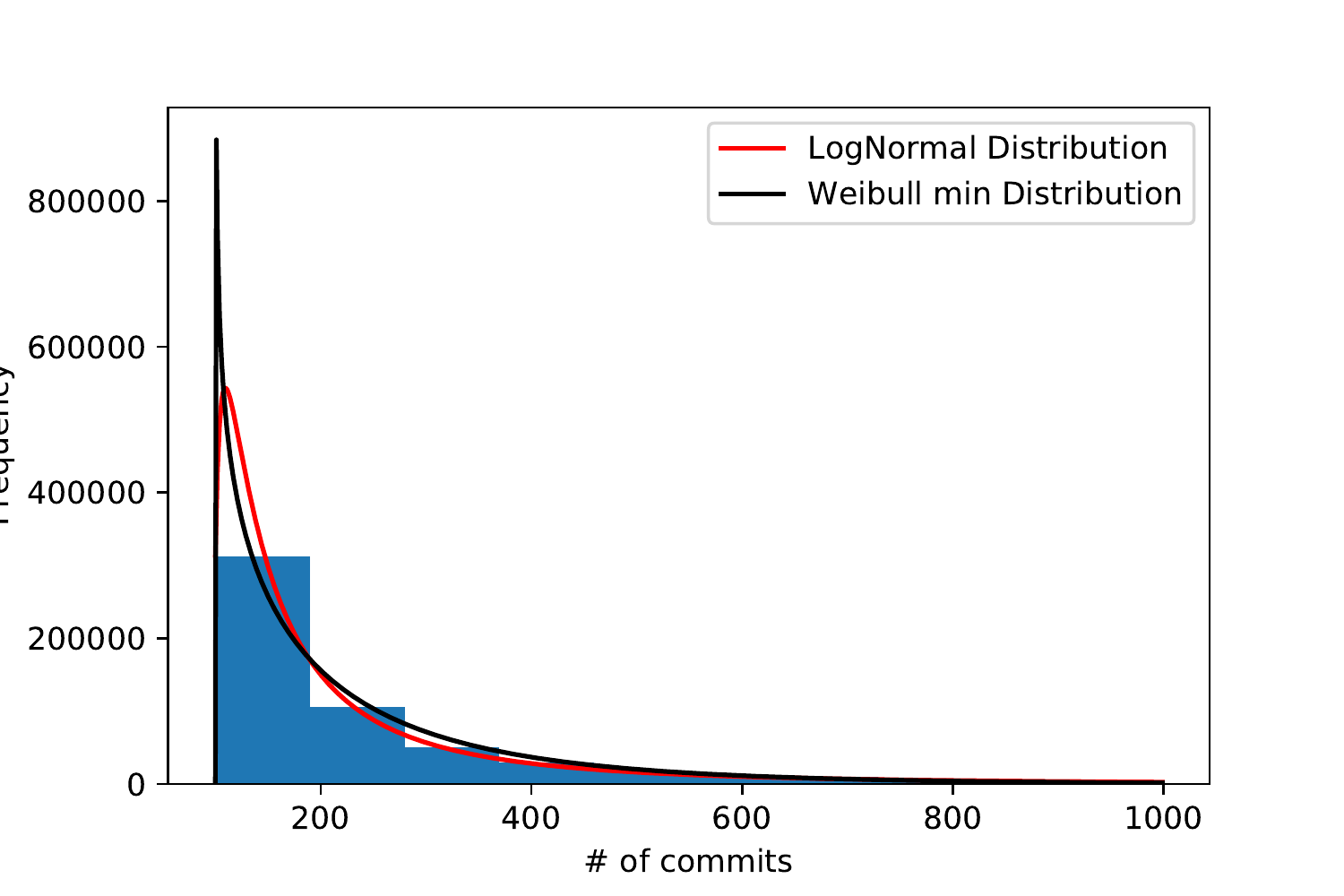}
    \caption{Histogram of number of commits between 100 and 1K}
    \label{fig:452M_hist_bw 100 & 1000}
\end{figure}

\begin{figure}
    \centering
    \includegraphics[width=\myfigwidth]{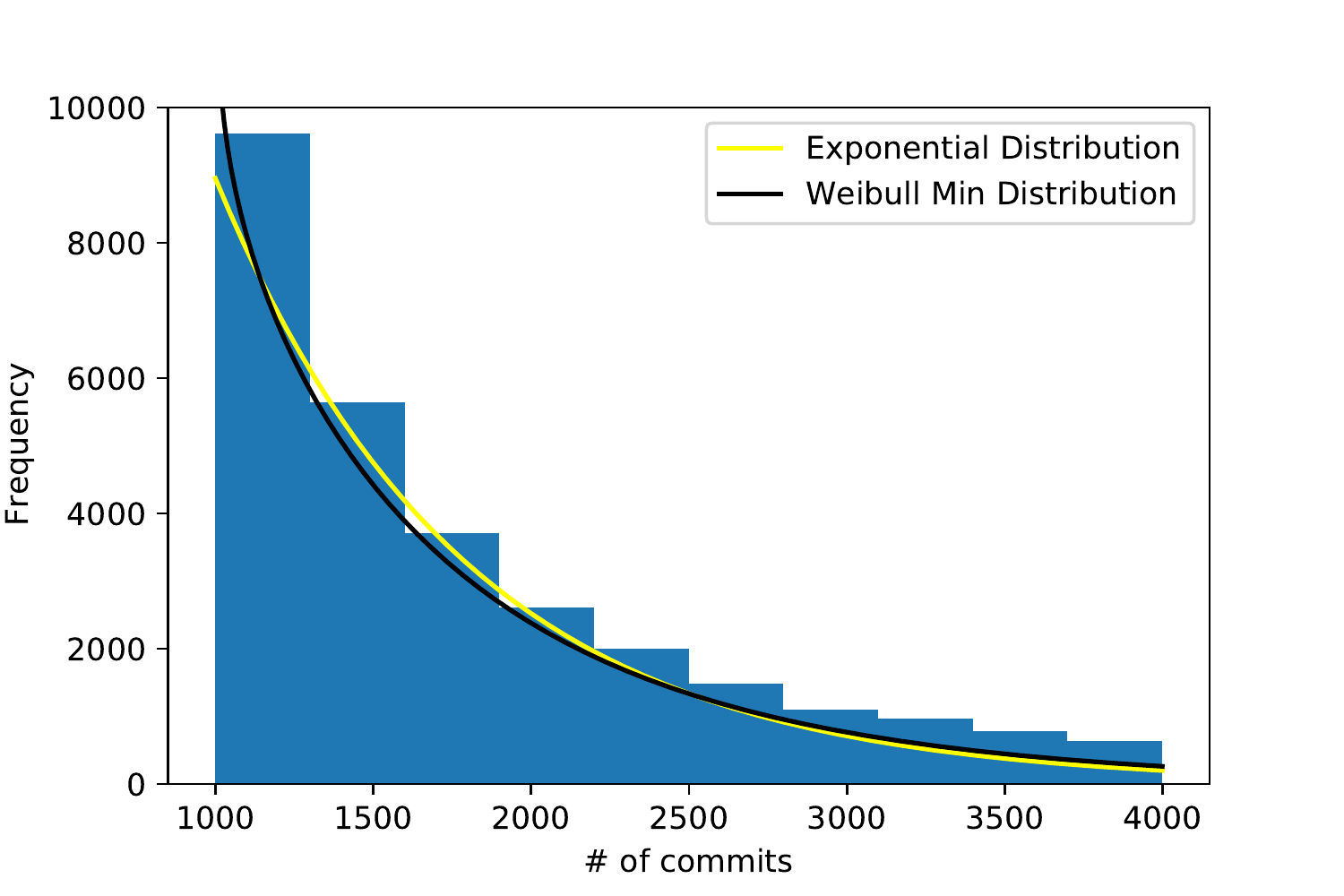}
    \caption{Histogram of number commits between 1K and 4K}
    \label{fig:452M_hist_bw 1000 & 4000}
\end{figure}

\begin{figure}
    \centering
    \includegraphics[width=\myfigwidth]{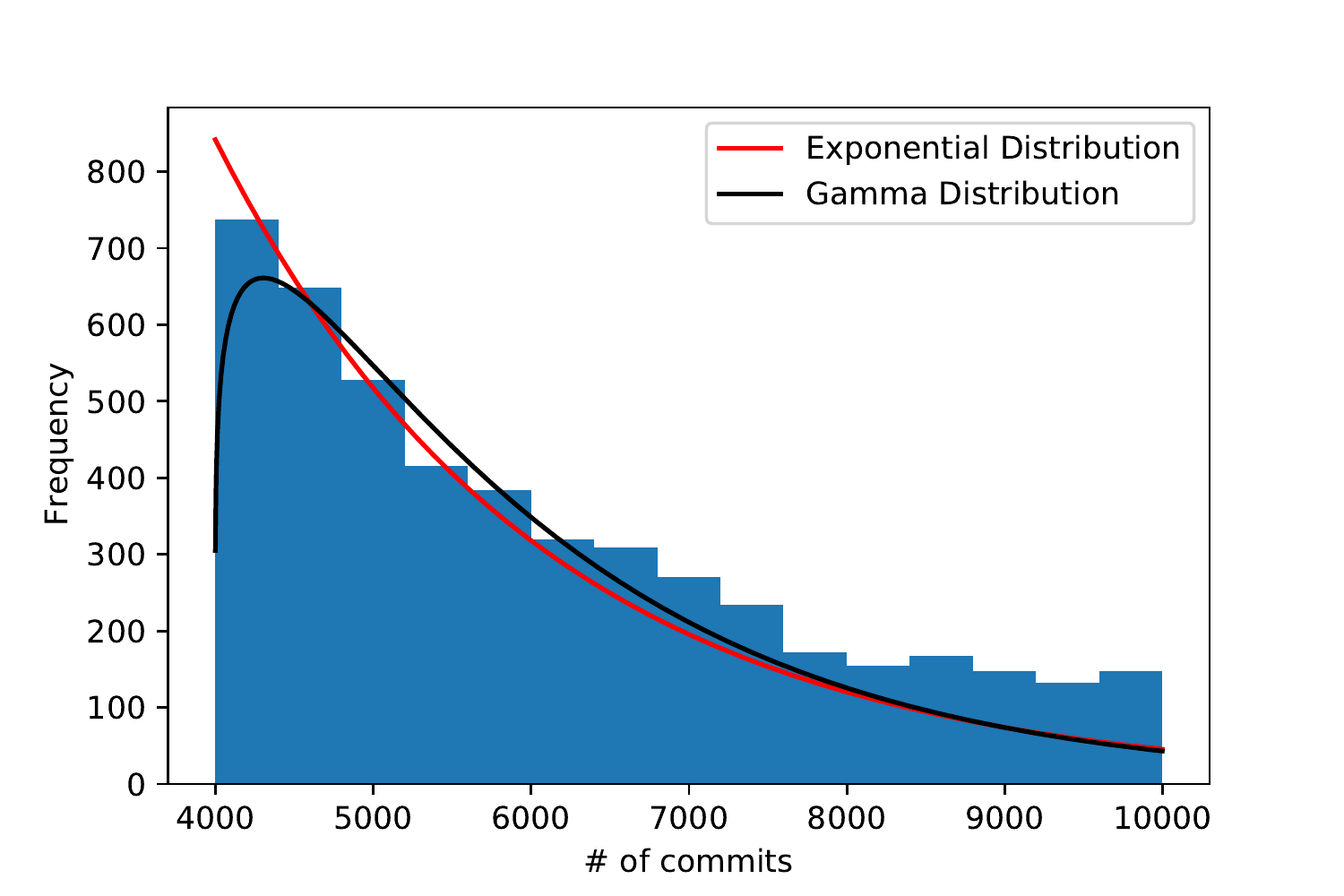}
    \caption{Histogram number of commits between 4K and 10K}
    \label{fig:452M_hist_bw 4000 & 10000}
\end{figure}

\begin{figure}
    \centering
    \includegraphics[width=\myfigwidth]{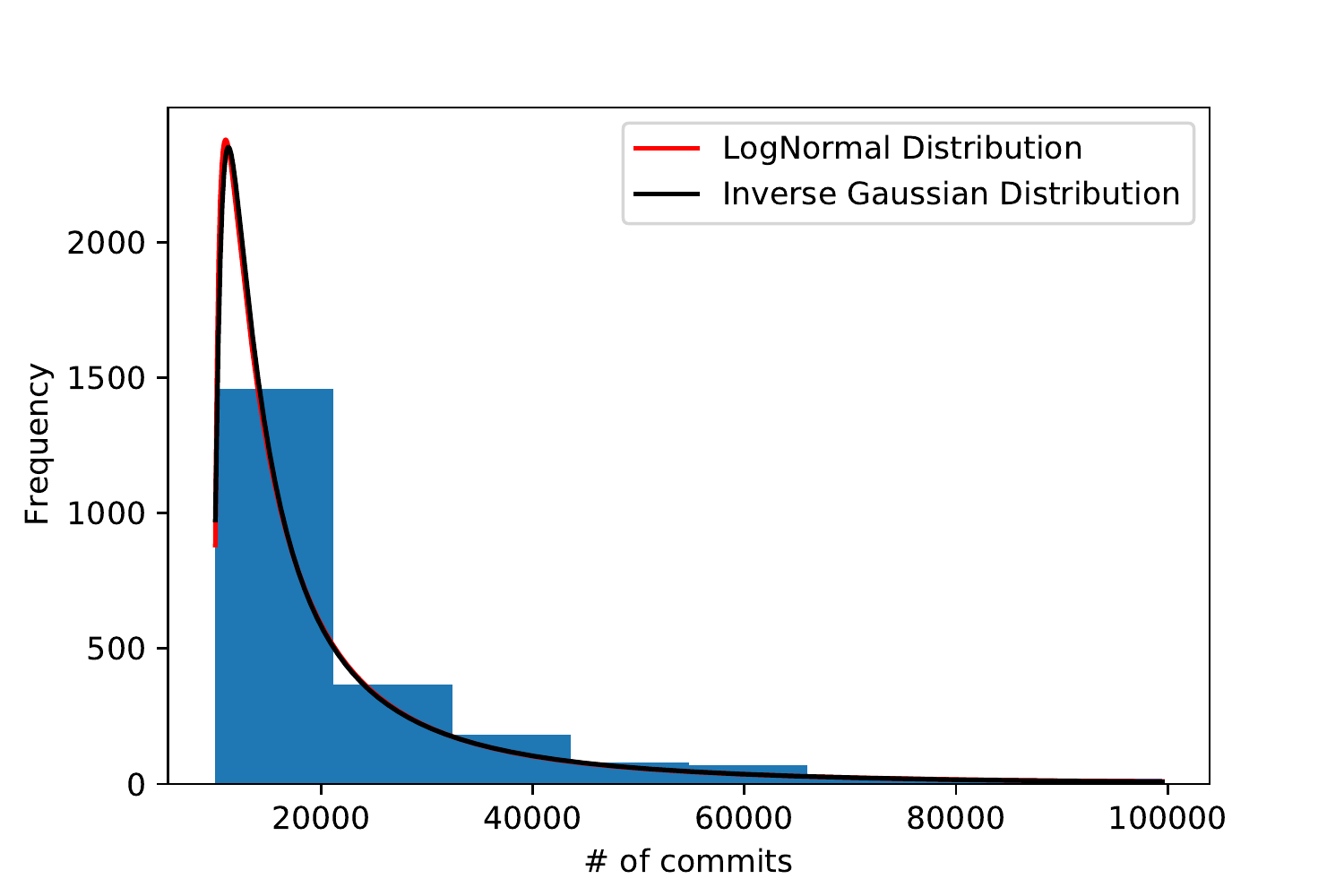}
    \caption{Histogram number of commits between 10K and 100K}
    \label{fig:452M_hist_bw 10000 & 100k}
\end{figure}

\begin{figure}
    \centering
    \includegraphics[width=\myfigwidth]{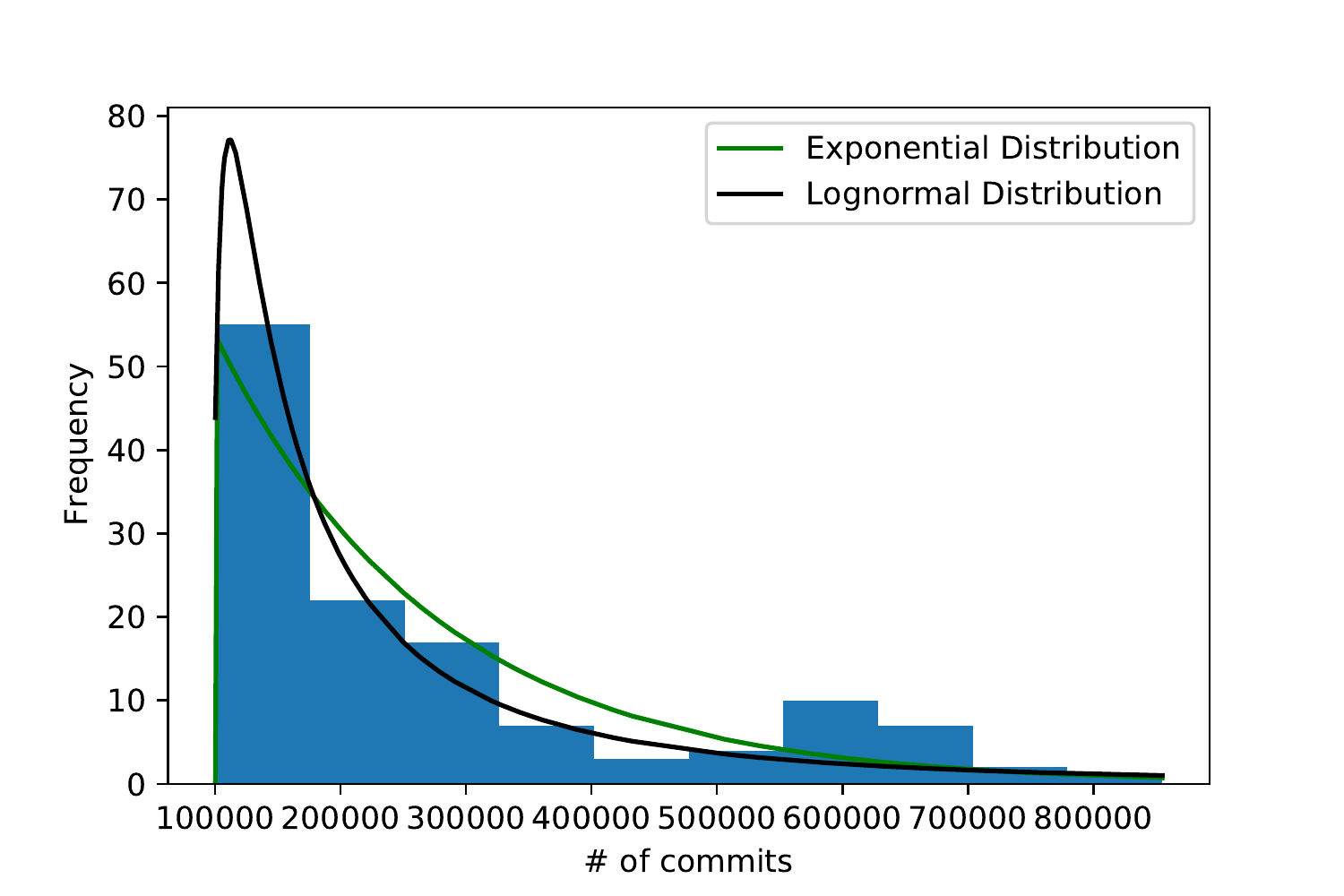}
    \caption{Histogram number of commits greater than 100K}
    \label{fig:452M_hist_greater than 100k}
\end{figure}

Figure~\ref{fig:452M_hist_bw 20 & 100} shows the histogram of number of commits between 20 and 100 (very low activity repositories). In this figure (and all other related figures), we overlay the Probability Density Functions (PDFs) of the best fitting distributions on the histogram.
Figure~\ref{fig:452M_hist_bw 100 & 1000} shows the histogram of number of commits between 100 and 1000 (low activity repositories).
Figure~\ref{fig:452M_hist_bw 1000 & 4000} shows the histogram of number of commits between 1000 and 4000 (medium activity repositories).
Figure~\ref{fig:452M_hist_bw 4000 & 10000} shows the histogram of number of commits between 4000 and 10000 (medium activity repositories).
Figure~\ref{fig:452M_hist_bw 10000 & 100k} shows the histogram of number of commits between 10000 and 100K (high activity repositories).
Figure~\ref{fig:452M_hist_greater than 100k} shows the histogram of number of commits greater than 100K (very high activity repositories).

\begin{figure}
    \centering
    \includegraphics[width=\myfigwidth]{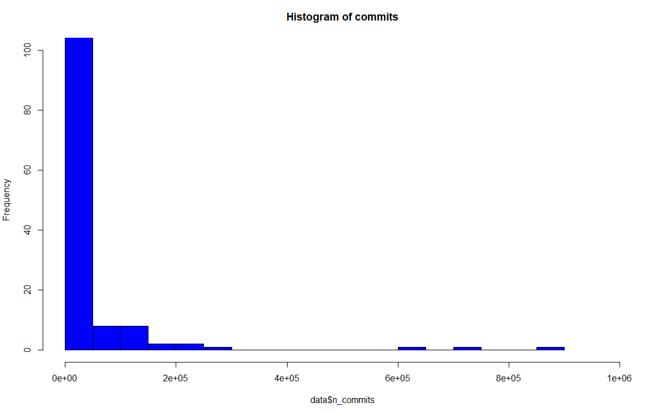}
    \caption{Histogram of all the commits to all repositories}
    \label{fig:All commits}
\end{figure}

Figure~\ref{fig:All commits} shows the histogram of all the commits.

\subsection{Distributions of Contributors to Repositories}

\begin{figure}
    \centering
    \includegraphics[width=\myfigwidth]{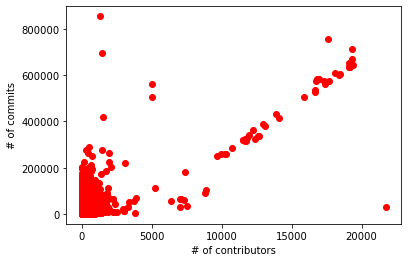}
    \caption{Number of Commits vs Number of Contributors}
    \label{fig:452M_CommitsVsContri}
\end{figure}

\begin{figure}
    \centering
    \includegraphics[width=\myfigwidth]{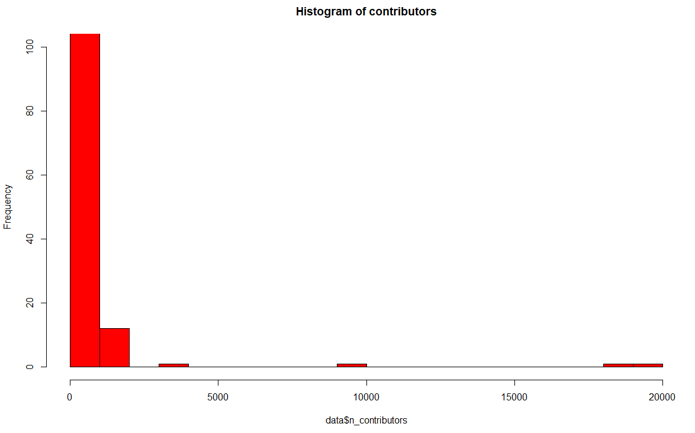}
    \caption{Histogram of the numbers of contributors to the repositories}
    \label{fig:All contributors}
\end{figure}

Figure~\ref{fig:452M_CommitsVsContri} shows the relationship between number of commits and number of contributors across all the repositories.
Figure~\ref{fig:All contributors} shows the histogram of all the contributors.

\section{StackOverFlow Developer Dataset}
Stack Overflow conducts an annual survey of developers across the globe every year. We used the 2019, 2020 and 2021 datasets to find the best fit distributions to the professional years of experience of coders. The datasets can be found through this link: \href{https://insights.stackoverflow.com/survey}{StackOverflow Dataset}

\subsection{Overview and Preprocessing of the Data}

\subsubsection{Preprocessing 2021 Dataset}

In the preprocessing stage, we dropped all rows that contain N/A values or coders with less than 1 year of experience or more than 50 years of experience. And to categorize the professionals from all coders, we used the column named ‘Main Branch’ and restricted the analysis to those participants who marked ‘I am a developer by profession’.

\subsubsection{Preprocessing 2020 Dataset}
We started our preprocessing of the data by converting ``Less than 1 year'' values to 0, ``More than 50 years'' to 51 and dropping ``NA'' values from the column ``YearsCodePro''. We also considered only ``Employed full-time'' entries for ``Employment'' column. Further, values in the range of 30-90 were considered for the column ``WorkWeekHrs''. 

\subsubsection{Preprocessing for Comparison of Datasets}

In order to be able to make comparison across the three years, we normalize the comparison by using the same preprocessing method as used for the 2021 dataset.  In other words, the preprocessing steps described earlier for the 2021 dataset are applied to the 2019 and 2020 datasets also, and the resulting data is used for direct comparison across the years.

\subsection{Professional Coder Distributions in 2021}

\begin{figure}
    \centering
    \includegraphics[width=\myfigwidth]{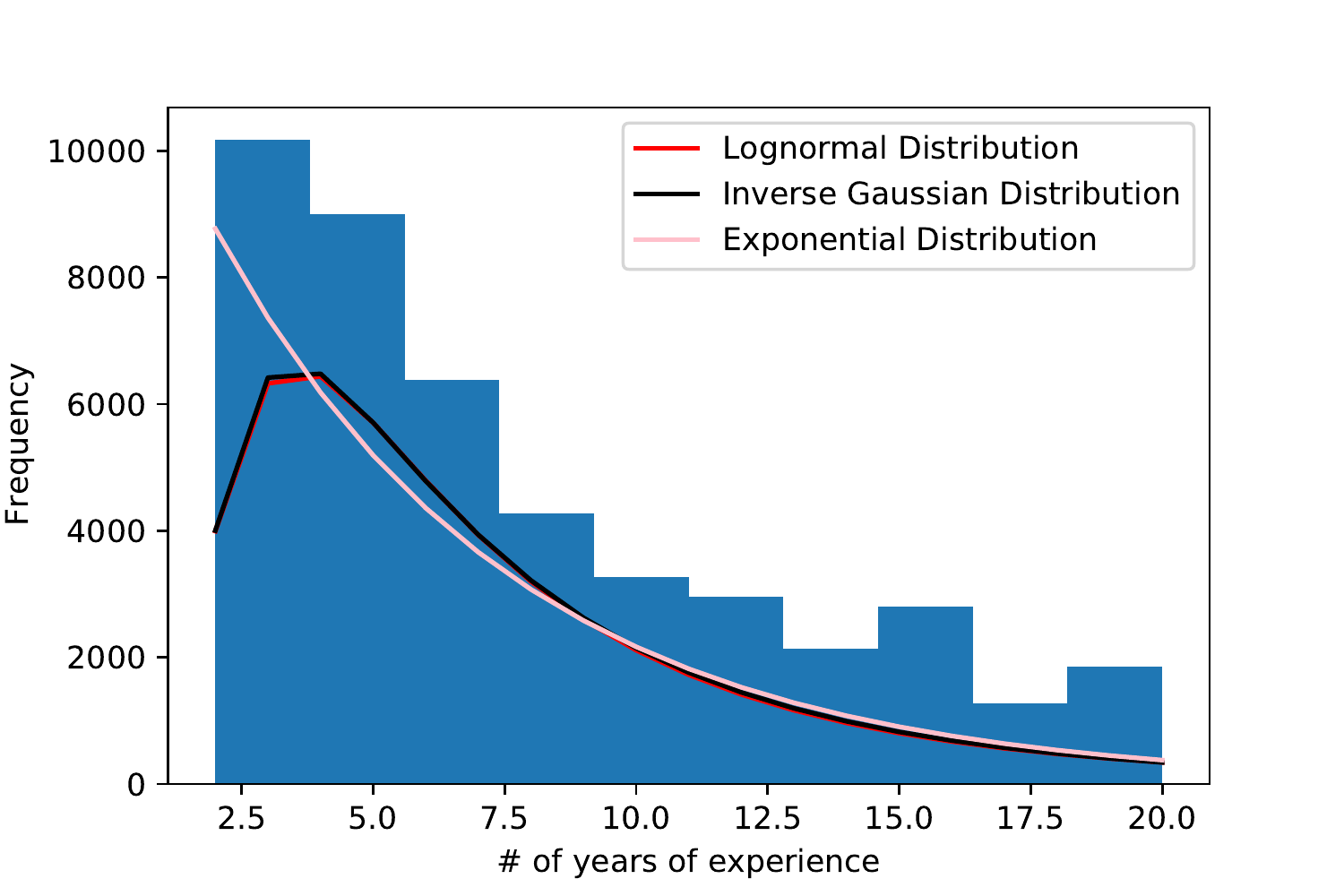}
    \caption{Histogram of professional coders between 2 and 20 years of experience}
    \label{fig:Coders_2and20}
\end{figure}

\begin{figure}
    \centering
    \includegraphics[width=\myfigwidth]{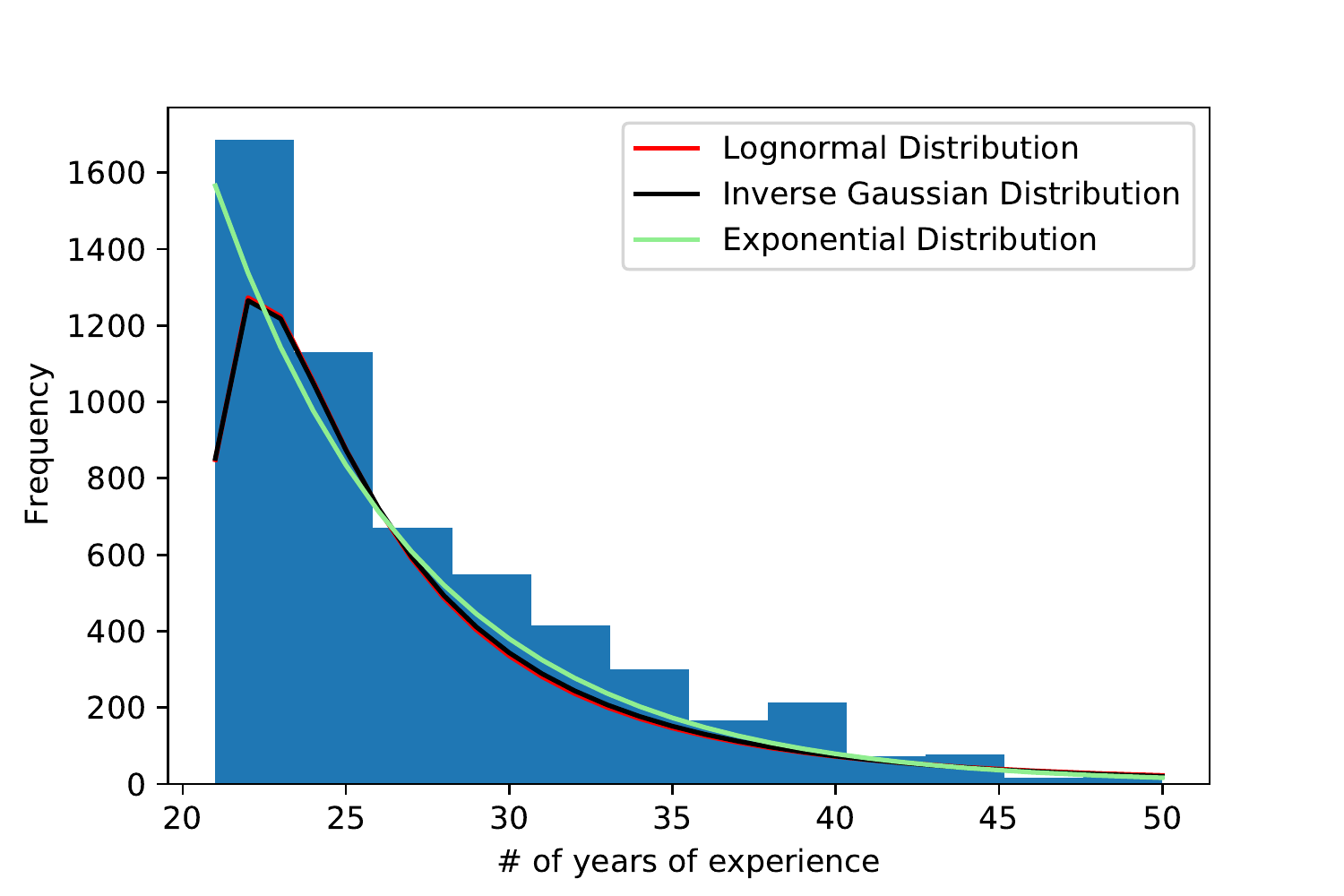}
    \caption{Histogram of professional coders with more than 20 years of experience}
    \label{fig:Coders_morethan20}
\end{figure}

\begin{figure}
    \centering
    \includegraphics[width=\myfigwidth]{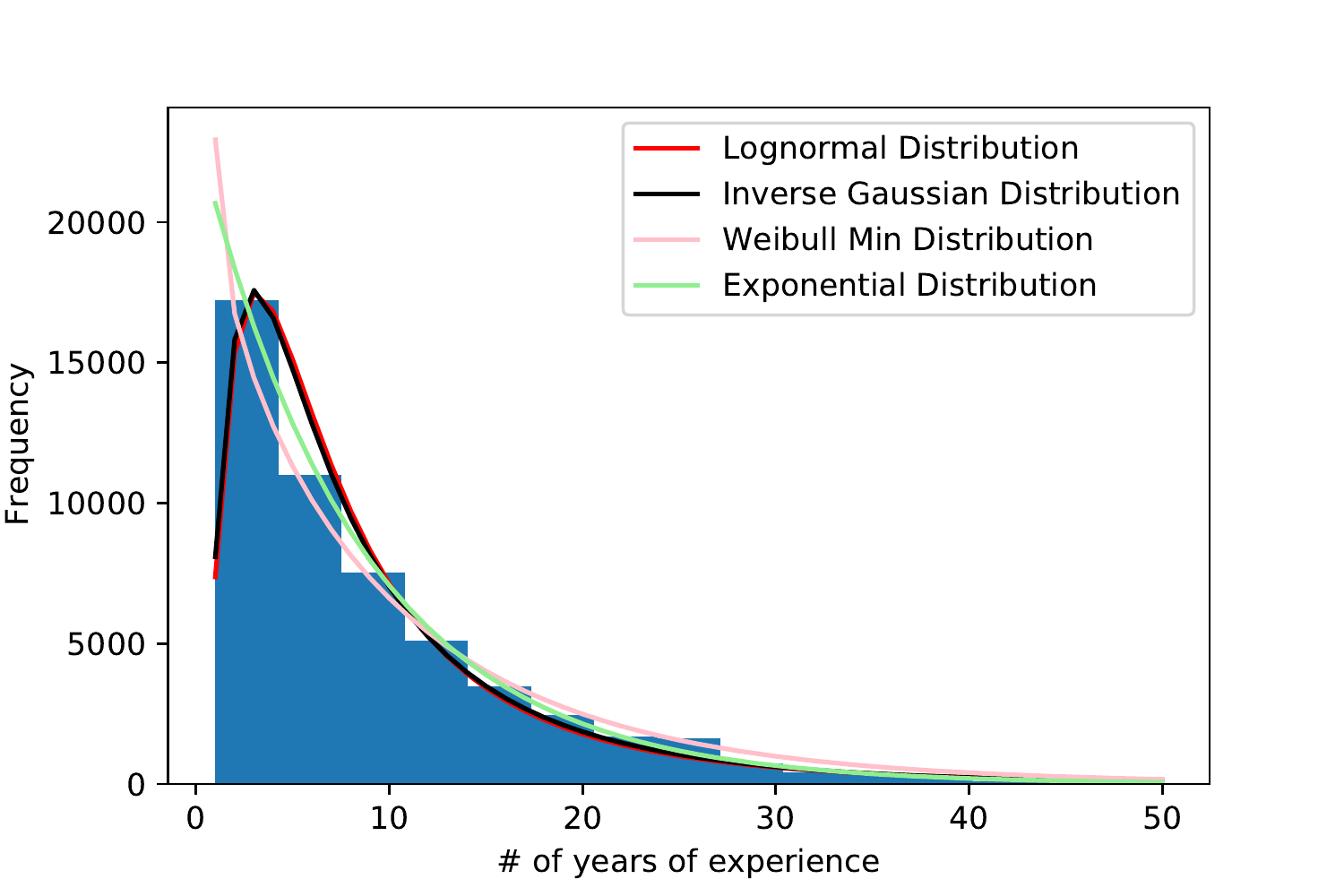}
    \caption{Histogram of all the professional coders}
    \label{fig:Professionals}
\end{figure}

\begin{figure}
    \centering
    \includegraphics[width=\myfigwidth]{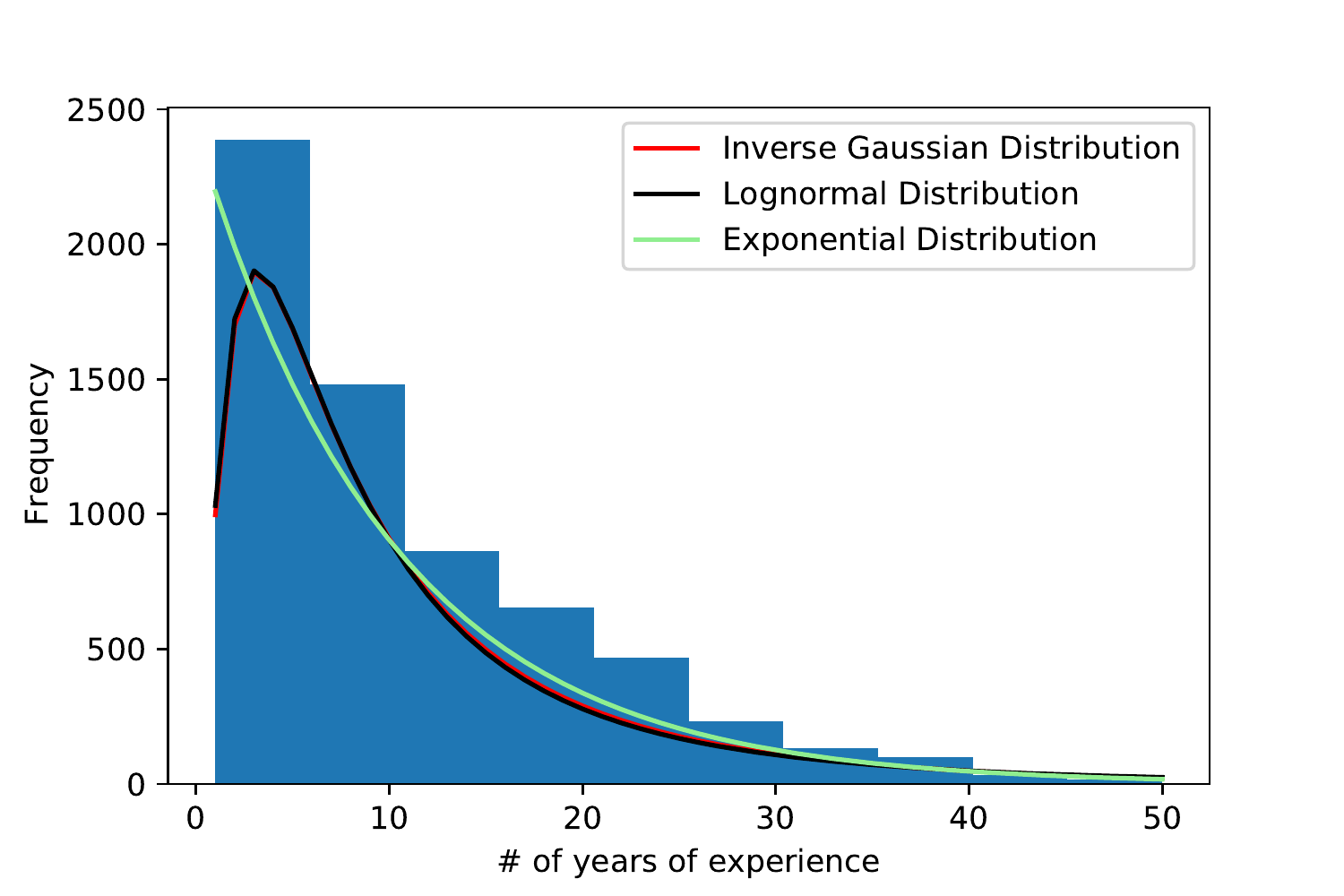}
    \caption{Histogram of non-professional coders}
    \label{fig:Non-Professionals}
\end{figure}

Figure~\ref{fig:Coders_2and20} shows the histogram of professional coders with years of experience between 2 and 20 years.
Figure~\ref{fig:Coders_morethan20} shows the histogram of professional coders with more than 20 years of experience.
Figure~\ref{fig:Professionals} shows the histogram of all the Professional coders with more than 1 year of experience.
Figure~\ref{fig:Non-Professionals} shows the histogram of all the Professional coders with more than 1 year of experience.
Table~\ref{tbl:stackoverflow-2021-coder-distributions} captures all the distributions from these figures into best-fit and second-best-fit distributions for all the groups.

\begin{table}[h!]
\caption{Distributions of coders in the StackOverflow 2021 dataset}
\label{tbl:stackoverflow-2021-coder-distributions}
\begin{tabular}{|C{0.3\textwidth}|C{0.3\textwidth}|C{0.3\textwidth}|}\hline
\textbf{Group} & \textbf{Best fit}& \textbf{Second best fit} \\\hline\hline
Professional coders bw 2 and 20 & Log-Normal (0.89, -1.29, 0.92) & Inverse Gaussian (0.77, -1.39, 1.8)\\\hline
Professional coders more than 20 & Inverse Gaussian (0.71, -1.42, 2.00) & Log-Normal (0.84, -1.32, 0.97)\\\hline
All professional coders & Log-Normal (0.91, -1.13, 0.78)  & Inverse Gaussian (0.89, -1.20, 1.36)  \\\hline
Non- professional coders & Exponential (-1.04,1.04)    & Beta: Visually looks weird so dropped \\\hline
\end{tabular}
\end{table}

\subsection{Professional Coder Distributions in 2020}

Here, we start with determining the frequency histograms of the coders which provides the visual indication of the potential distributions to attempt in fitting the data. This is described in Section~\ref{sec:SO2020-freqhist}.

For the rest of the analyses in this section, the \texttt{fitdistrplus} package in the R statistical analysis software is used.

We compute the skewness-kurtosis measures in order to uncover specific candidate distributions to further narrow down the potential distributions.  This is achieved using the \texttt{descdist} function of the R package.  The results from this step are described in Section~\ref{sec:SO2020-skewness}.

The dataset is then tested against fitted distributions (i.e., Normal, Poisson and Negative Binomial) using the \texttt{fitdist} function, which provides the Log-likelihood, Akaike information criterion (AIC), and Bayesian information criterion (BIC) values. This step is described in Section~\ref{sec:SO2020-fitofdist}.

Following this, we generated the P-P plots for theoretical probabilities against the distribution of the empirical data (this is achieved using the \texttt{ppcomp} function).
The CDF plots for empirical cumulative distribution are generated and plotted against fitted distribution functions (this is achieved using the \texttt{cdfcomp} function).  These plots are presented in Section~\ref{sec:SO2020-closenessoffit}.

The density plots for the histogram are fitted against density functions (this is performed using \texttt{denscomp} function). This is described in Section~\ref{sec:SO2020-densityoverlay}.

Finally, considering all the statistical analyses, we select the best fitting distribution for each professional coding year range.  This is described in Section~\ref{sec:SO2020-bestfits}.
 
\subsubsection{Frequency Histograms of Professional Coders}
\label{sec:SO2020-freqhist}

\begin{figure}
    \centering
    \includegraphics[width=12 cm, height= 7 cm]{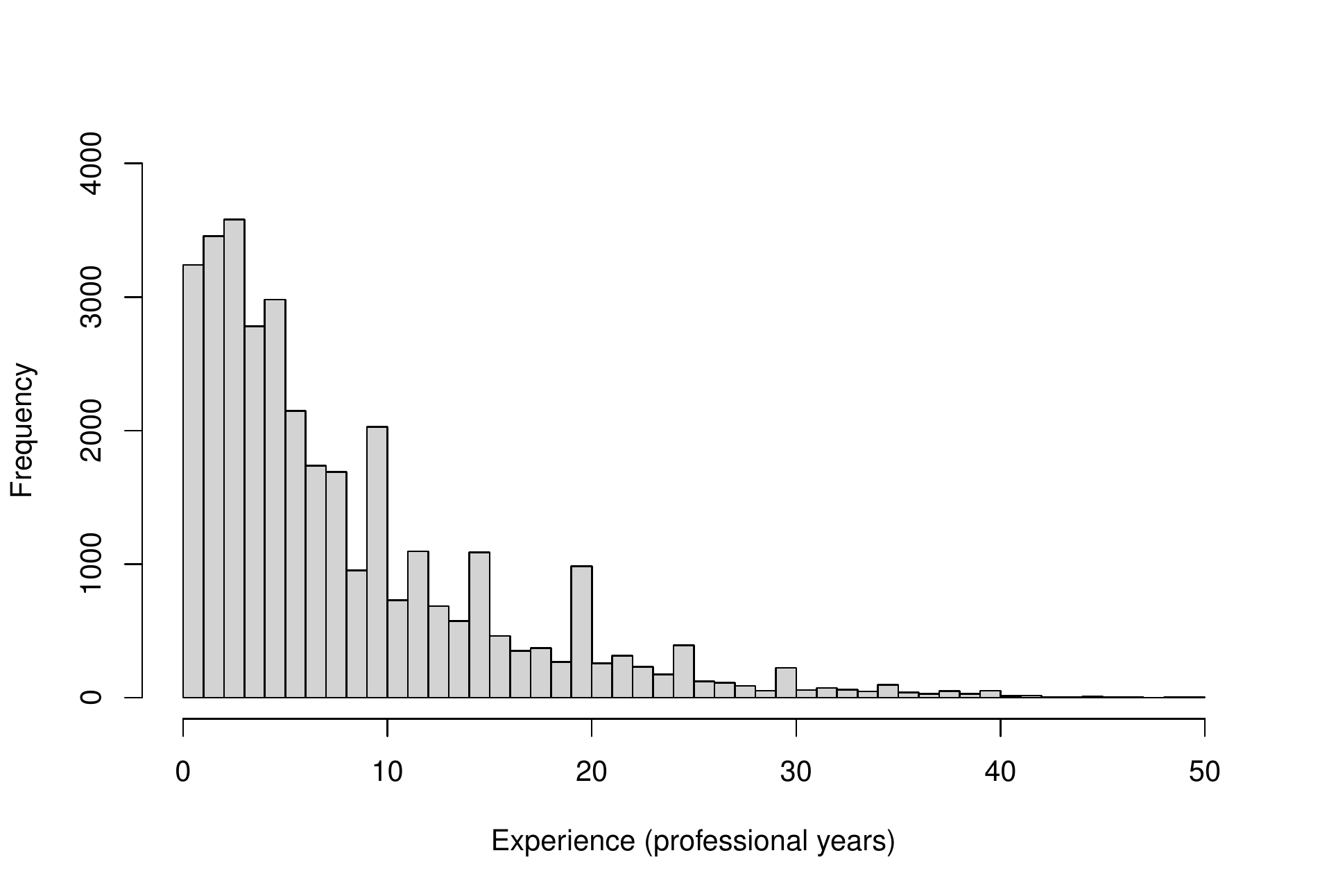}
    \caption{Frequency histogram of professional coding years (0-50)}
    \label{fig:freq_hist_0_50}
\end{figure}

\begin{figure}
    \centering
    \includegraphics[width=12 cm, height= 8 cm]{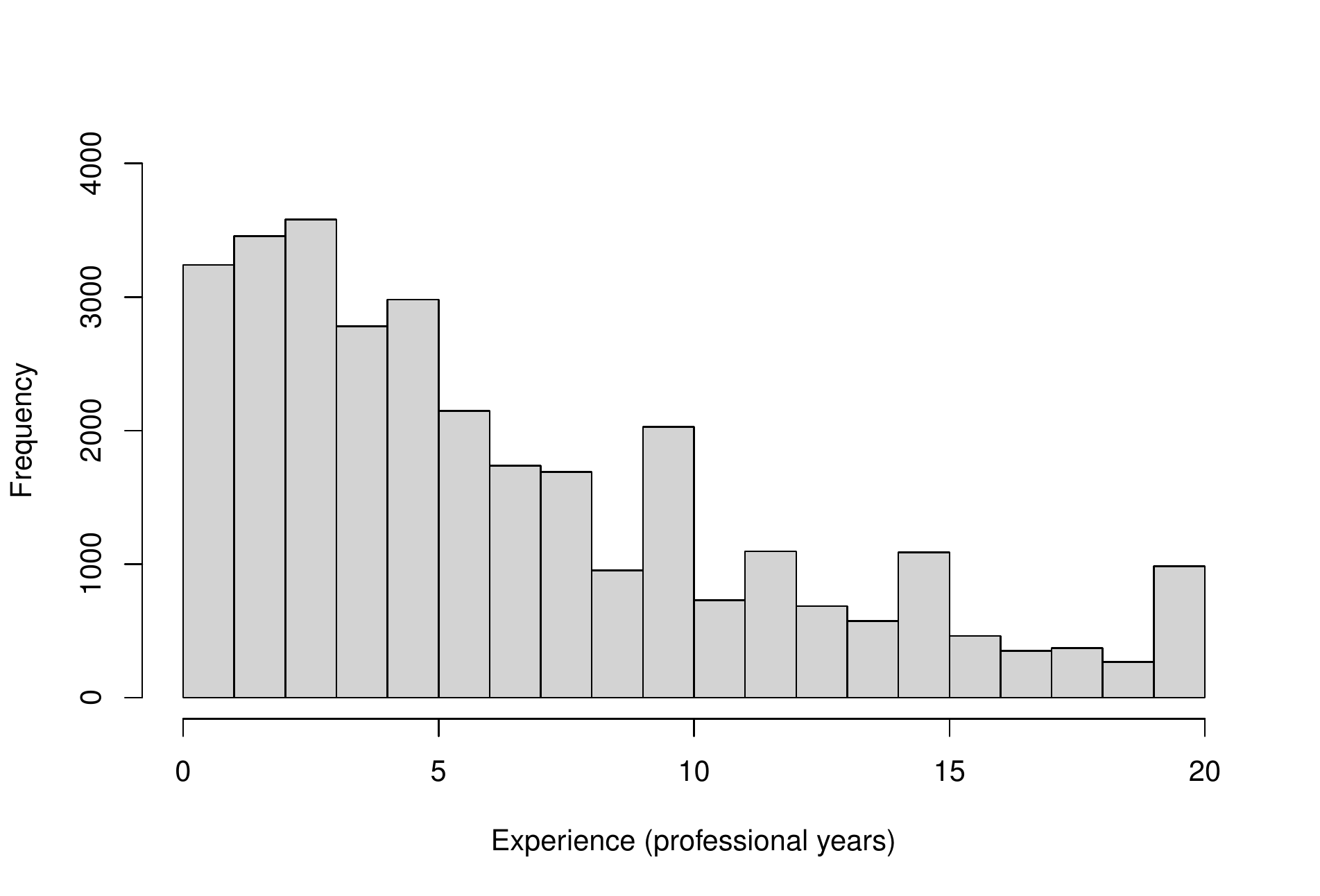}
    \caption{Frequency histogram of professional coding years (0-20)}
    \label{fig:hist_freq_0_20}
\end{figure}

\begin{figure}
    \centering
    \includegraphics[width=12 cm, height= 8 cm]{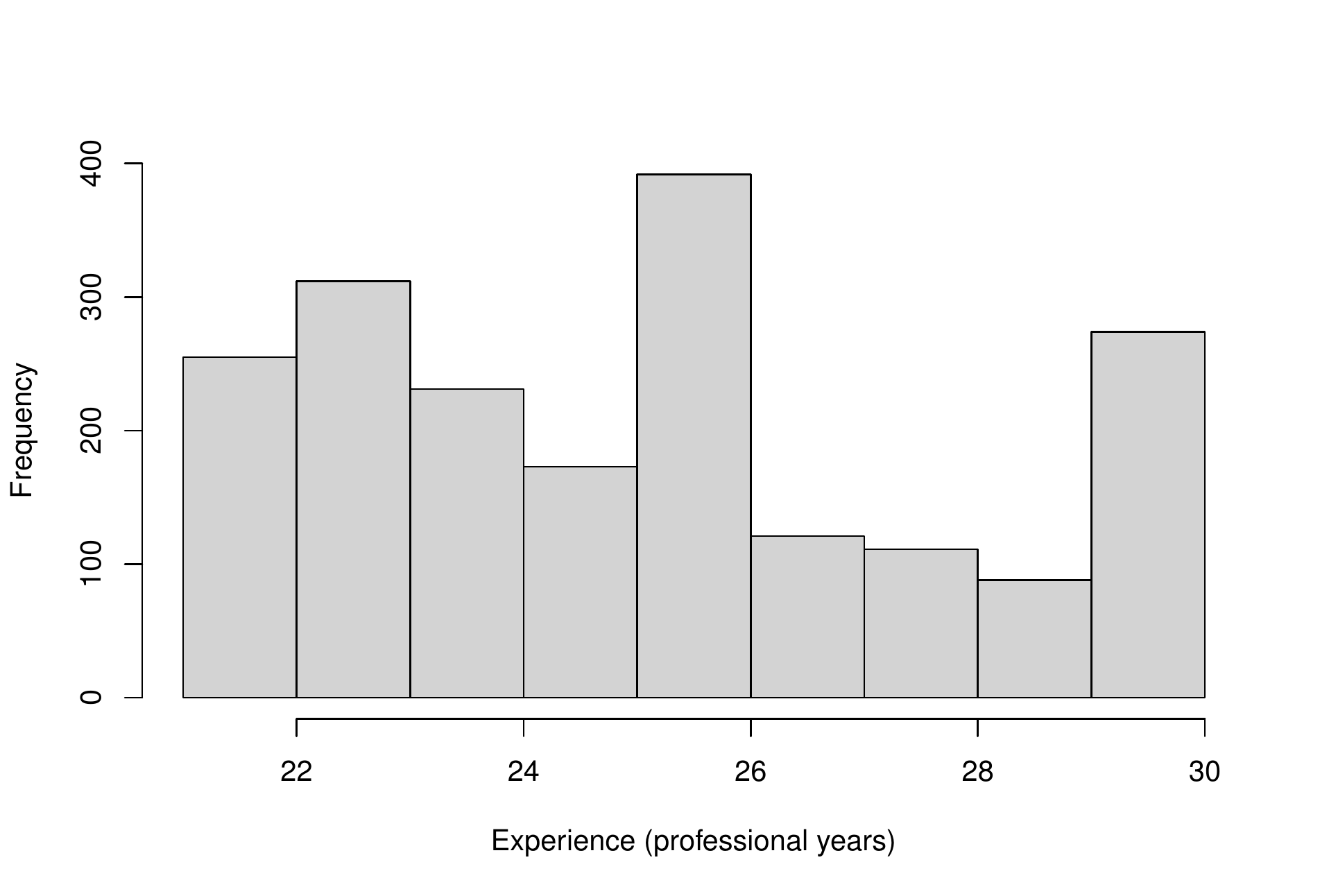}
    \caption{Frequency histogram of professional coding years (21-30)}
    \label{fig:hist_freq_21_30}
\end{figure}

\begin{figure}
    \centering
    \includegraphics[width=12 cm, height= 8 cm]{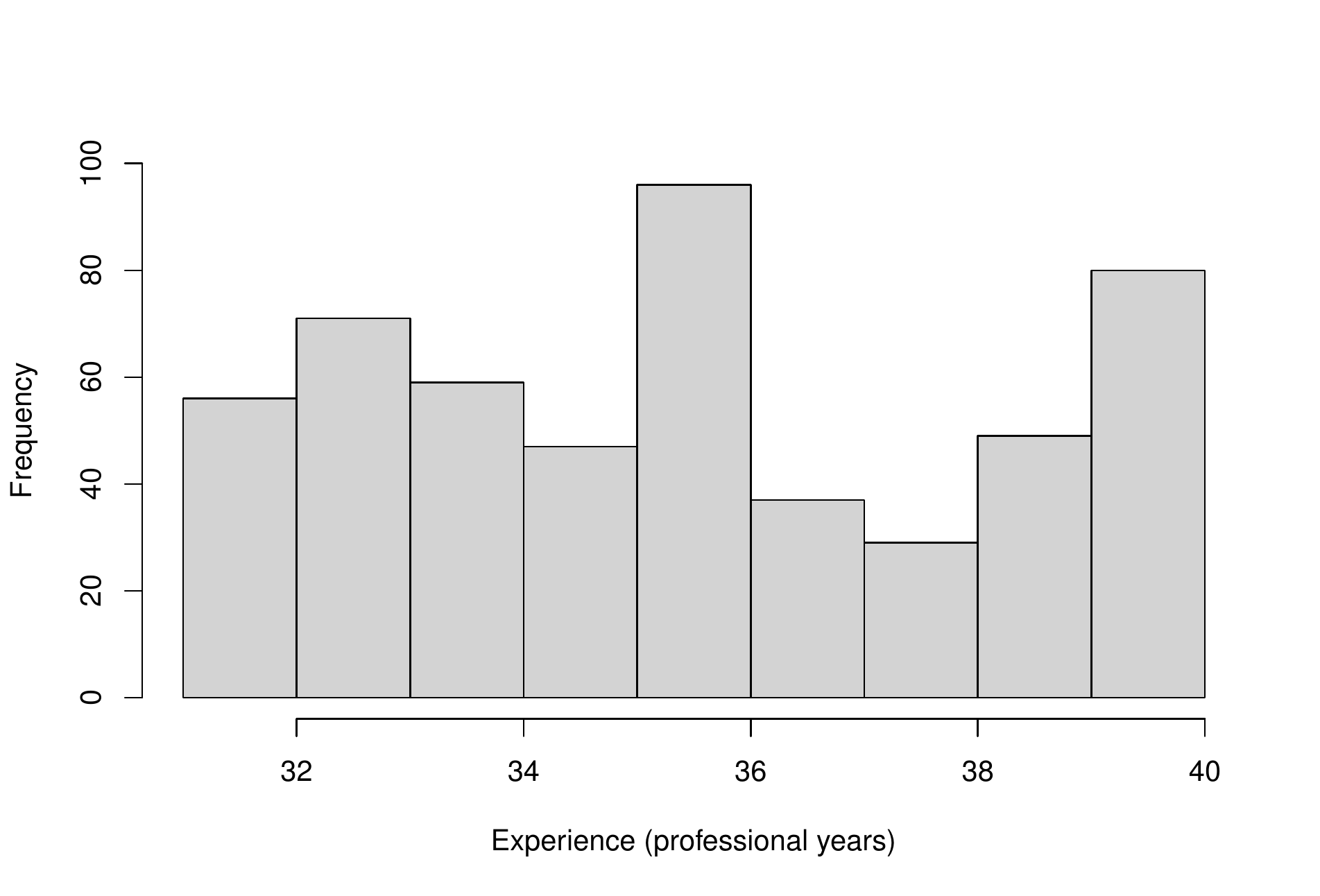}
    \caption{Frequency histogram of professional coding years (31-40)}
    \label{fig:hist_freq_31_40}
\end{figure}

Figure~\ref{fig:hist_freq_0_20},
Figure~\ref{fig:hist_freq_21_30}, and
Figure~\ref{fig:hist_freq_31_40}
show the frequency histogram of professional coders with
0-20, 21-30, and 31-40 years of working experience, respectively.
Figure~\ref{fig:freq_hist_0_50} captures the same for the broader range of 0-50 years working experience. 

\begin{figure}
    \centering
    \includegraphics[width=12 cm, height= 7.5 cm]{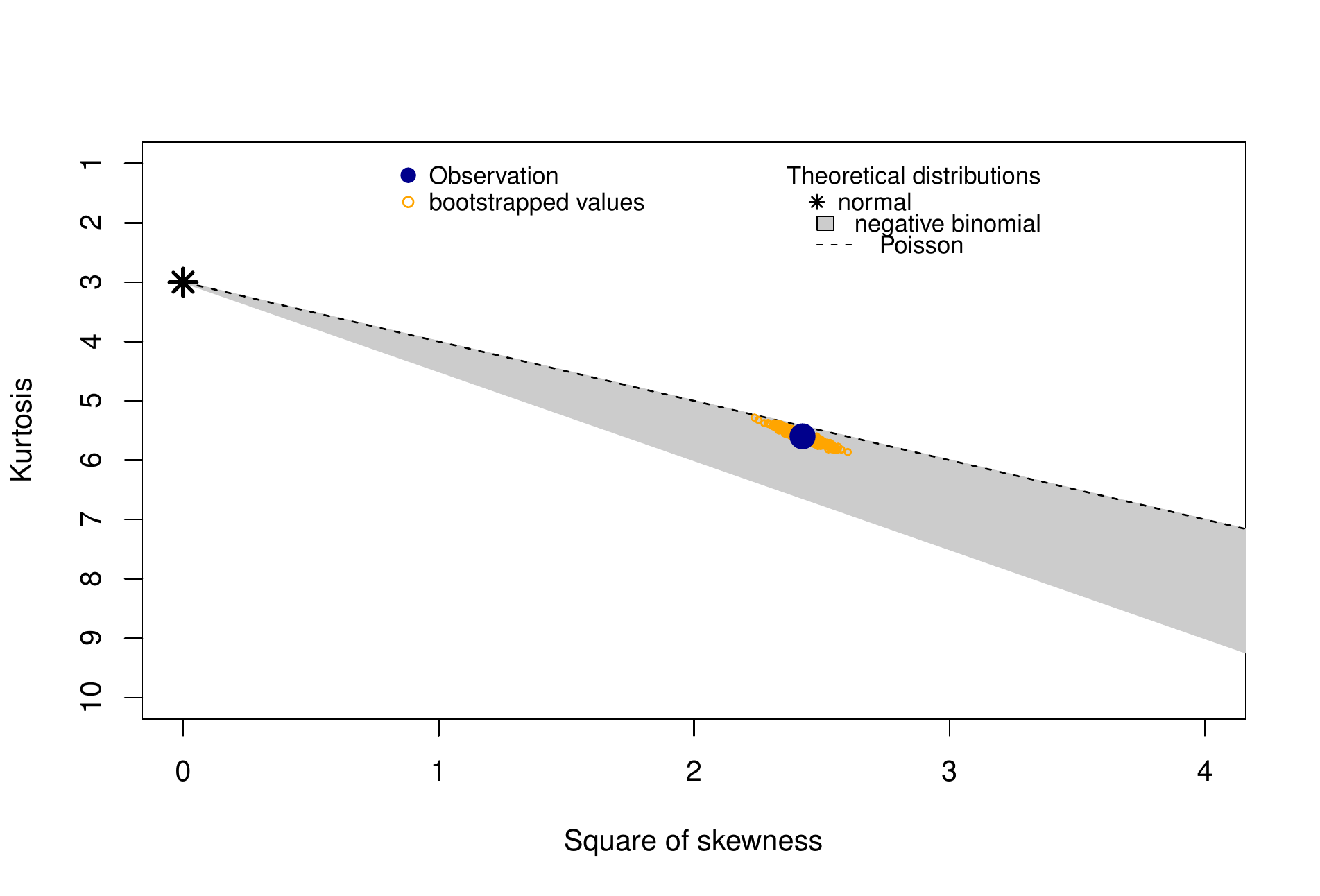}
    \caption{Skewness-kurtosis plot of professional coding years (0-50)}
    \label{fig:cullen_frey_0_50}
\end{figure}

\subsubsection{Determining the Potential Matching Distributions}
\label{sec:SO2020-skewness}

\begin{figure}
    \centering
    \includegraphics[width=12 cm, height= 7 cm]{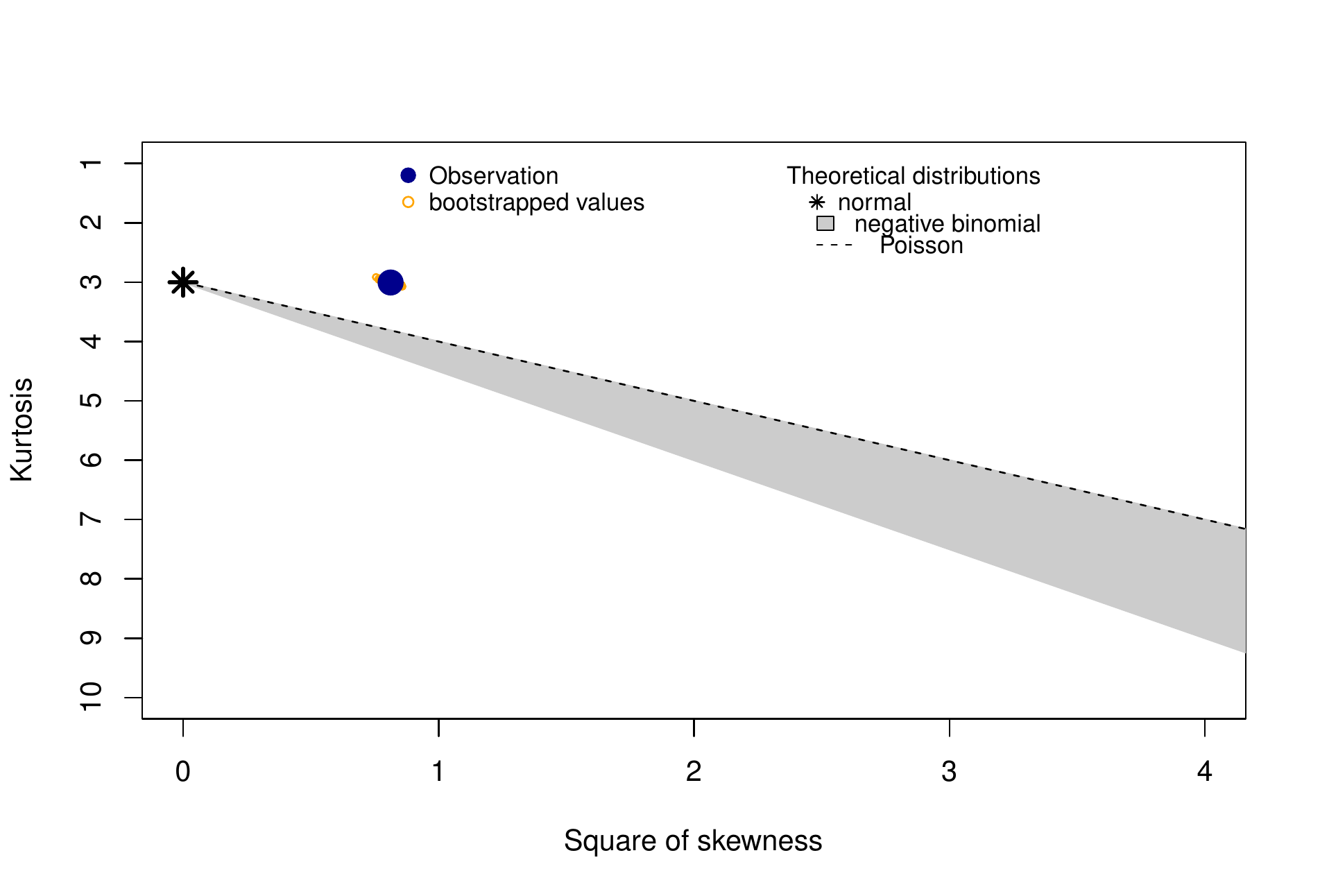}
    \caption{Skewness-kurtosis plot of professional coding years (0-20)}
    \label{fig:cullen_frey_0_20}
\end{figure}

\begin{figure}
    \centering
    \includegraphics[width=12 cm, height= 7 cm]{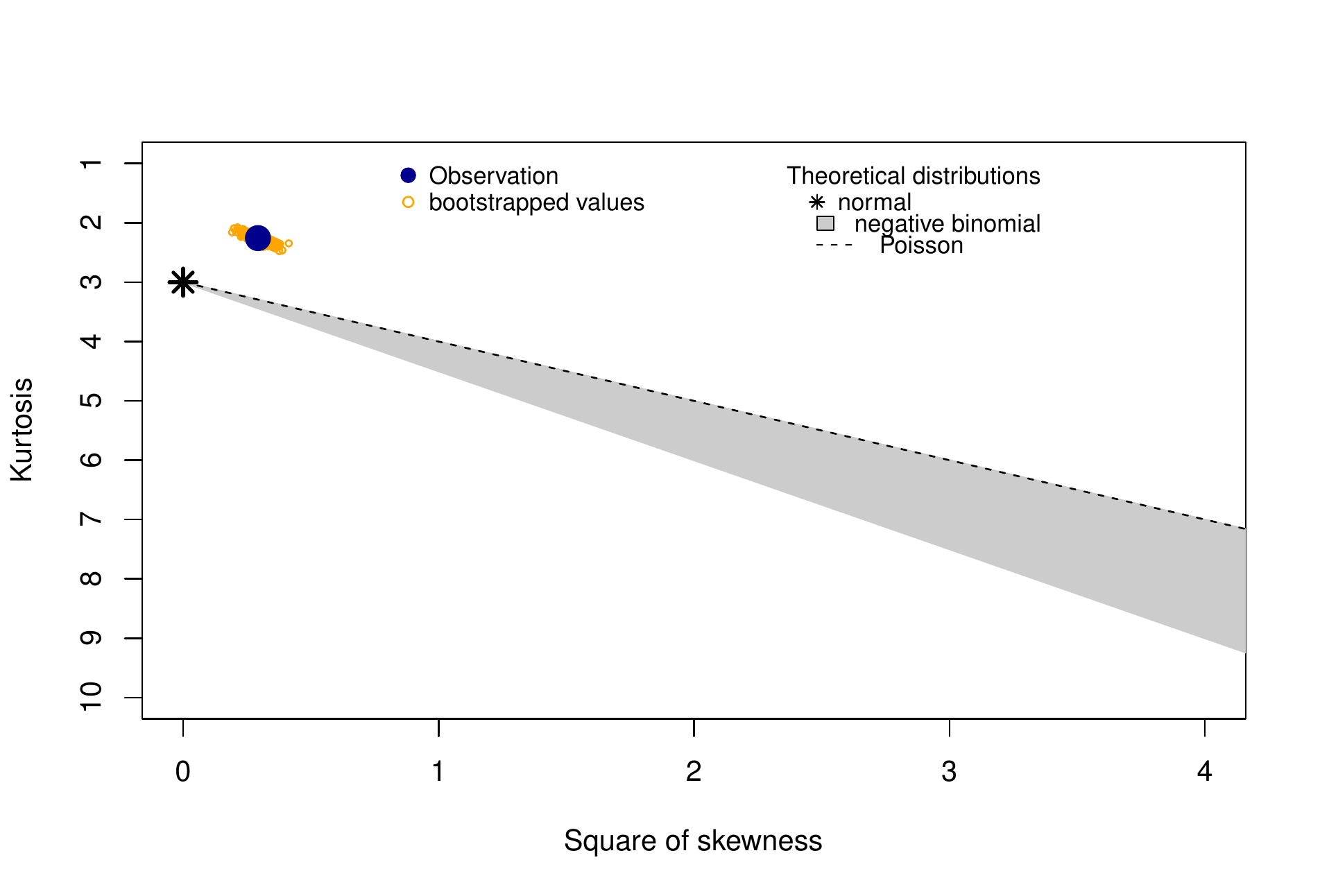}
    \caption{Skewness-kurtosis plot of professional coding years (21-30)}
    \label{fig:cullen_frey_21_30}
\end{figure}

\begin{figure}
    \centering
    \includegraphics[width=12 cm, height= 8 cm]{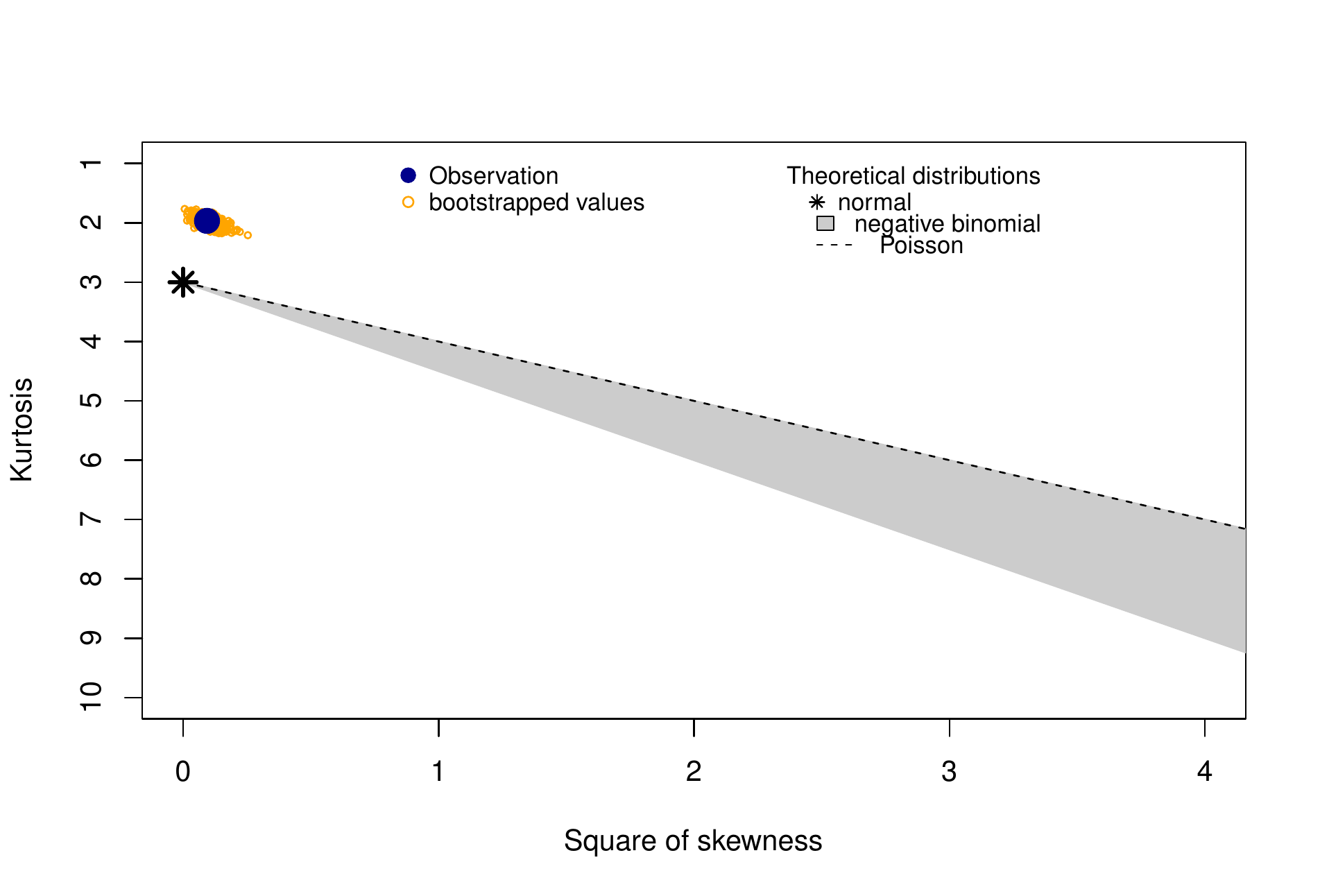}
    \caption{Skewness-kurtosis plot of professional coding years (31-40)}
    \label{fig:cullen_frey_31_40}
\end{figure}

For every group representing the coding experience range of the coders, the skewness-kurtosis plot of professional coders in that experience range is computed. This plot helps to visually determine which discrete distribution the data set follows more closely.
Figure~\ref{fig:cullen_frey_0_20},
Figure~\ref{fig:cullen_frey_21_30}, and
Figure~\ref{fig:cullen_frey_31_40} 
show the skewness-kurtosis plots of professional coders with 
0-20, 21-30, and 31-40 years of working experience, respectively.
Figure~\ref{fig:cullen_frey_0_50} captures the same for the broader range from 0 to 50 years of experience.

Table~\ref{tbl:2020-StackOverflow-Summary} shows the frequency of coders and the best distribution for each professional coding year range along with parameters. 
\begin{table}
    \centering
    \begin{tabular}{|C{0.25\textwidth}|C{0.25\textwidth}|C{0.40\textwidth}|}\hline
         Coding Year Range &  Count of Coders &  Best Fit \\\hline\hline
         0--50 &  33,734 & Neg. Binomial (1.59,8.33)\\\hline
         0--20 &  31,202 & Neg. Binomial (2.2,6.8)\\\hline
         21--30 &  1,957 & Normal (24.66,2.81)\\\hline
         31--40 &  524 & Normal (35.02,2.82)\\\hline
    \end{tabular}
    \caption{Best-fit distributions for coding year ranges}
    \label{tbl:2020-StackOverflow-Summary}
\end{table}

\subsubsection{Fit of Distributions}
\label{sec:SO2020-fitofdist}

\begin{figure}
    \centering
    \includegraphics[width=10 cm, height= 7 cm]{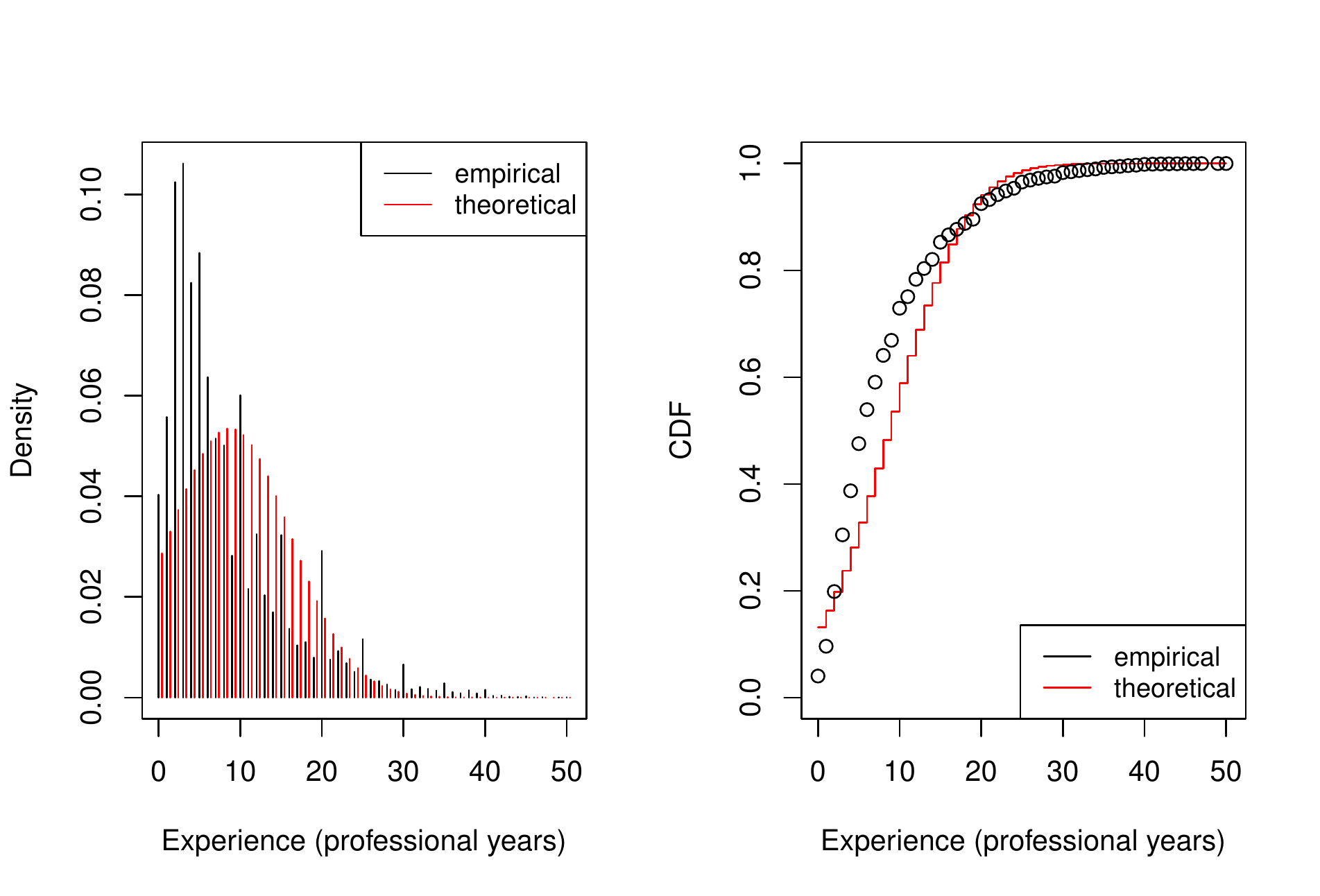}
    \caption{Density (left) and CDF (right) of empirical (data set) and Normal distribution for professional coding years 0-50}
    \label{fig:normal_0_50}
\end{figure}

\begin{figure}
    \centering
    \includegraphics[width=12 cm, height= 7.5 cm]{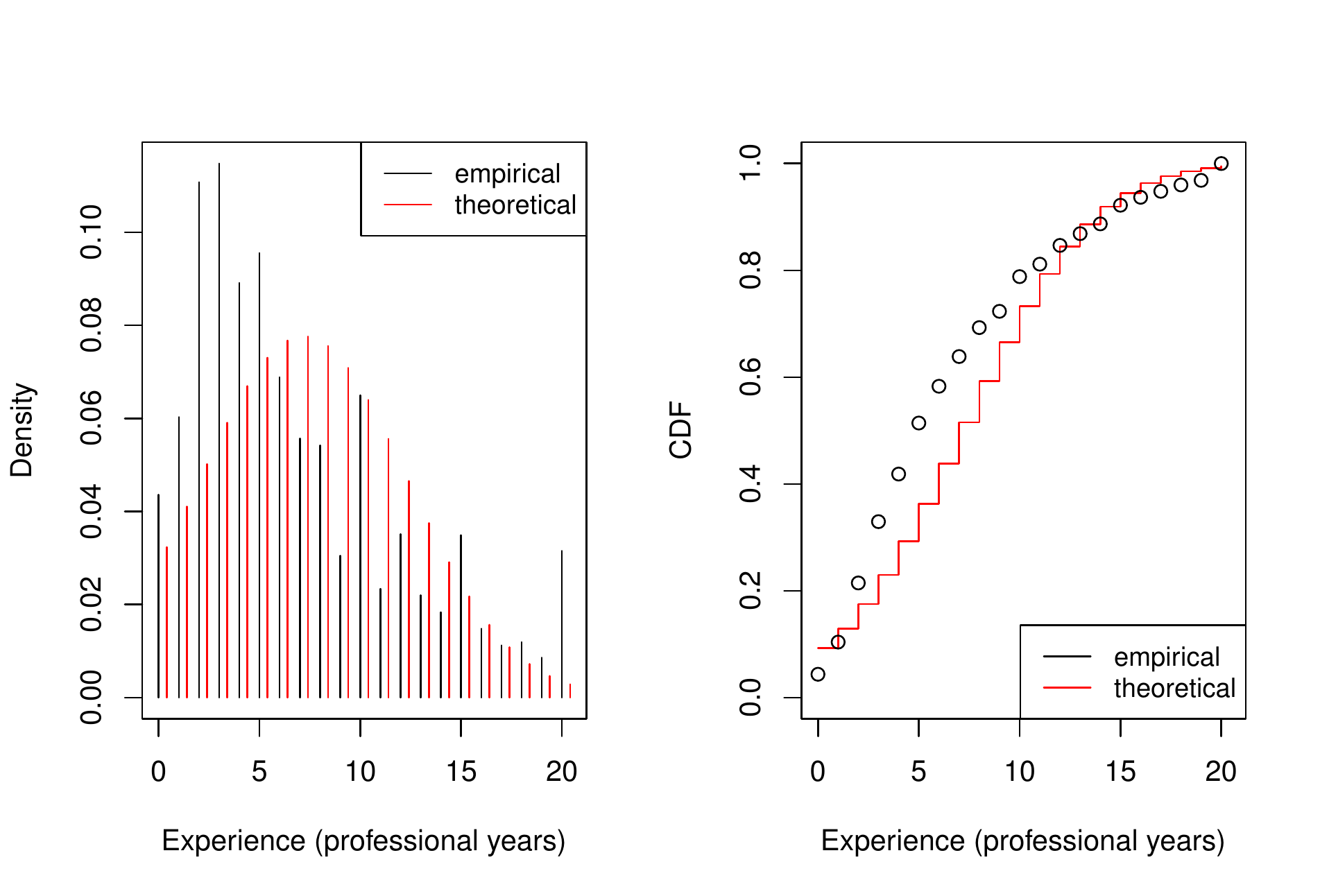}
    \caption{Density (left) and CDF (right) of empirical (data set) and Normal distribution for professional coding years 0-20}
    \label{fig:normal_0_20}
\end{figure}

\begin{figure}
    \centering
    \includegraphics[width=12 cm, height= 7.8 cm]{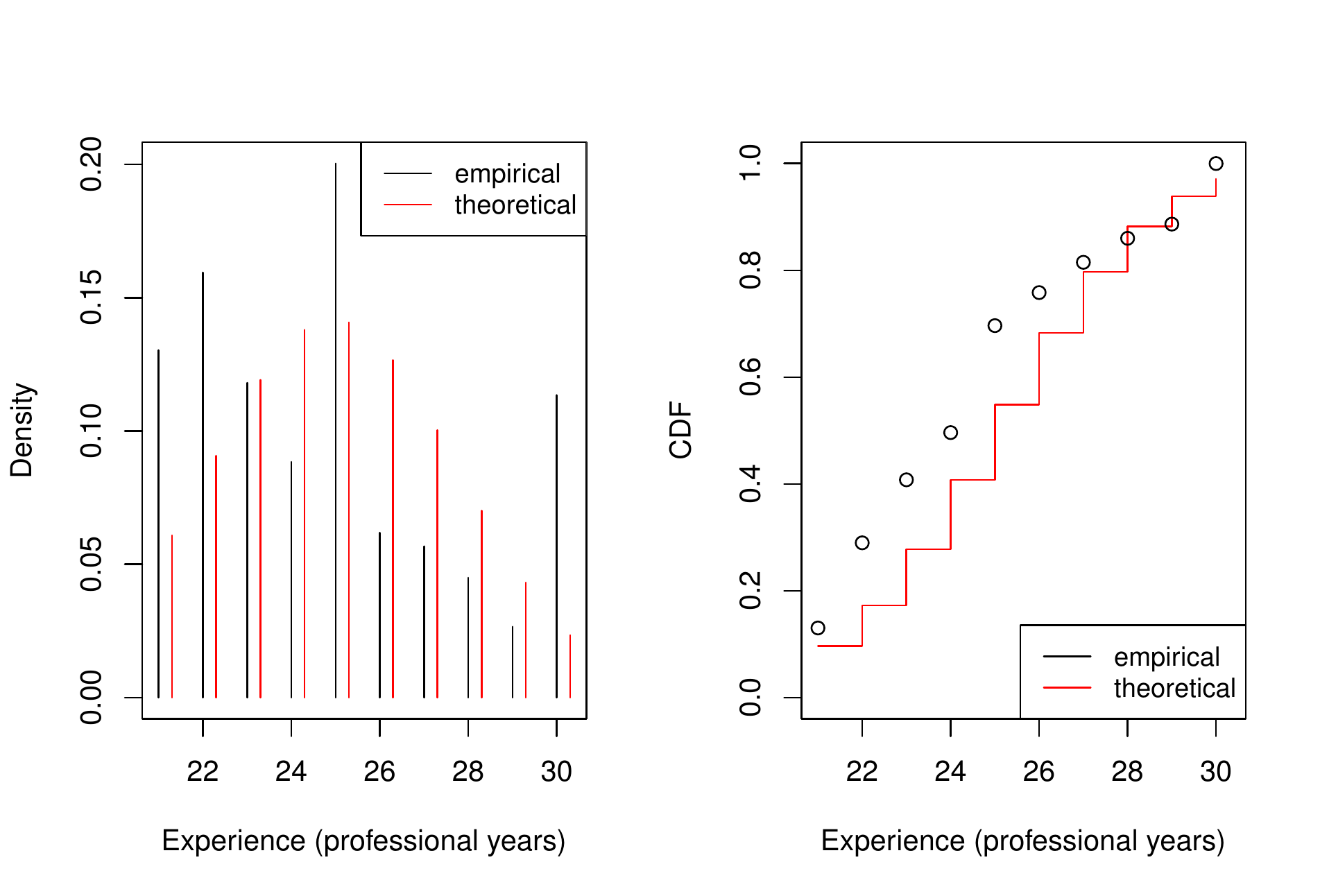}
    \caption{Density (left) and CDF (right) of empirical (data set) and Normal distribution for professional coding years 21-30}
    \label{fig:normal_21_30}
\end{figure}

\begin{figure}
    \centering
    \includegraphics[width=12 cm, height= 7.8 cm]{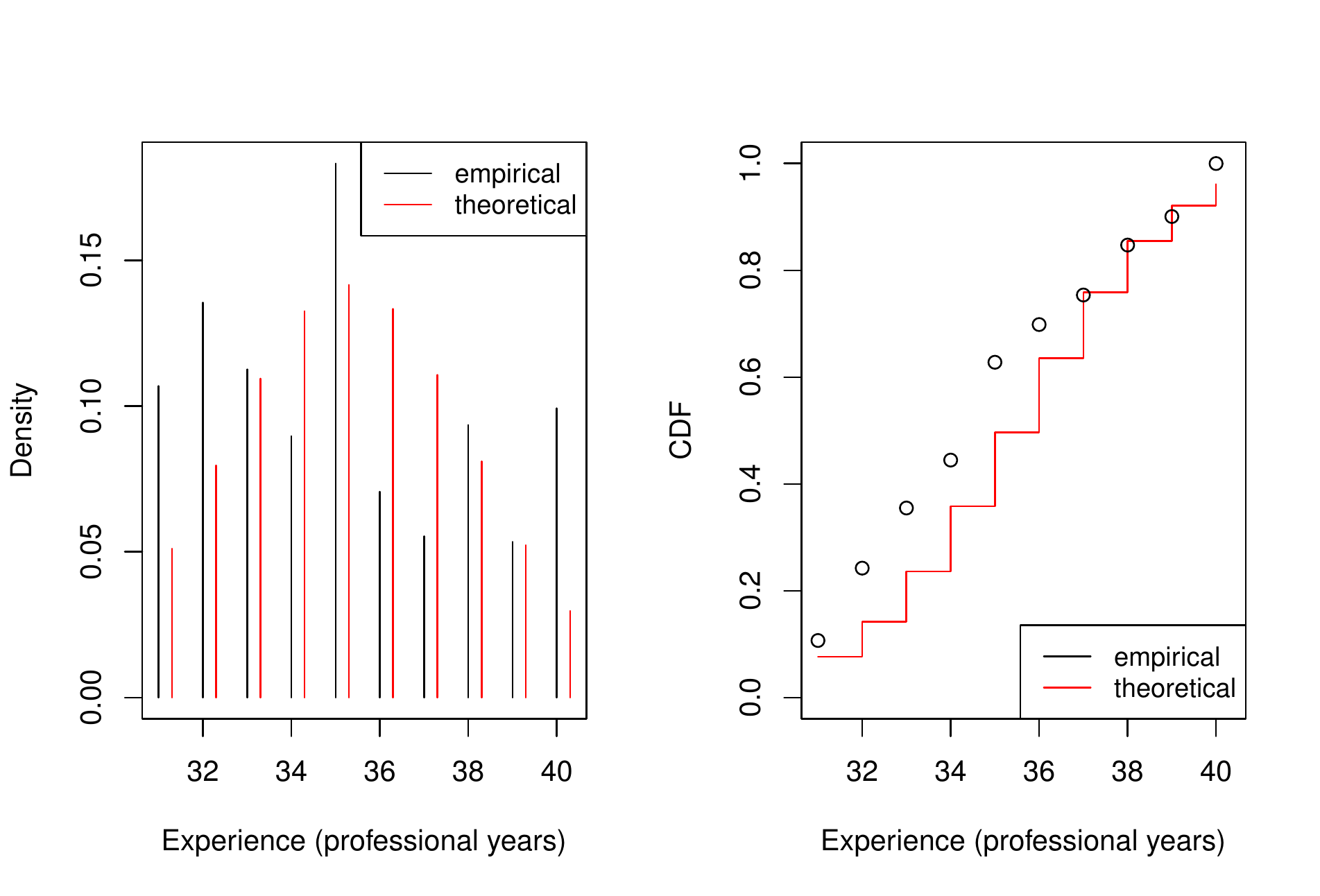}
    \caption{Density (left) and CDF (right) of empirical (data set) and Normal distribution for professional coding years 31-40}
    \label{fig:normal_31_40}
\end{figure}

Figure~\ref{fig:normal_0_20} (left), Figure~\ref{fig:normal_21_30} (left), and Figure~\ref{fig:normal_31_40} (left)
represent density plots that compare between empirical (data set) and Normal distribution of professional coders with 0-20, 21-30, and 31-40 years of working experience, respectively. Figure~\ref{fig:normal_0_50} (left) captures the same for the broader range of 0-50 years working experience.

\begin{figure}
    \centering
    \includegraphics[width=12 cm, height= 8 cm]{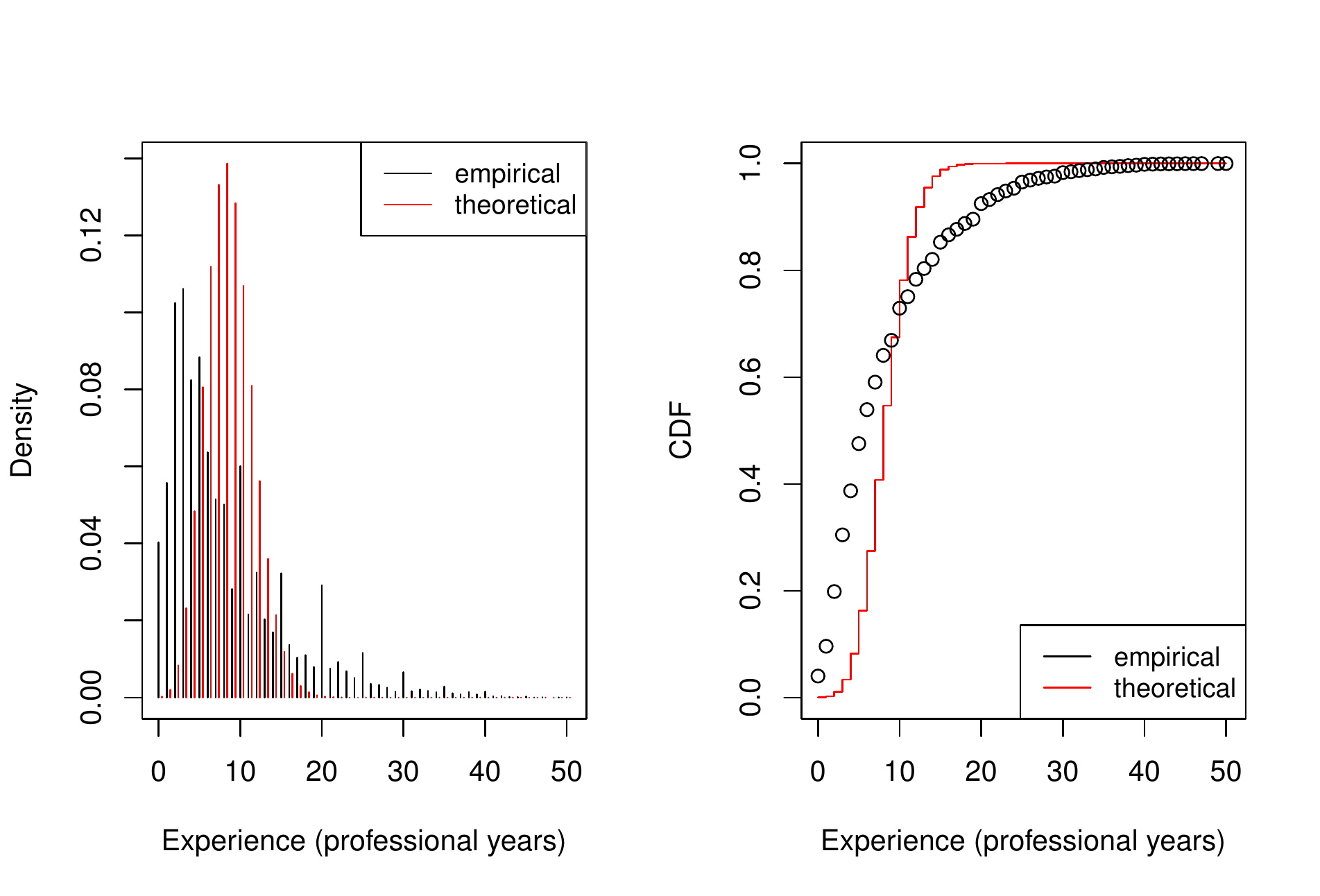}
    \caption{Density (left) and CDF (right) of empirical (data set) and Poisson distribution for professional coding years 0-50}
    \label{fig:poisson_0_50}
\end{figure}

\begin{figure}
    \centering
    \includegraphics[width=12 cm, height= 7 cm]{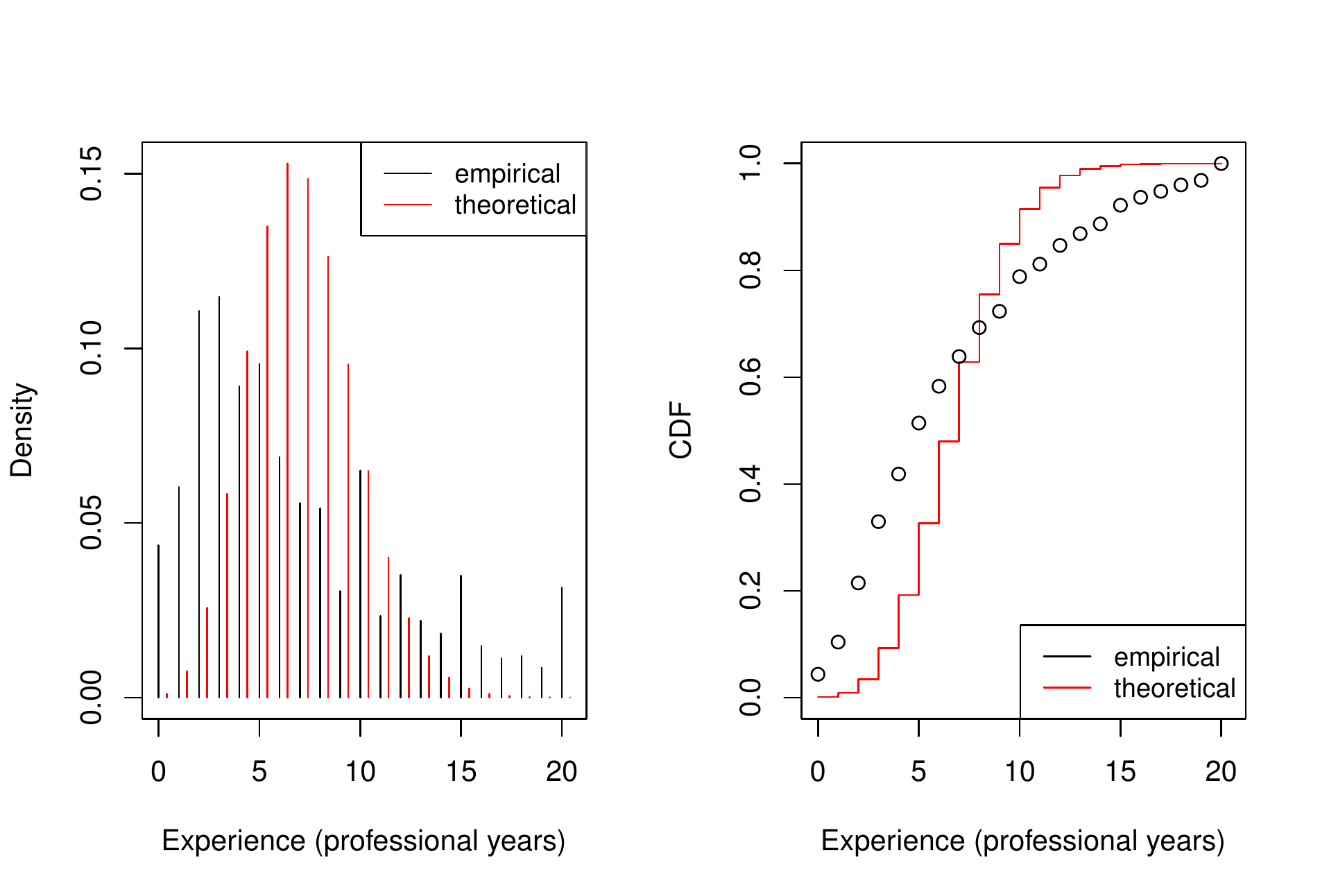}
    \caption{Density (left) and CDF (right) of empirical (data set) and Poisson distribution for professional coding years 0-20}
    \label{fig:poisson_0_20}
\end{figure}

\begin{figure}
    \centering
    \includegraphics[width=12 cm, height= 7.8 cm]{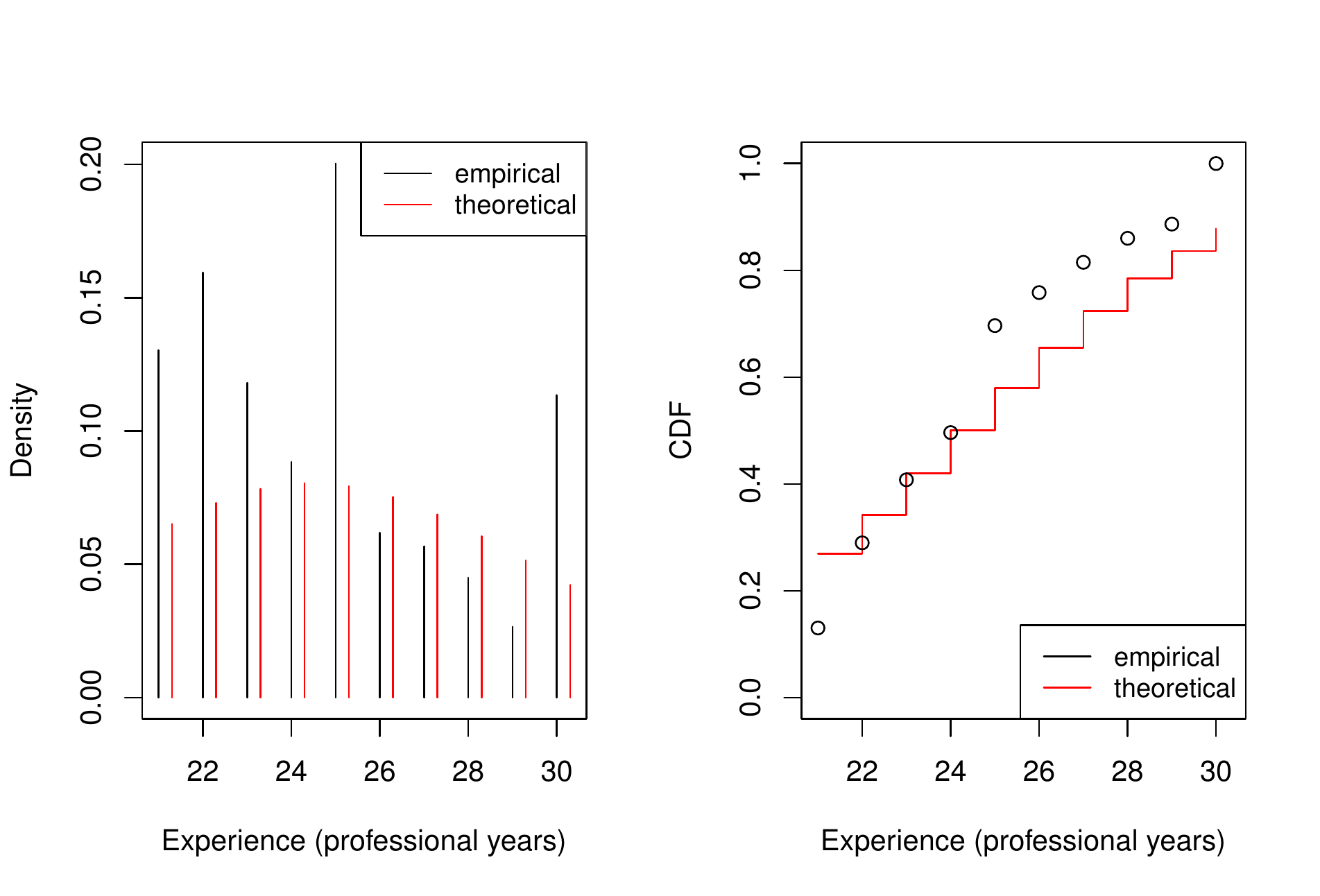}
    \caption{Density (left) and CDF (right) of empirical (data set) and Poisson distribution for professional coding years 21-30}
    \label{fig:poisson_21_30}
\end{figure}

\begin{figure}
    \centering
    \includegraphics[width=12 cm, height= 7.8 cm]{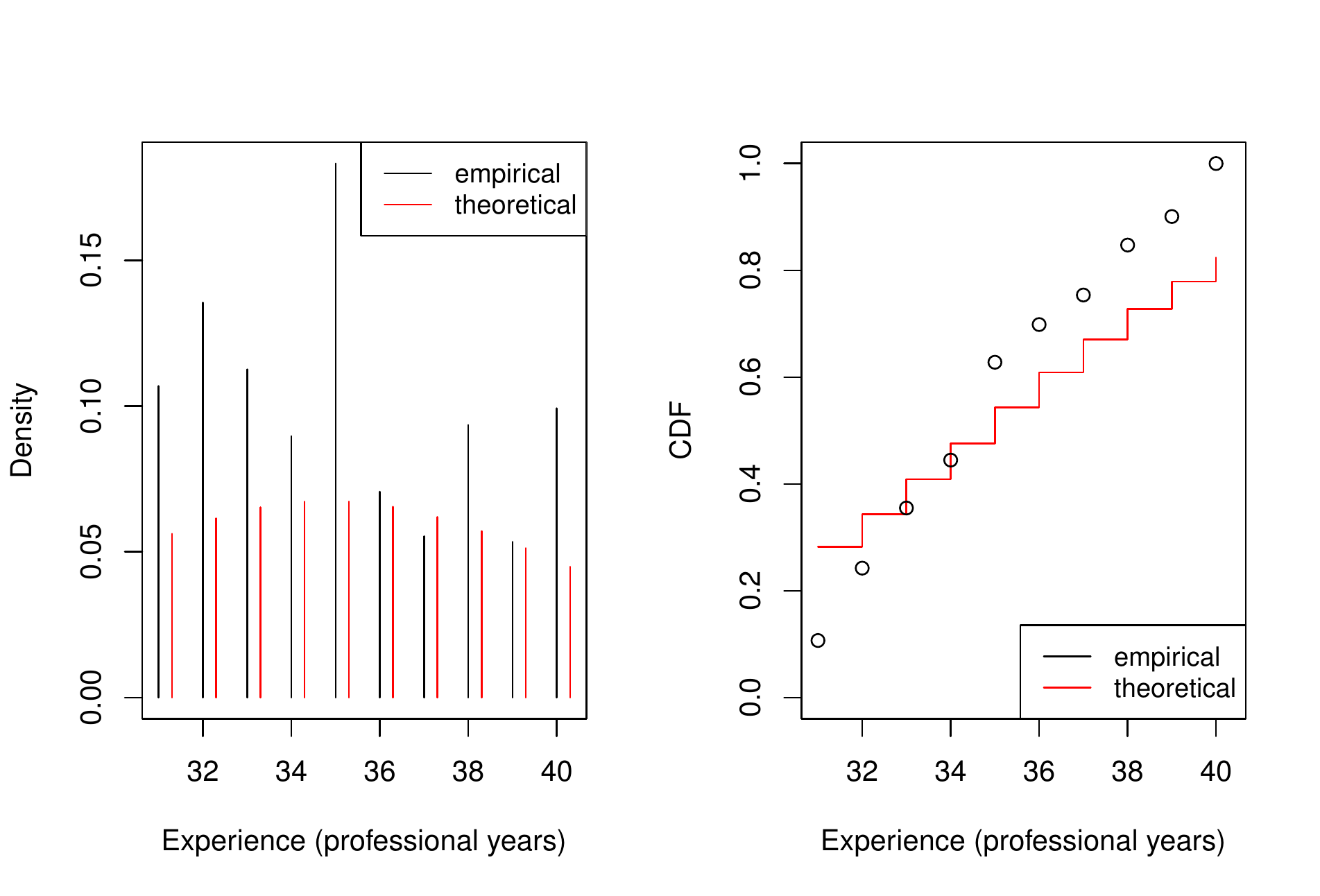}
    \caption{Density (left) and CDF (right) of empirical (data set) and Poisson distribution for professional coding years 31-40}
    \label{fig:poisson_31_40}
\end{figure}

Figure~\ref{fig:poisson_0_20} (left), Figure~\ref{fig:poisson_21_30} (left), and Figure~\ref{fig:poisson_31_40} (left)
represent density plots that compare between empirical (data set) and Poisson distribution of professional coders with 0-20, 21-30, and 31-40 years of working experience, respectively. Figure~\ref{fig:poisson_0_50} (left) captures the same for the broader range of 0-50 years working experience.

\begin{figure}
    \centering
    \includegraphics[width=12 cm, height= 8.2 cm]{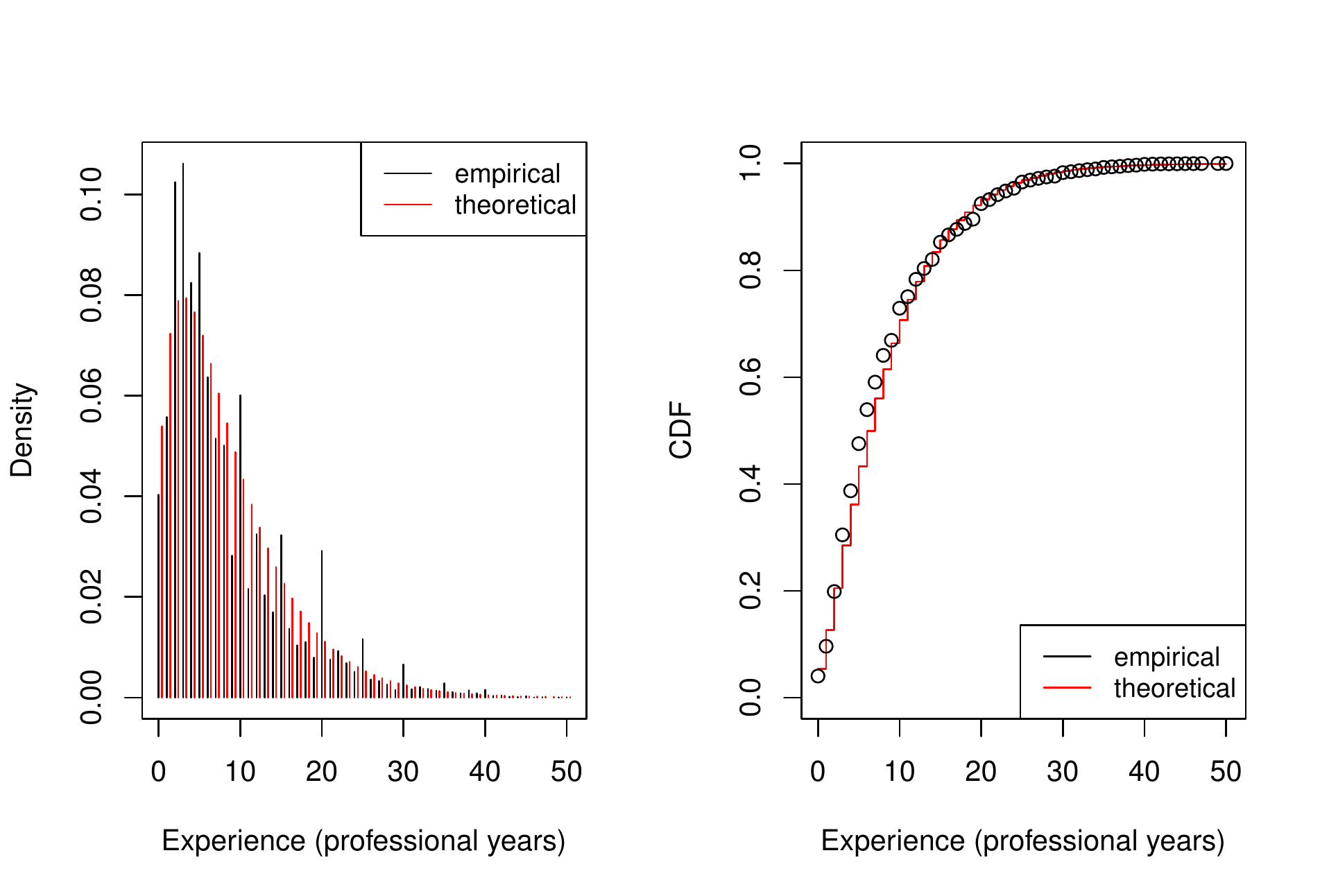}
    \caption{Density (left) and CDF (right) of empirical (data set) and Negative Binomial distribution for professional coding years 0-50}
    \label{fig:n. binomial_0_50}
\end{figure}

\begin{figure}
    \centering
    \includegraphics[width=12 cm, height= 8 cm]{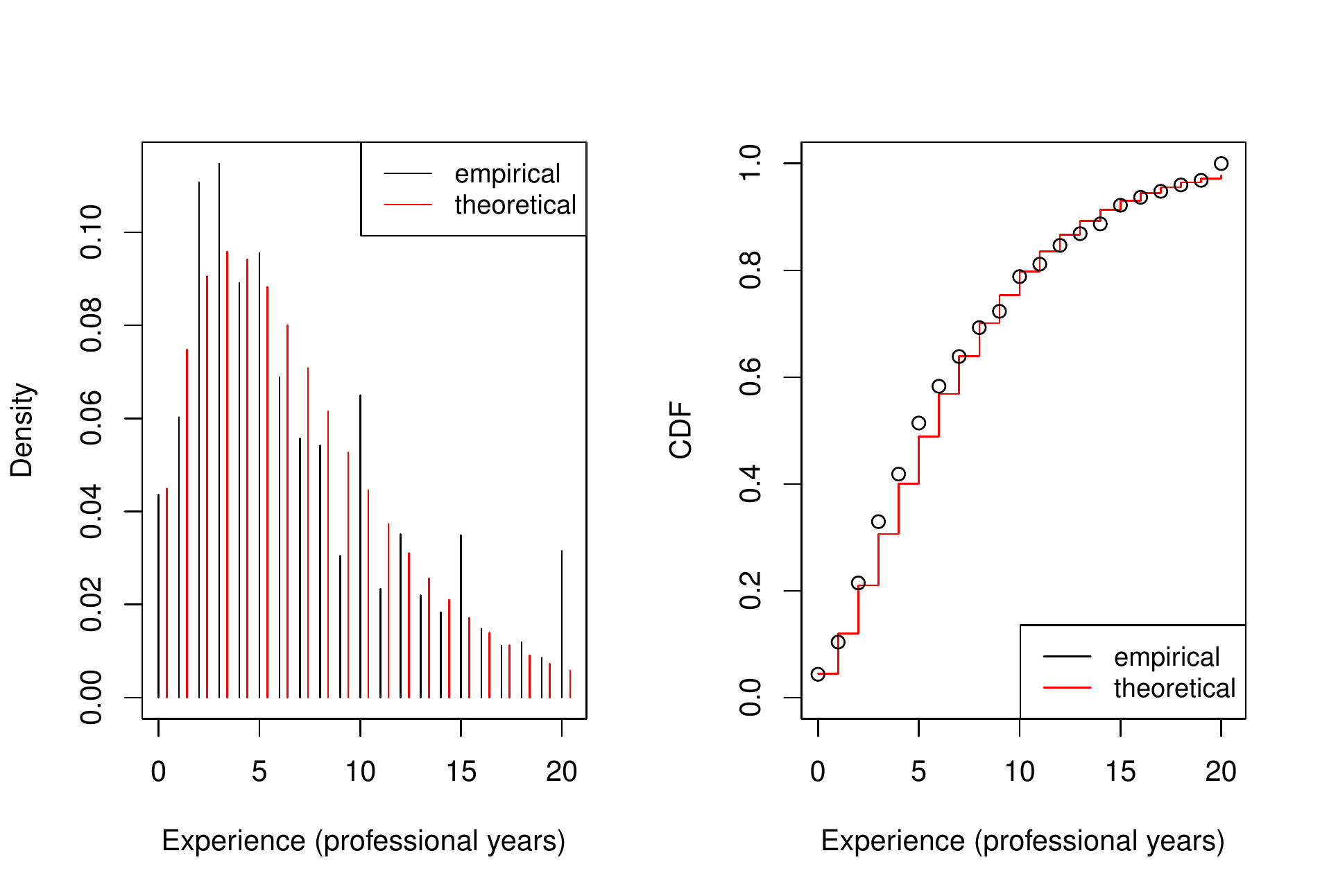}
    \caption{Density (left) and CDF (right) of empirical (data set) and Negative Binomial distribution for professional coding years 0-20}
    \label{fig:n. binomial_0_20}
\end{figure}

\begin{figure}
    \centering
    \includegraphics[width=12 cm, height= 7.8 cm]{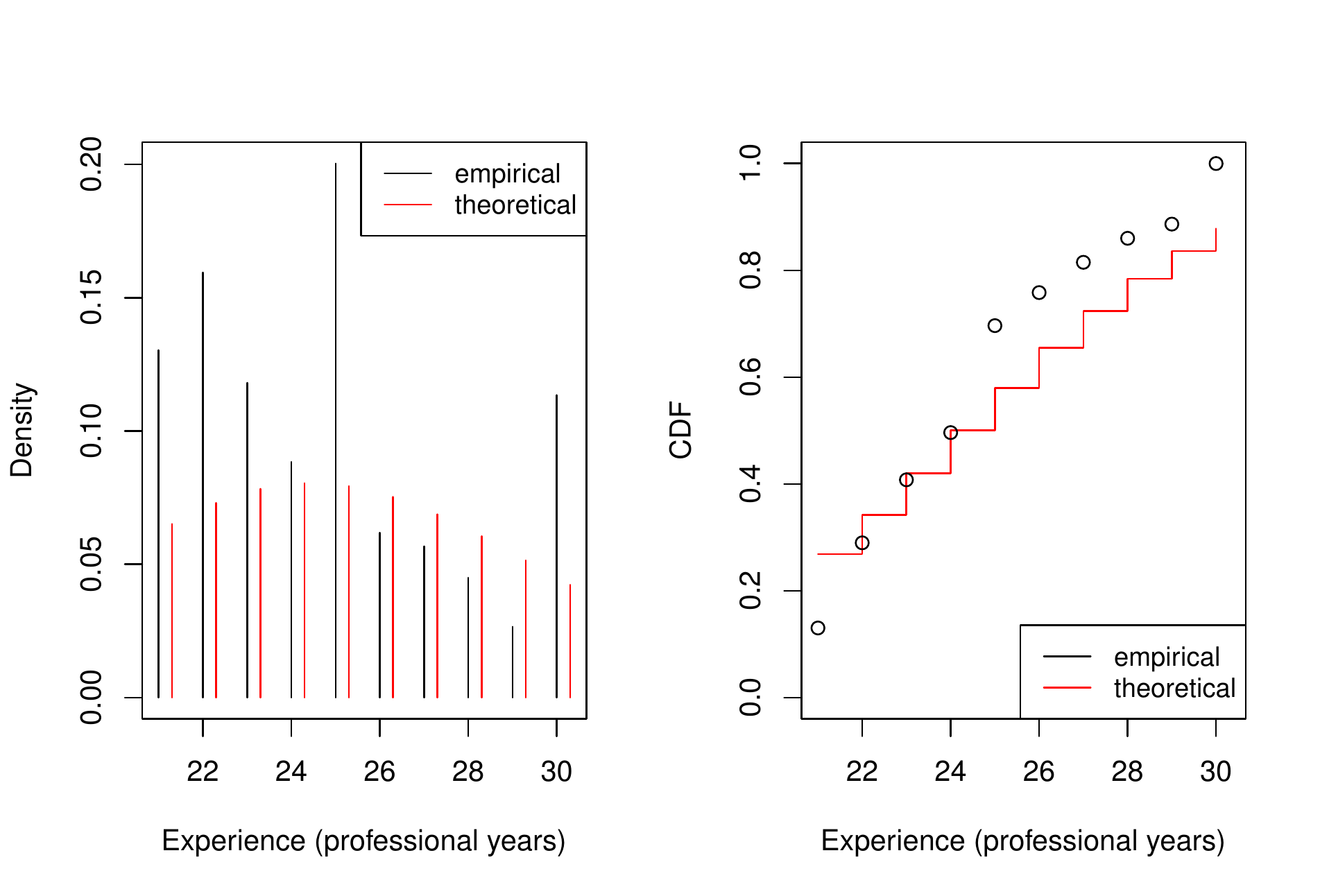}
    \caption{Density (left) and CDF (right) of empirical (data set) and Negative Binomial distribution for professional coding years 21-30}
    \label{fig:n. binomial_21_30}
\end{figure}

\begin{figure}
    \centering
    \includegraphics[width=12 cm, height= 7.8 cm]{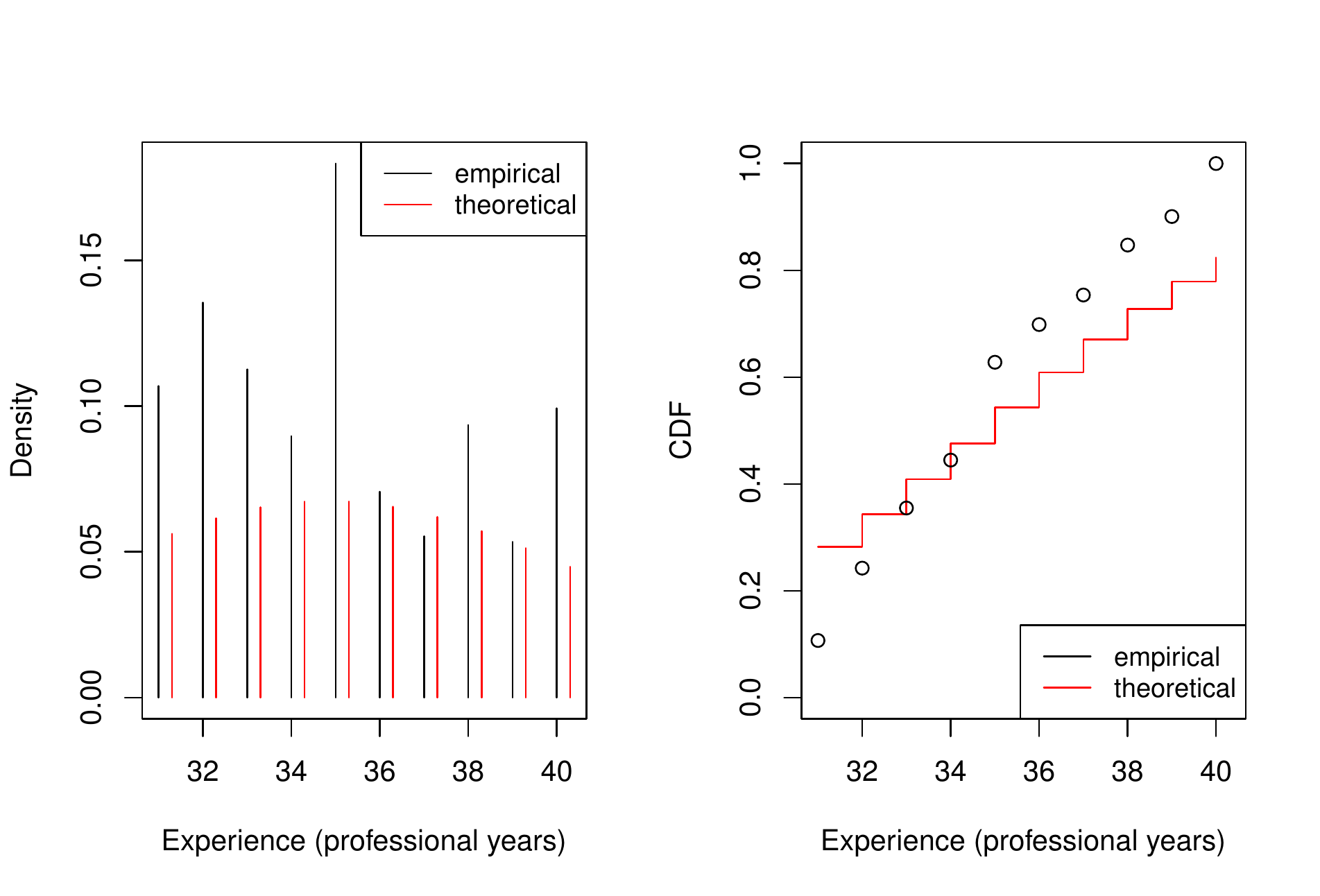}
    \caption{Density (left) and CDF (right) of empirical (data set) and Negative Binomial distribution for professional coding years 31-40}
    \label{fig:n. binomial_31_40}
\end{figure}

Figure~\ref{fig:n. binomial_0_20} (left), Figure~\ref{fig:n. binomial_21_30} (left), and Figure~\ref{fig:n. binomial_31_40} (left)
represent density plots that compare between empirical (data set) and Negative Binomial distribution of professional coders with 0-20, 21-30, and 31-40 years of working experience, respectively. Figure~\ref{fig:n. binomial_0_50} (left) captures the same for the broader range of 0-50 years working experience.

Figure~\ref{fig:normal_0_20} (right), Figure~\ref{fig:normal_21_30} (right), and Figure~\ref{fig:normal_31_40} (right)
represent CDF plots that compare between empirical (data set) and Normal distribution of professional coders with 0-20, 21-30, and 31-40 years of working experience, respectively. Figure~\ref{fig:normal_0_50} (right) captures the same for the broader range of 0-50 years working experience.

Figure~\ref{fig:poisson_0_20} (right), Figure~\ref{fig:poisson_21_30} (right), and Figure~\ref{fig:poisson_31_40} (right)
represent CDF plots that compare between empirical (data set) and Poisson distribution of professional coders with 0-20, 21-30, and 31-40 years of working experience, respectively. Figure~\ref{fig:poisson_0_50} (right) captures the same for the broader range of 0-50 years working experience.

Figure~\ref{fig:n. binomial_0_20} (right), Figure~\ref{fig:n. binomial_21_30} (right), and Figure~\ref{fig:n. binomial_31_40} (right)
represent CDF plots that compare between empirical (data set) and Negative Binomial distribution of professional coders with 0-20, 21-30, and 31-40 years of working experience, respectively. Figure~\ref{fig:n. binomial_0_50} (right) captures the same for the broader range of 0-50 years working experience.

\subsubsection{Closeness of Fit of Cumulative Distribution Functions}
\label{sec:SO2020-closenessoffit}

The fit of each distribution can be evaluated by comparing their cumulative distribution function curves against the the perfect linear match representing the actual data on a plot of distribution probability versus empirical probability (P-P plot).

\begin{figure}
    \centering
    \includegraphics[width=12 cm, height= 7.5 cm]{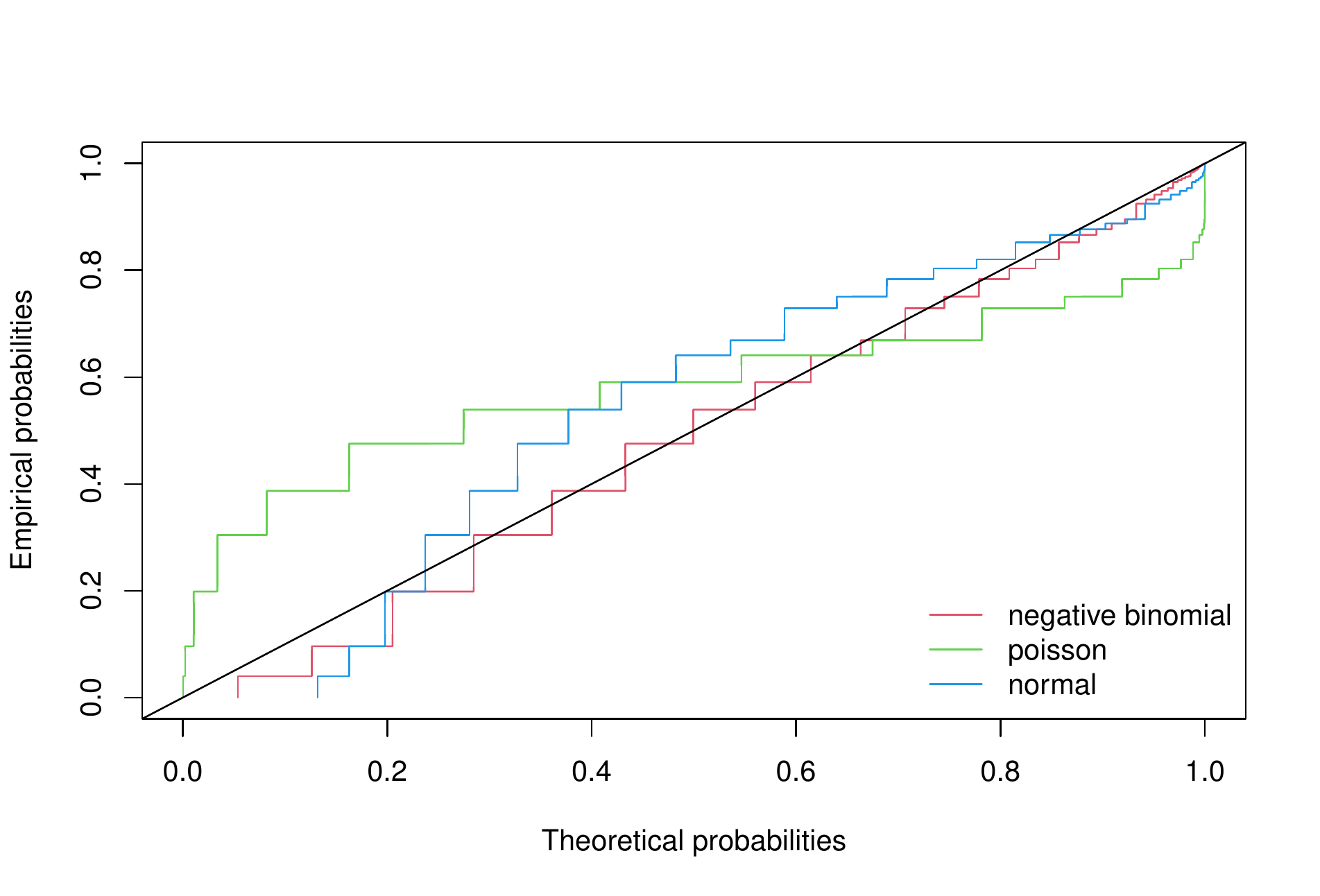}
    \caption{P-P plot of Normal, Poisson and Negative Binomial distribution along with empirical distribution (data set) for professional coding years 0-50}
    \label{fig:P_P_0_50}
\end{figure}

\begin{figure}
    \centering
    \includegraphics[width=12 cm, height= 7.5 cm]{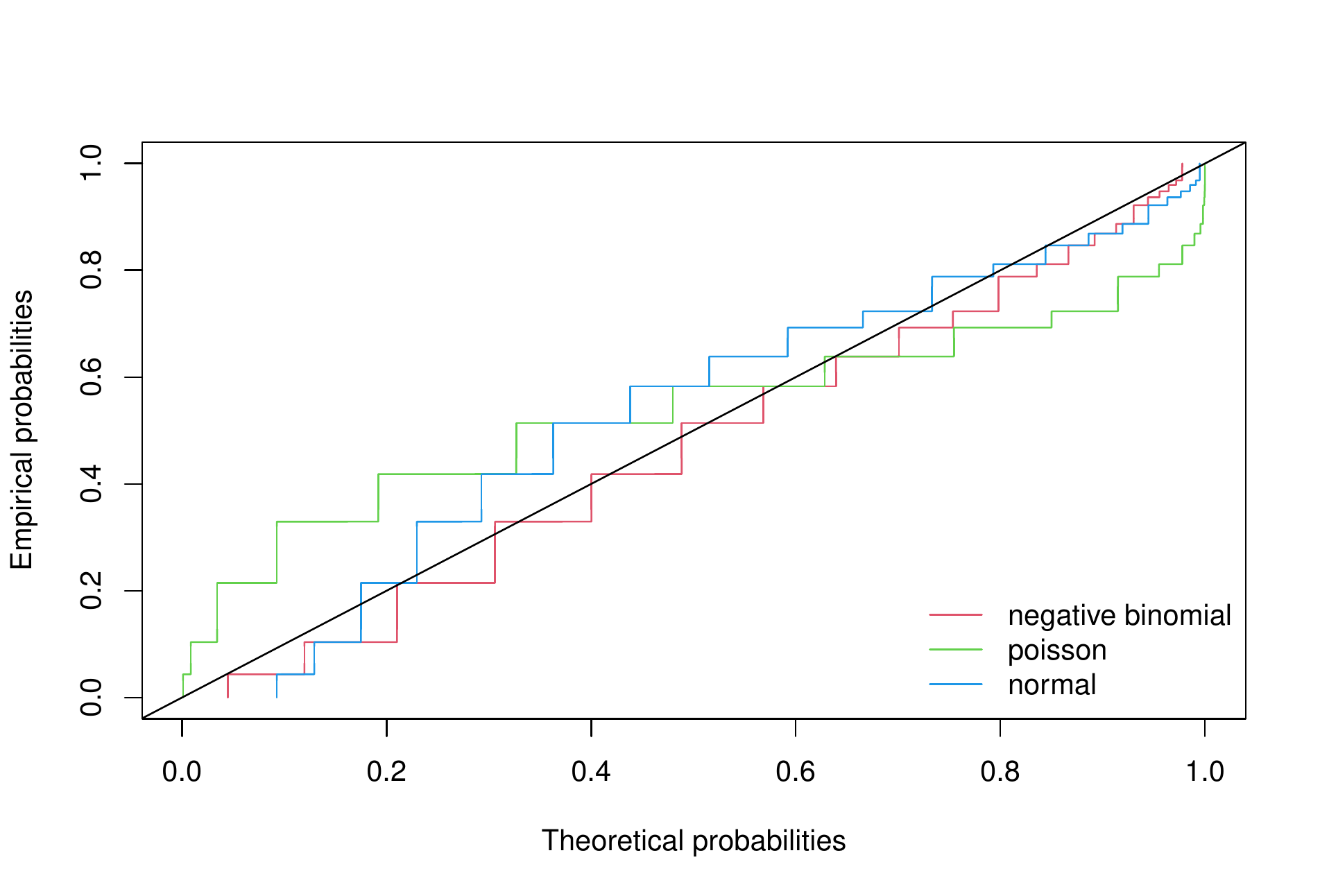}
    \caption{P-P plot of Normal, Poisson and Negative Binomial distribution along with empirical distribution (data set) for professional coding years 0-20}
    \label{fig:P_P_0_20}
\end{figure}

\begin{figure}
    \centering
    \includegraphics[width=12 cm, height= 7.6 cm]{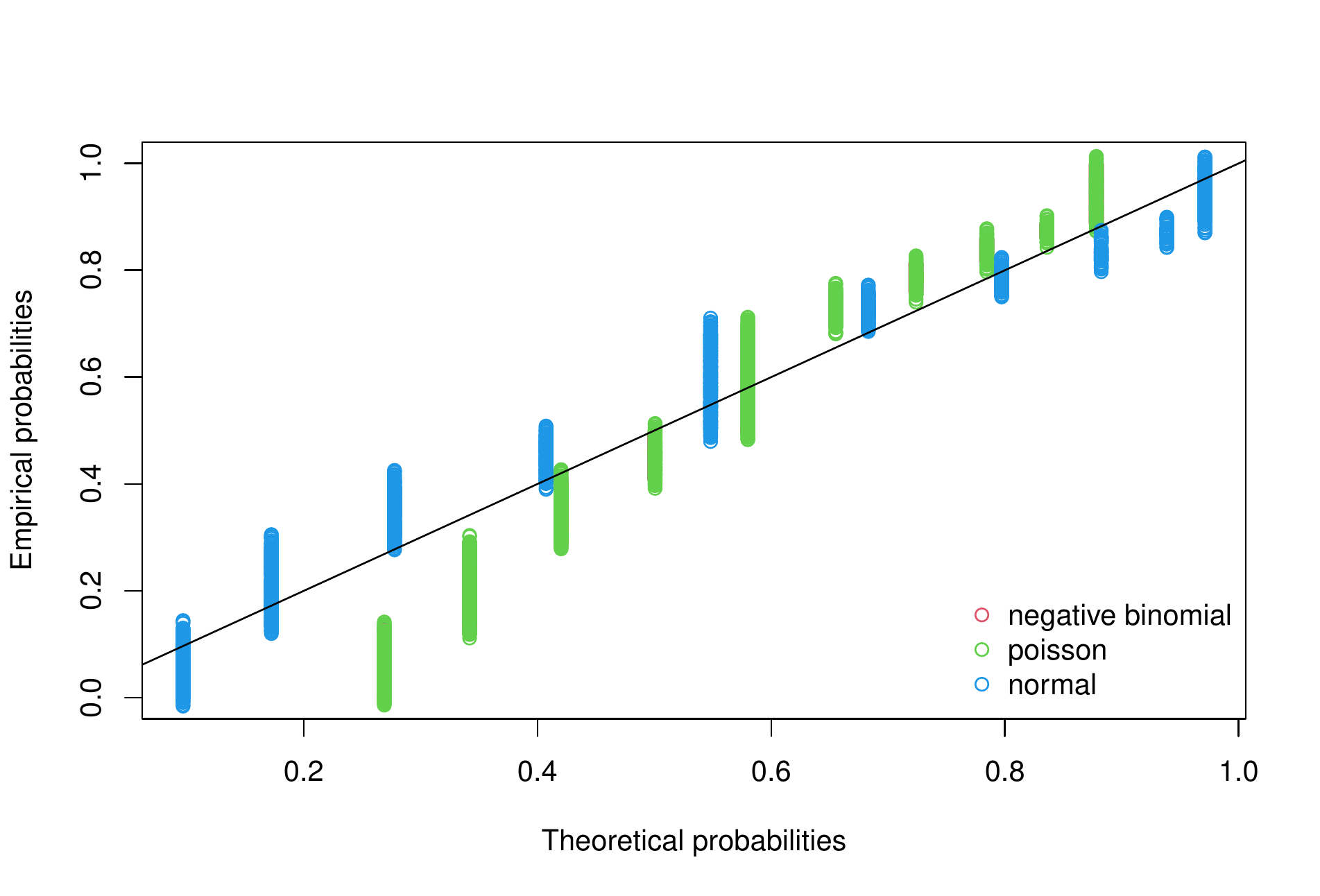}
    \caption{P-P plot of Normal, Poisson and Negative Binomial distribution along with empirical distribution (data set) for professional coding years 21-30}
    \label{fig:P_P_21_30}
\end{figure}

\begin{figure}
    \centering
    \includegraphics[width=12 cm, height= 7.6 cm]{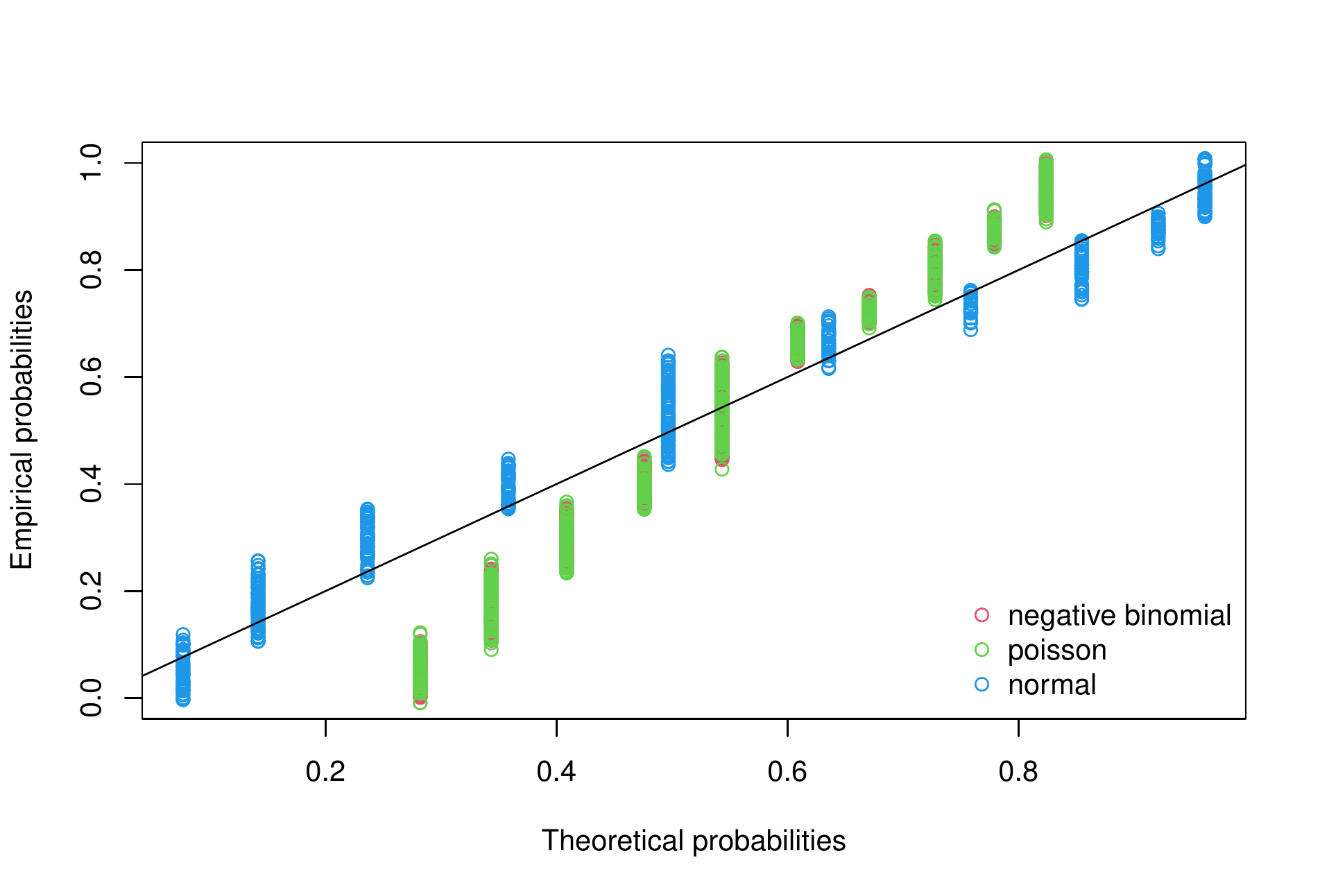}
    \caption{P-P plot of Normal, Poisson and Negative Binomial distribution along with empirical distribution (data set) for professional coding years 31-40}
    \label{fig:P_P_31_40}
\end{figure}

Figure~\ref{fig:P_P_0_20}, 
Figure~\ref{fig:P_P_21_30}, and
Figure~\ref{fig:P_P_31_40}
show the P-P plots for the Normal, Poisson and Negative Binomial distribution with respect to empirical distribution (data set) of professional coders with
0-20, 21-30, and 31-40 years of working experience, respectively.
Figure~\ref{fig:P_P_0_50} captures the same for the broader range of 0-50 years working experience.

\begin{figure}
    \centering
    \includegraphics[width=12 cm, height= 7.5 cm]{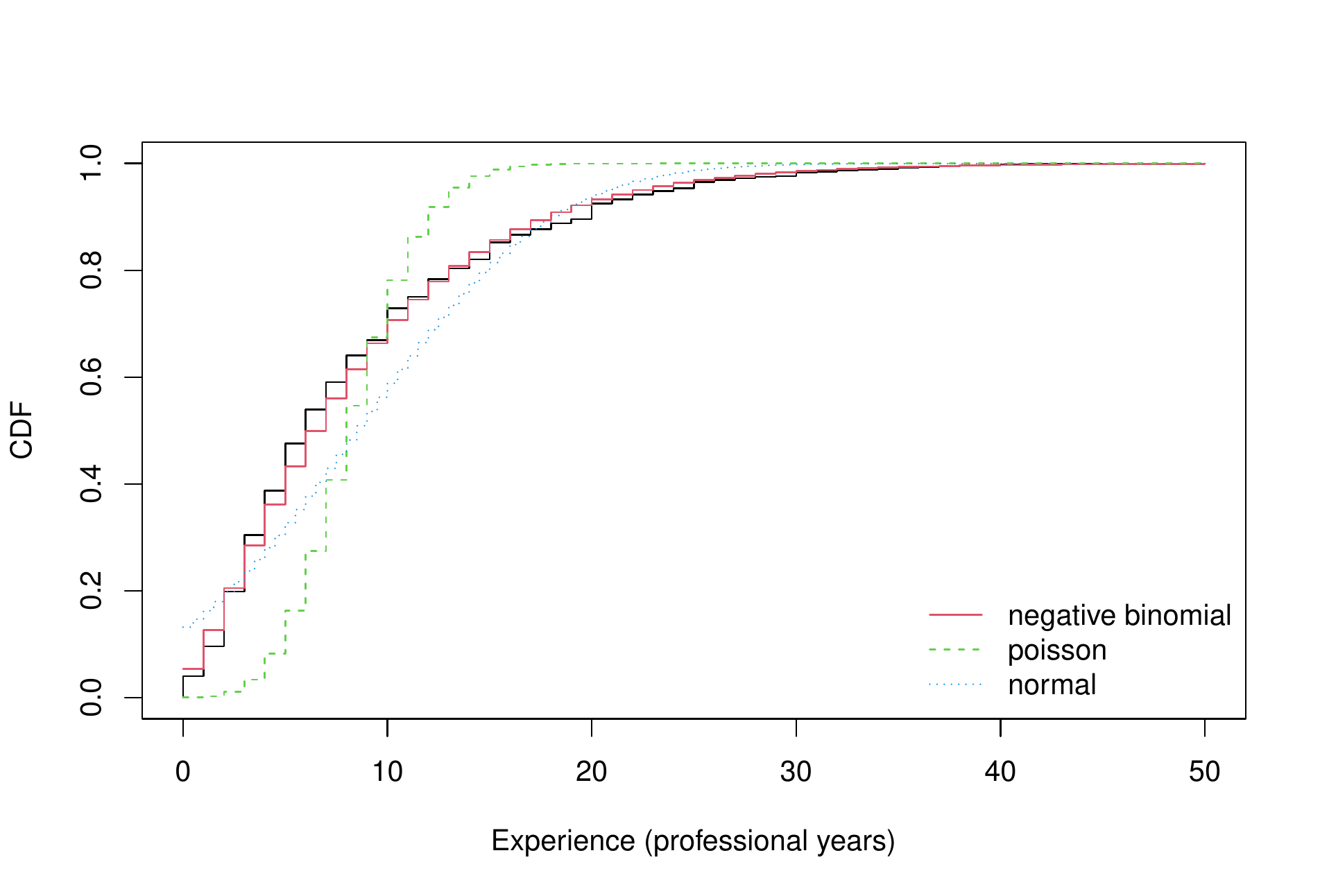}
    \caption{CDF plot of Normal, Poisson and Negative Binomial distribution along with empirical distribution (data set) for professional coding years 0-50}
    \label{fig:CDF_0_50}
\end{figure}

\begin{figure}
    \centering
    \includegraphics[width=12 cm, height= 7.5 cm]{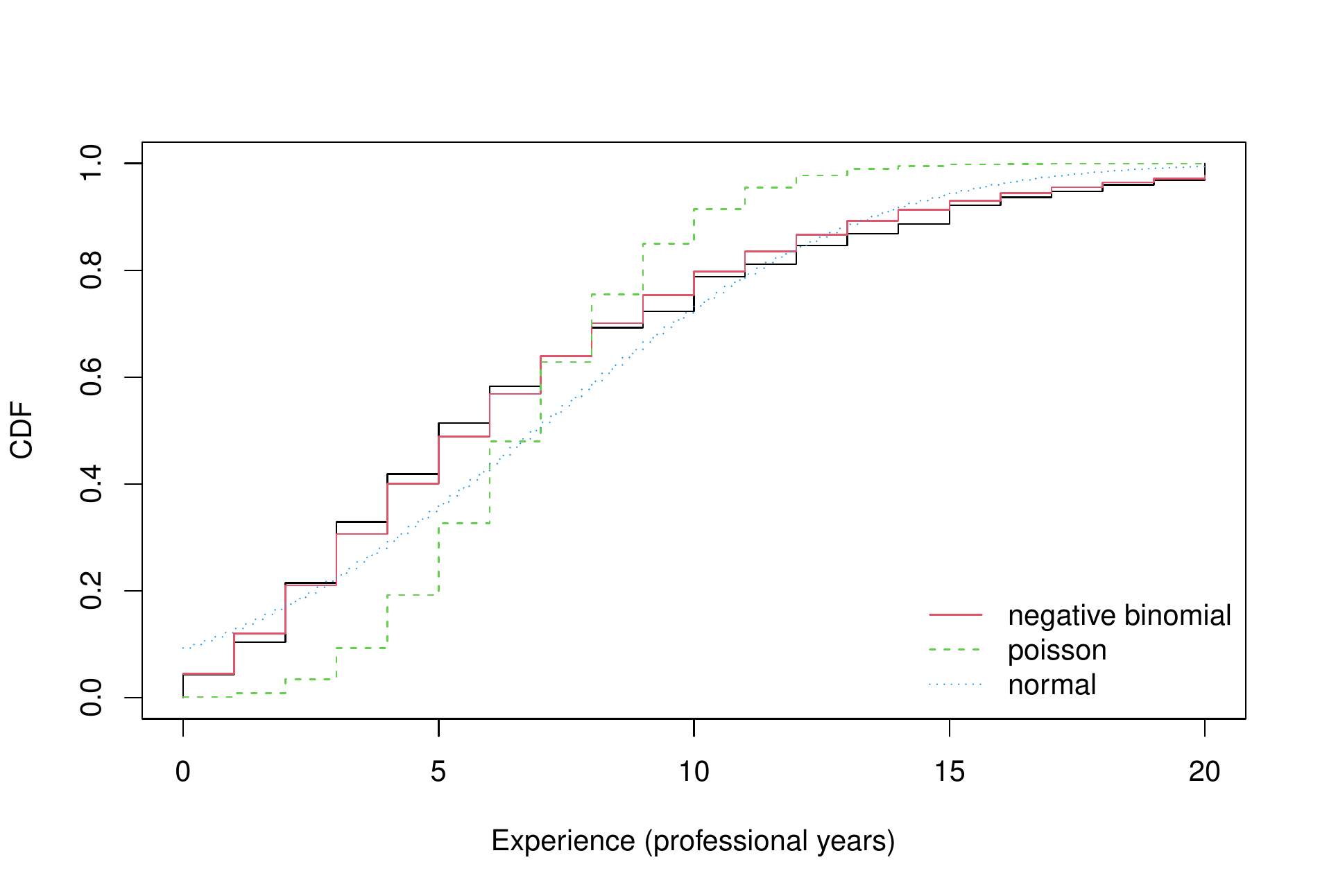}
    \caption{CDF plot of Normal, Poisson and Negative Binomial distribution along with empirical distribution (data set) for professional coding years 0-20}
    \label{fig:CDF_0_20}
\end{figure}

\begin{figure}
    \centering
    \includegraphics[width=12 cm, height= 7.5 cm]{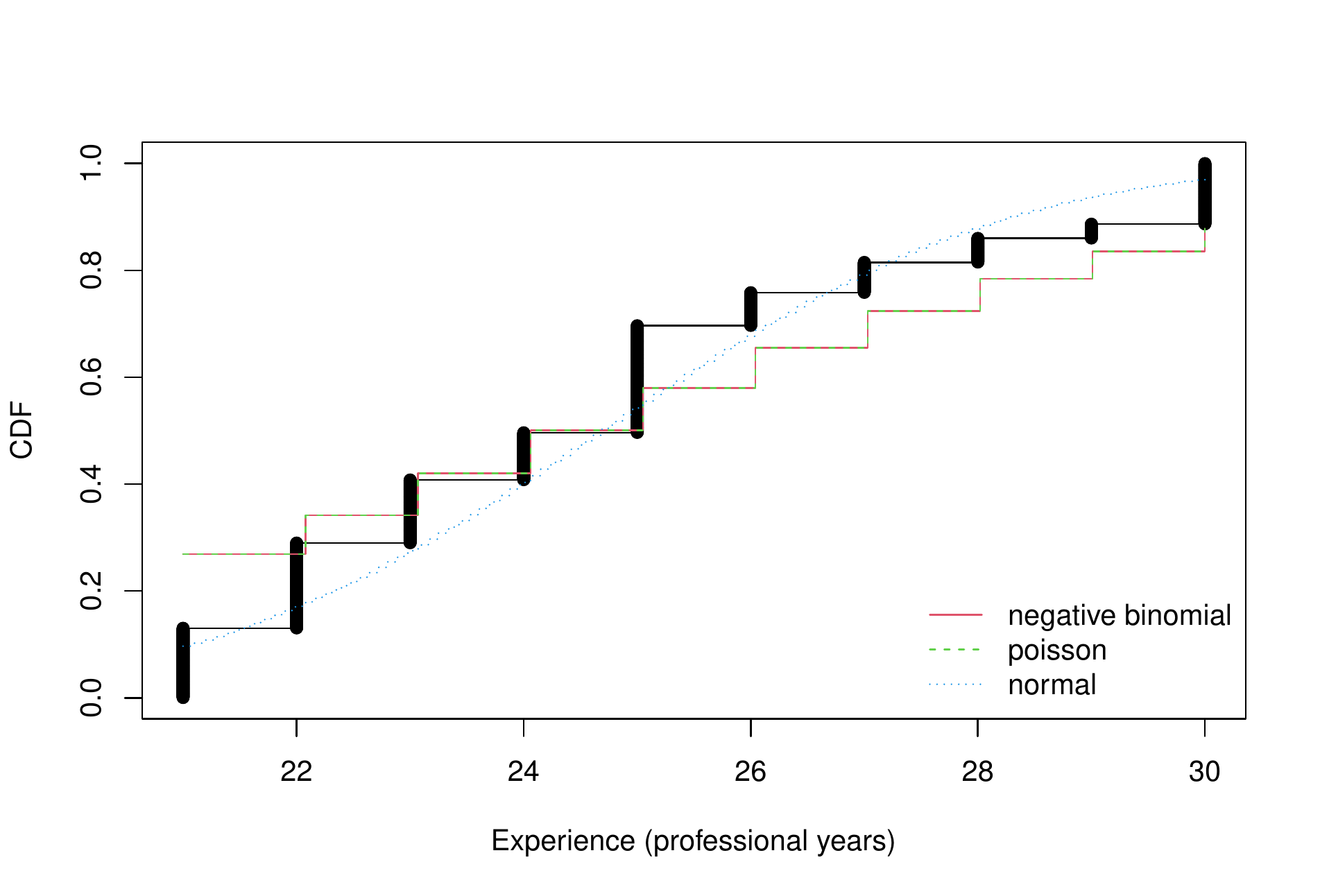}
    \caption{CDF plot of Normal, Poisson and Negative Binomial distribution along with empirical distribution (data set) for professional coding years 21-30}
    \label{fig:CDF_21_30}
\end{figure}

\begin{figure}
    \centering
    \includegraphics[width=12 cm, height= 7.5 cm]{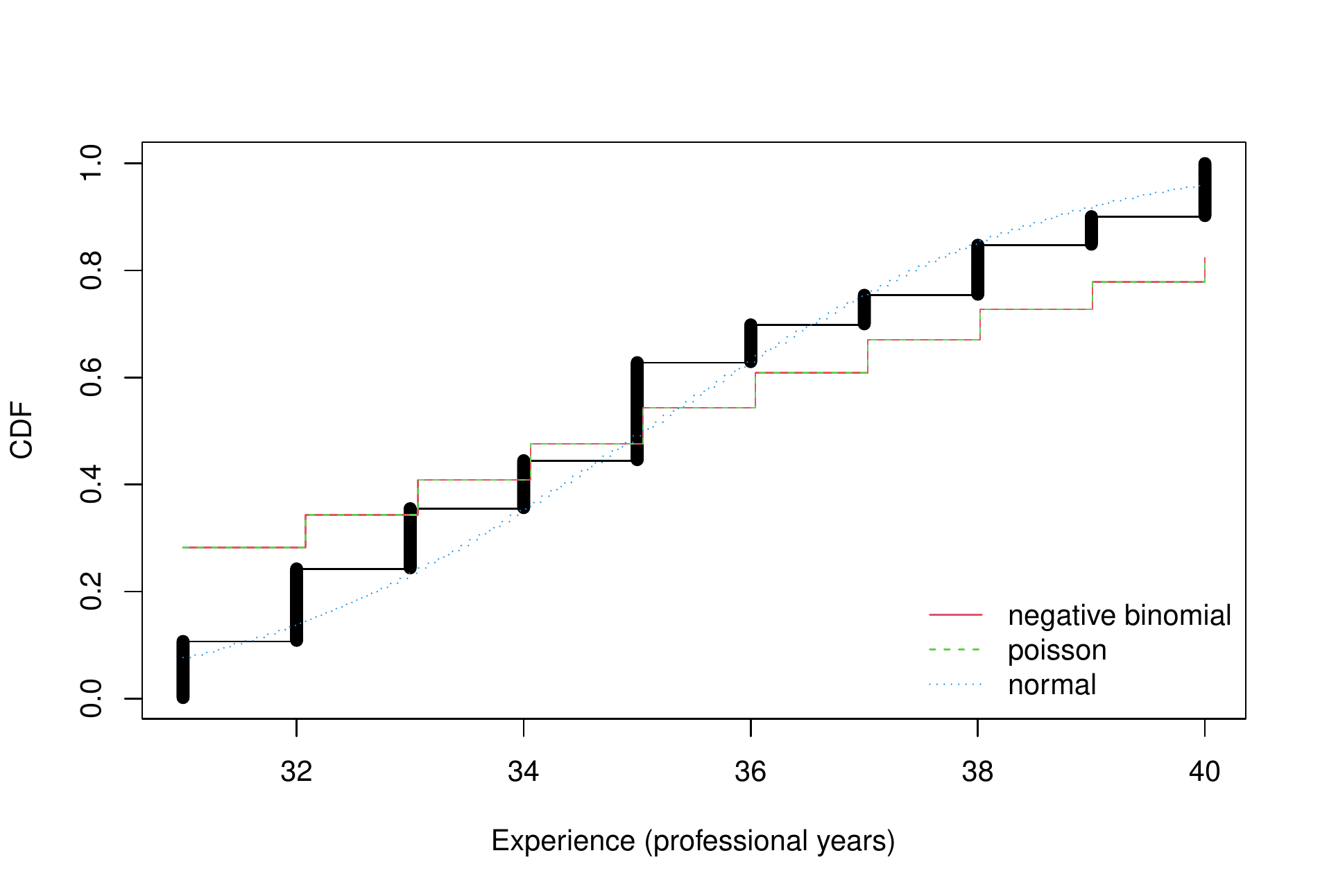}
    \caption{CDF plot of Normal, Poisson and Negative Binomial distribution along with empirical distribution (data set) for professional coding years 31-40}
    \label{fig:CDF_31_40}
\end{figure}

Figure~\ref{fig:CDF_0_20}, 
Figure~\ref{fig:CDF_21_30}, and
Figure~\ref{fig:CDF_31_40}
show the CDF plots for the Normal, Poisson and Negative Binomial distribution with respect to empirical distribution (data set) of professional coders with
0-20, 21-30, and 31-40 years of working experience respectively.
Figure~\ref{fig:CDF_0_50} captures the same for the broader range of 0-50 years working experience.

\subsubsection{Density Overlay for Different Distributions}
\label{sec:SO2020-densityoverlay}

\begin{figure}
    \centering
    \includegraphics[width=12 cm, height= 8 cm]{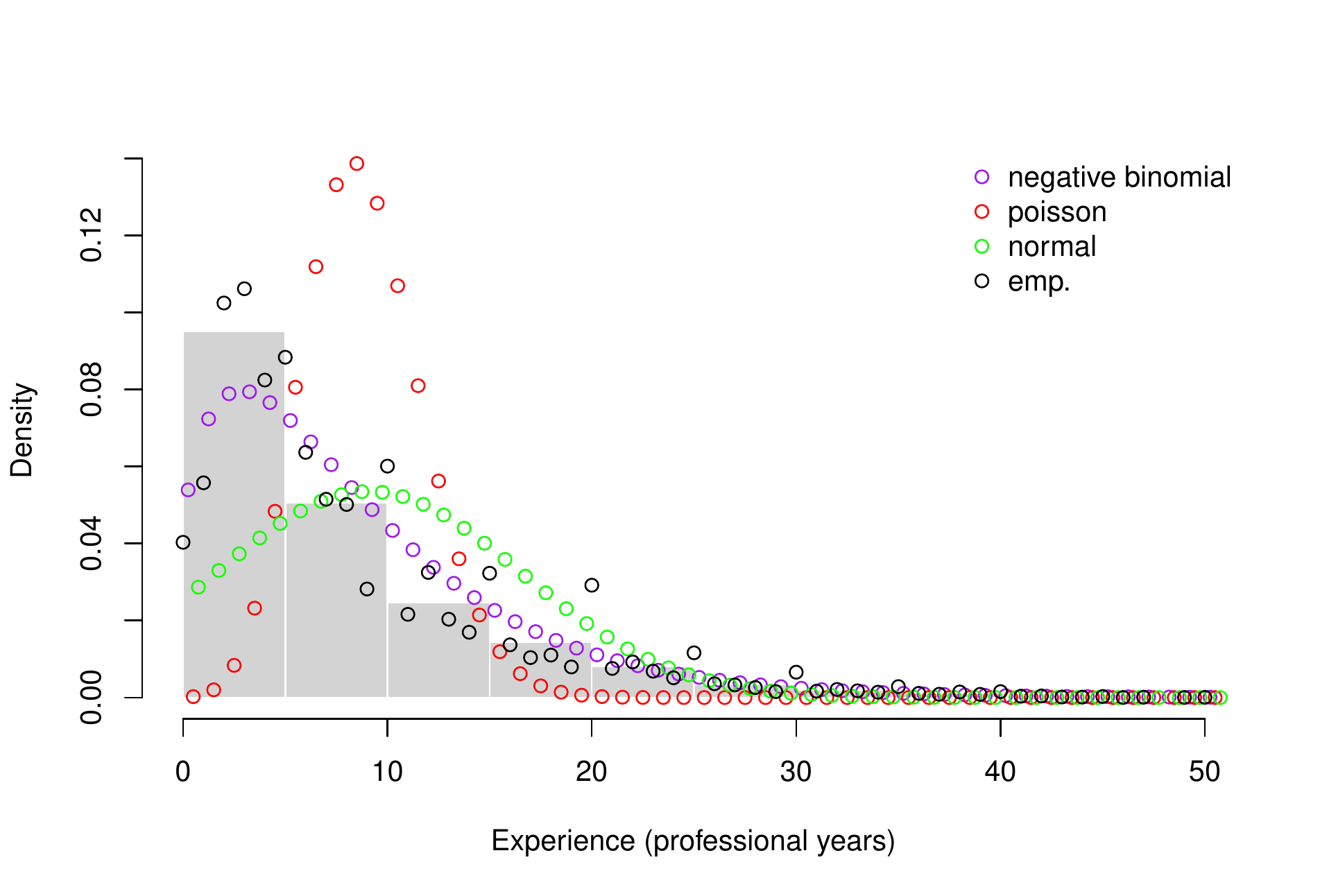}
    \caption{Histogram and distribution densities for professional coding years 0-50}
    \label{fig:den_overlay_def_bin}
\end{figure}

\begin{figure}
    \centering
    \includegraphics[width=12 cm, height= 7 cm]{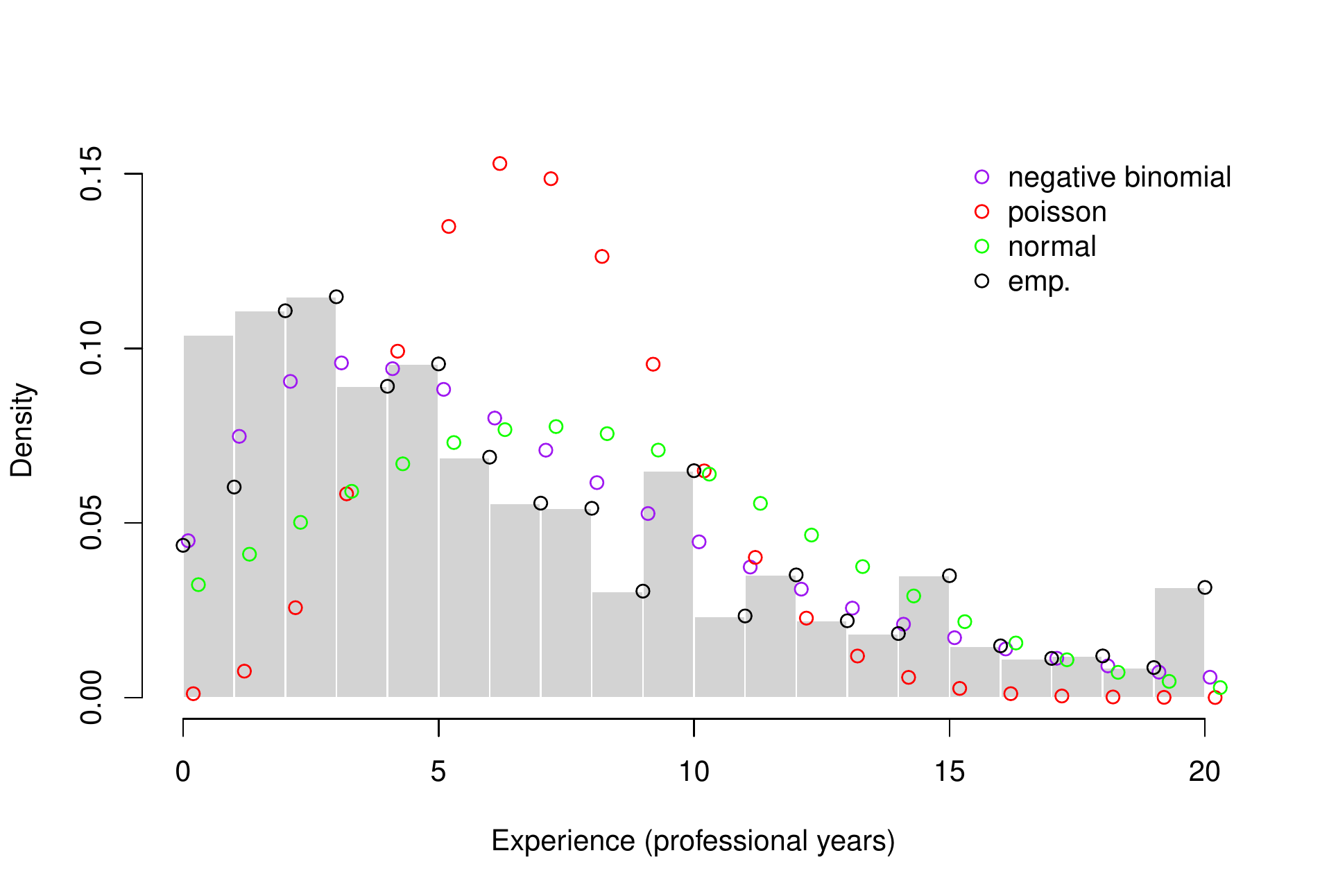}
    \caption{Histogram and distribution densities for professional coding years 0-20}
    \label{fig:den_overlay_def_bin_0_20}
\end{figure}

\begin{figure}
    \centering
    \includegraphics[width=12 cm, height= 7 cm]{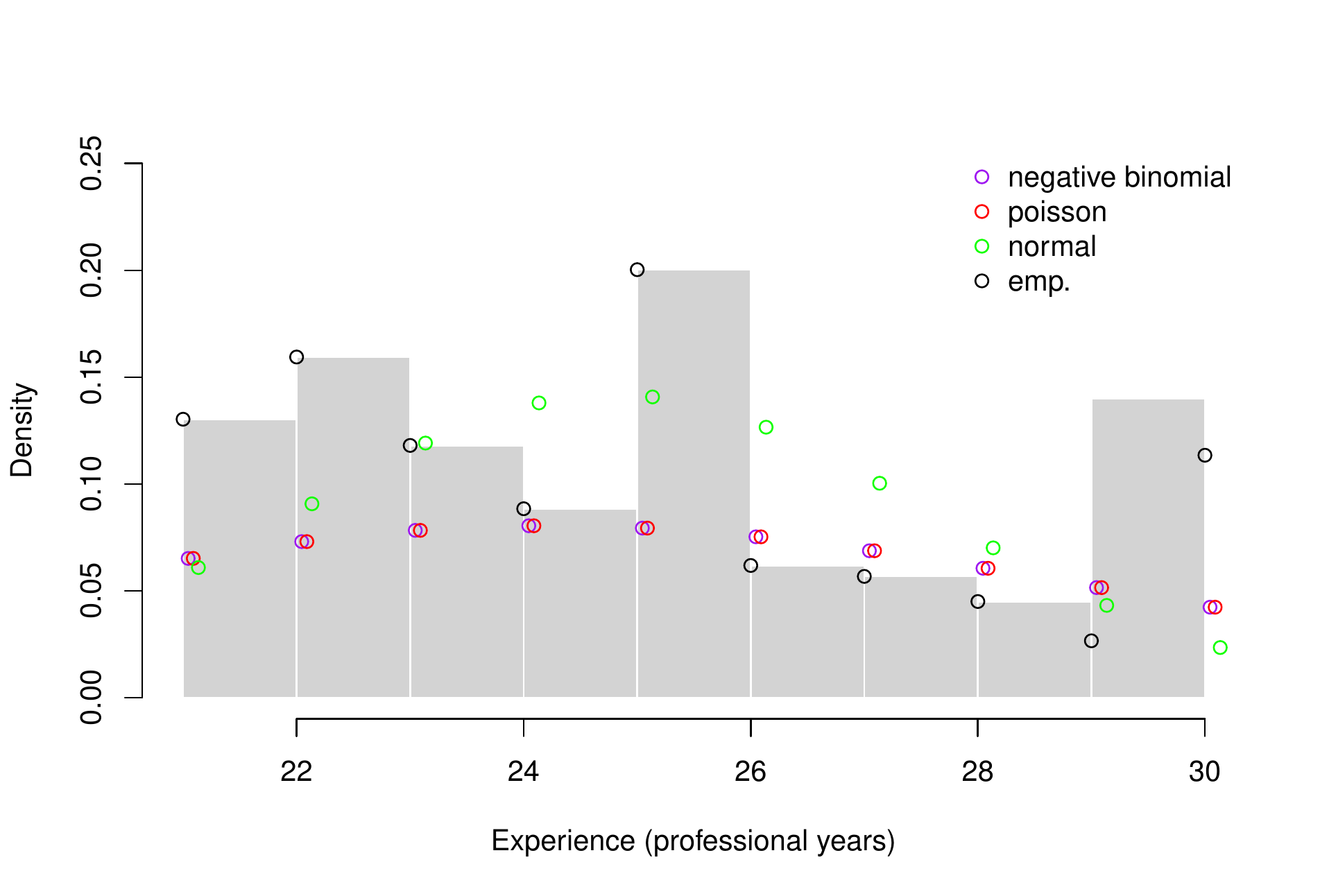}
    \caption{Histogram and distribution densities for professional coding years 21-30}
    \label{fig:den_overlay_def_bin_21_30}
\end{figure}

\begin{figure}
    \centering
    \includegraphics[width=12 cm, height= 7 cm]{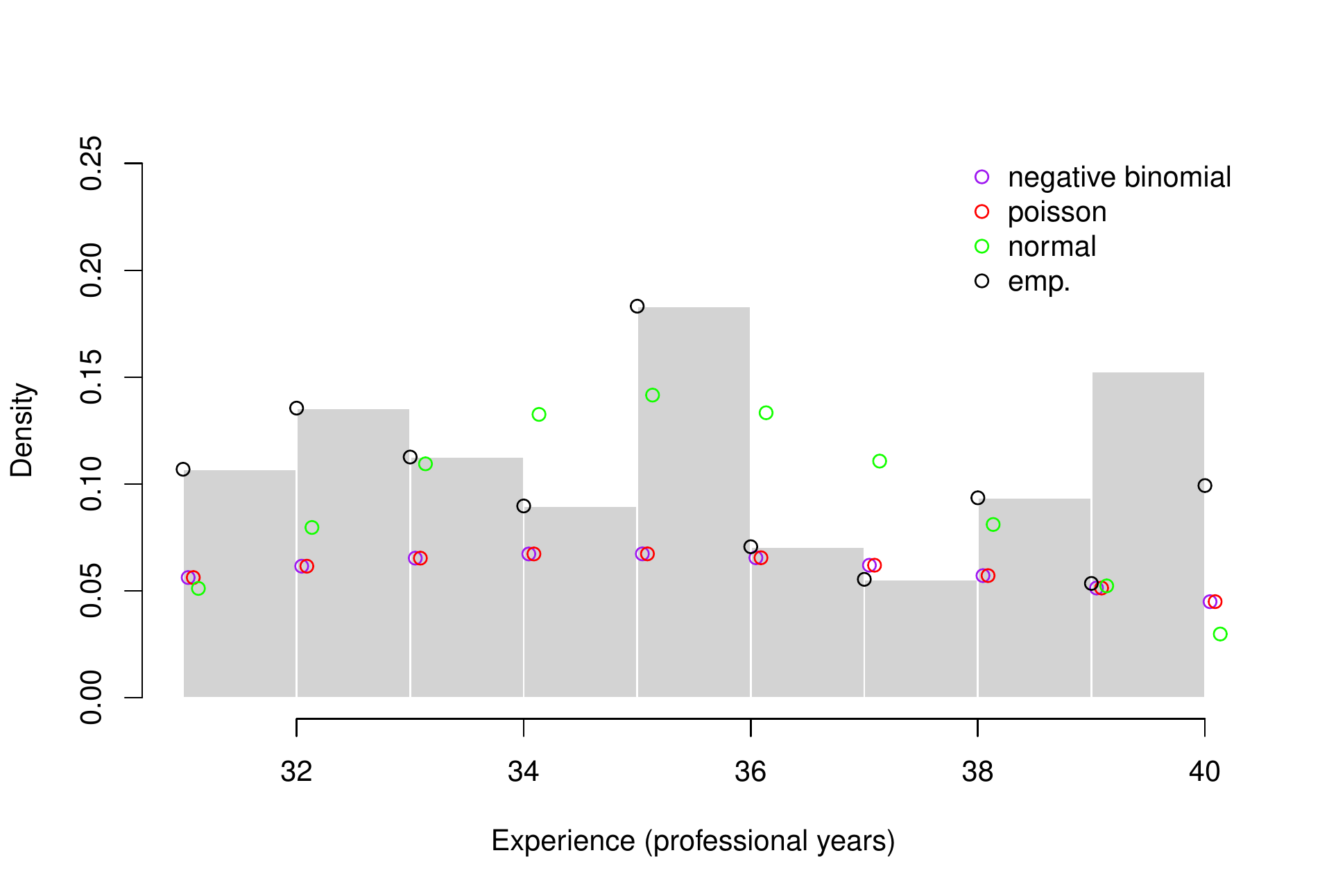}
    \caption{Histogram and distribution densities for professional coding years 31-40}
    \label{fig:den_overlay_def_bin_31_40}
\end{figure}

Figure~\ref{fig:den_overlay_def_bin_0_20}, Figure~\ref{fig:den_overlay_def_bin_21_30}, and Figure~\ref{fig:den_overlay_def_bin_31_40}
show the histogram (default bin size) against fitted Normal, Poisson and Negative Binomial density functions of professional coders with
0-20, 21-30, and 31-40 years of working experience respectively.
Figure~\ref{fig:den_overlay_def_bin} captures the same for the broader range of 0-50 years working experience, respectively.

\subsubsection{Best Fit Distributions}
\label{sec:SO2020-bestfits}

Table~\ref{tab:0_20},
Table~\ref{tab:21_30}, and
Table~\ref{tab:31_40} exhibits summary of fitting different distributions considering parameters, log-likelihood, Akaike's Information Criteria (AIC) and Bayesian Information Criteria (BIC) for professional coders with
0-20, 21-30, and 31-40 years of working experience respectively.
Table~\ref{tab:0_50} exhibits the same for the broader range of 0-50 years working experience.

\begin{table}
  \centering
  \caption{Fits of distributions for professional coding years 0-50}
    \begin{tabular}{|c|p{5.5em}|C{0.14\textwidth}|c|c|c|}\hline
    \multicolumn{2}{|c|}{\textbf{Distribution}} & \textbf{Log-likelihood} & \textbf{AIC} & \textbf{BIC} & \textbf{Best Fit} \\\hline\hline
    Normal & mean = 8.33\newline{}sd = 7.45 & -115,628 & 231,260 & 231,277 &  \\\hline
    Poisson & lambda=\newline{}8.33 & -159,718 & 319,438 & 319,447 & \\\hline
    N. Binomial & size = 1.59\newline{}mu = 8.33 & -105,885 & 211,774 & 211,790 &  \checkmark{} \\\hline
    \end{tabular}
  \label{tab:0_50}
\end{table}

\begin{table}
  \centering
  \caption{Fits of distributions for professional coding years 0-20}
    \begin{tabular}{|c|p{5.5em}|C{0.14\textwidth}|c|c|c|}\hline
    \multicolumn{2}{|c|}{\textbf{Distribution}} & \textbf{Log-likelihood} & \textbf{AIC} & \textbf{BIC} & \textbf{Best Fit} \\\hline\hline
    Normal & mean = 6.80\newline{}sd = 5.13 & -95,336 & 190,676 & 190,693 &  \\\hline
    Poisson & lambda=\newline{}6.80 & -113,532 & 227,066 & 227,074 & \\\hline
    N. Binomial & size = 2.20\newline{}mu = 6.80 & -90,475 & 180,955 & 180,972 & \checkmark{} \\\hline
    \end{tabular}
  \label{tab:0_20}
\end{table}


\begin{table}
  \centering
  \caption{Fits of distributions for professional coding years 21-30}
    \begin{tabular}{|c|p{5.5em}|C{0.14\textwidth}|c|c|c|}\hline
    \multicolumn{2}{|c|}{\textbf{Distribution}} & \textbf{Log-likelihood} & \textbf{AIC} & \textbf{BIC} & \textbf{Best Fit} \\\hline\hline
    Normal & mean = 24.66\newline{}sd = 2.81 & -4,801 & 9,607 & 9,618 &  \checkmark{} \\\hline
    Poisson & lambda=\newline{}24.66 & -5,244 & 10,490 & 10,496 & \\\hline
    N. Binomial & size = 1.08e+08\newline{}mu = 24.66 & -5,244 & 10,492 & 10,503 &  \\\hline
    \end{tabular}
  \label{tab:21_30}
\end{table}

\begin{table}
  \centering
  \caption{Fits of distributions for professional coding years 31-40}
    \begin{tabular}{|c|p{5.5em}|C{0.14\textwidth}|c|c|c|}\hline
    \multicolumn{2}{|c|}{\textbf{Distribution}} & \textbf{Log-likelihood} & \textbf{AIC} & \textbf{BIC} & \textbf{Best Fit} \\\hline\hline
    Normal & mean = 35.02\newline{}sd = 2.82 & -1,286 & 2,576 & 2,585 &  \checkmark{} \\\hline
    Poisson & lambda=\newline{}35.02 & -1,472 & 2,947 & 2,951 & \\\hline
    N. Binomial & size = 1.39e+08\newline{}mu = 35.02 & -1,472 & 2,949 & 2,957 &  \\\hline
    \end{tabular}
  \label{tab:31_40}
\end{table}

\subsection{Comparison of Changes across Years}

\begin{figure}
    \centering
    \includegraphics[width=\myfigwidth]{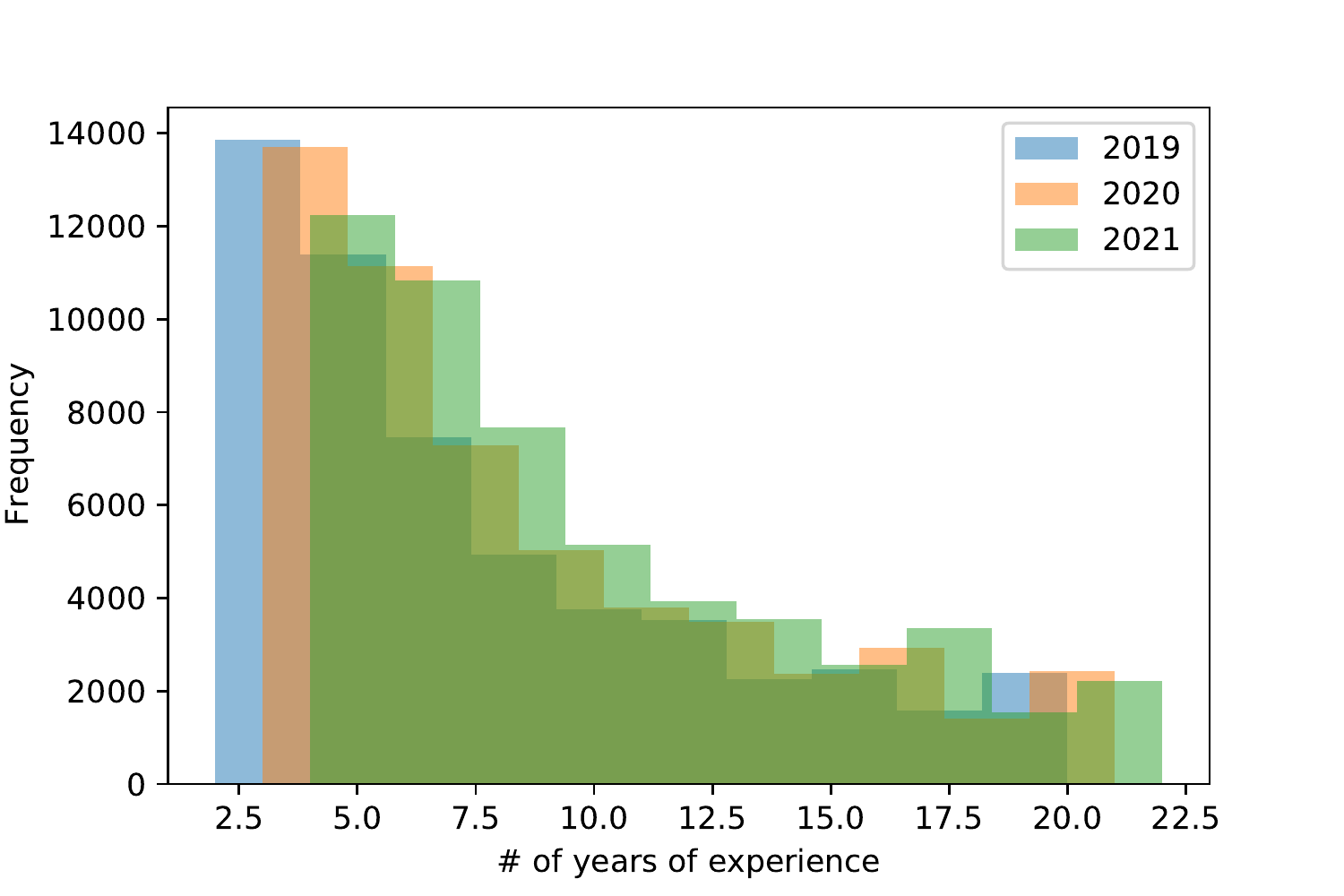}
    \caption{Histogram of professional coders in 2019, 2020 and 2021 with upto 20 years of experience}
    \label{fig:AllYears_Upto20}
\end{figure}

\begin{figure}
    \centering
    \includegraphics[width=\myfigwidth]{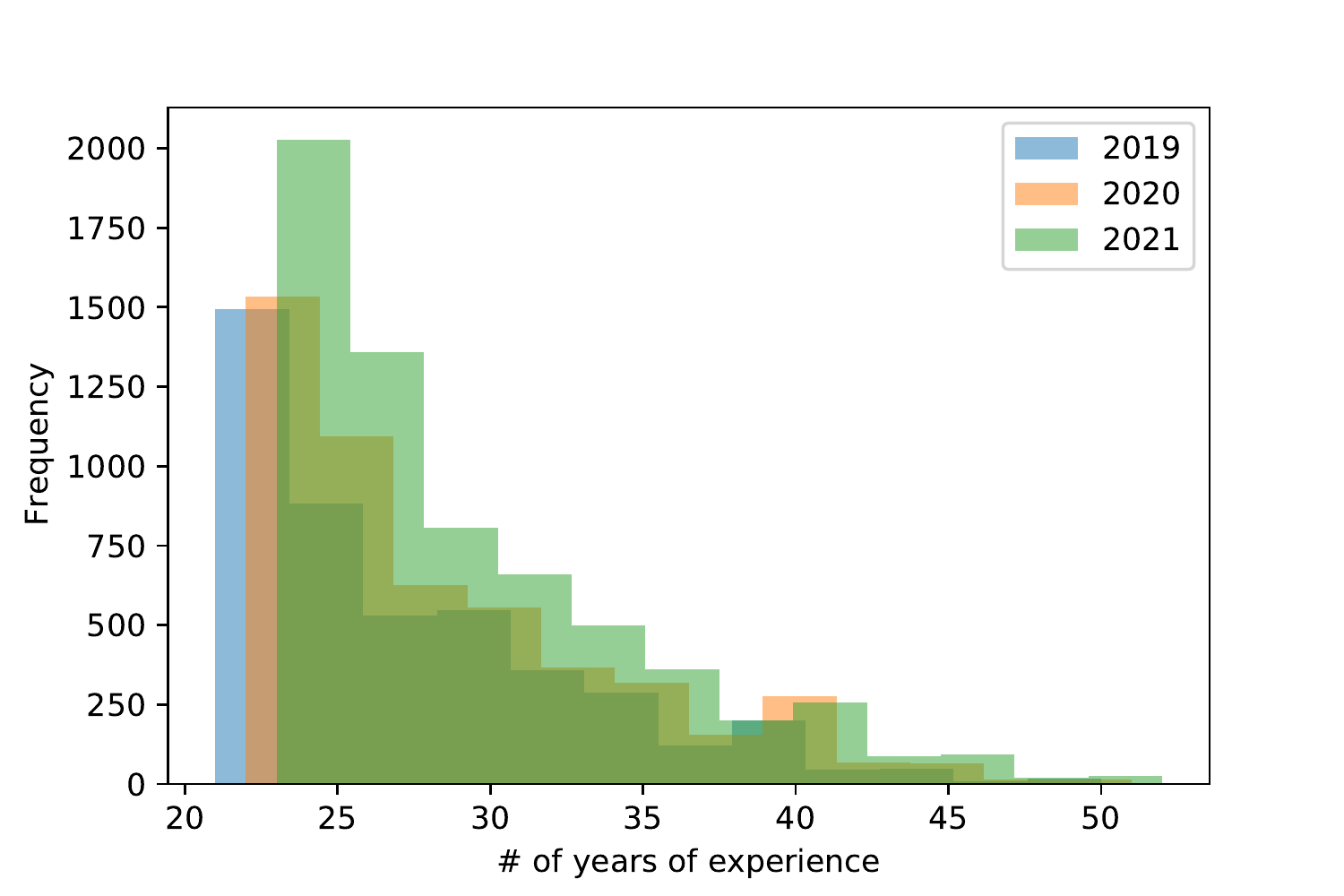}
    \caption{Histogram of professional coders in 2019, 2020 and 2021 between 20 and 50 years of experience}
    \label{fig:AllYears_Bw20and50}
\end{figure}

\begin{figure}
    \centering
    \includegraphics[width=\myfigwidth]{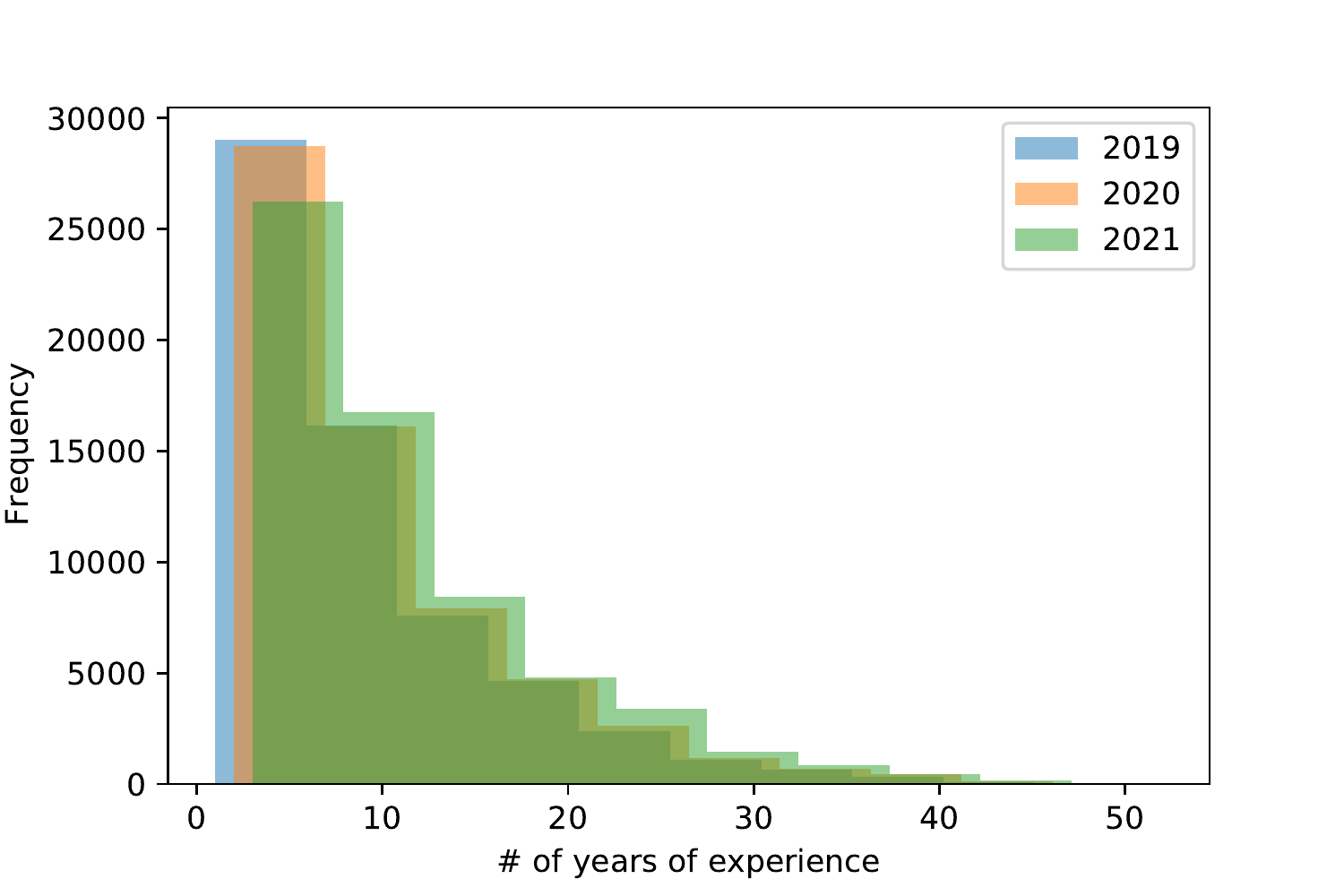}
    \caption{Histogram of all professional coders in 2019, 2020 and 2021}
    \label{fig:AllYears_AllProfessionals}
\end{figure}

\begin{figure}
    \centering
    \includegraphics[width=\myfigwidth]{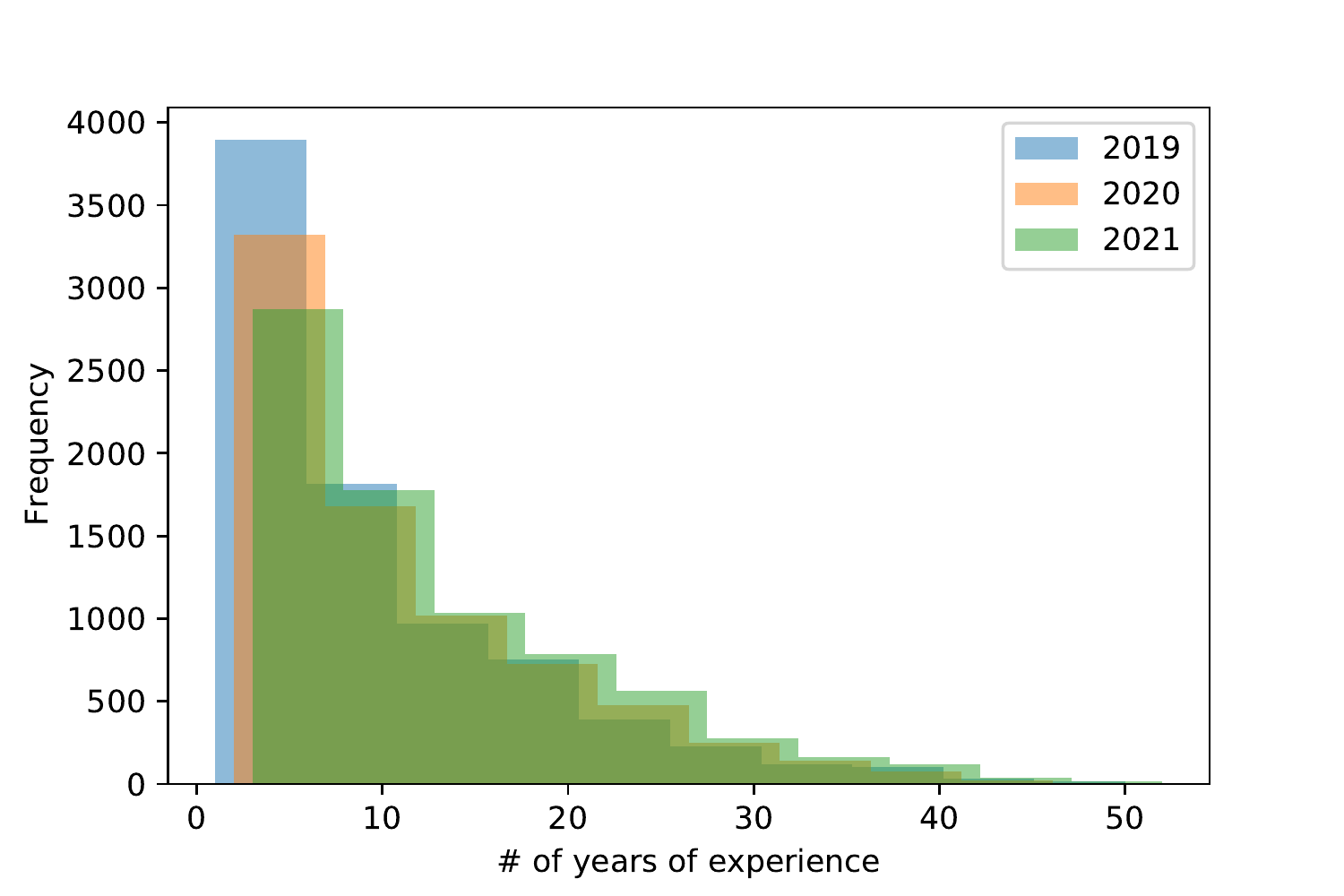}
    \caption{Histogram of all non-professional coders in 2019, 2020 and 2021}
    \label{fig:AllYears_AllNonProfessionals}
\end{figure}

Figure~\ref{fig:AllYears_Upto20} shows histogram of professional coders in 2019, 2020 and 2021 with 2 to 20 years of experience.
Figure~\ref{fig:AllYears_Bw20and50} shows histogram of professional coders in 2019, 2020 and 2021 with 20 to 50 years of experience.
Figure~\ref{fig:AllYears_AllProfessionals} shows histogram of all professional coders in 2019, 2020 and 2021 with more than 1 year of experience.
Figure~\ref{fig:AllYears_AllNonProfessionals} shows histogram of all non-professional coders in 2019, 2020 and 2021.

\section{TravisTorrent Dataset }
\subsection{Dataset Overview}

The TravisTorrent dataset contains metadata from \texttt{git} repositories for 1283 Java and Ruby projects hosted at Github up to January 27, 2017. It  consists of an \texttt{SQLite} table containing following columns:
  \begin{itemize}
      \item \textbf{project}: project name on Github, in the form 'owner/project'
      \item \textbf{sha}: the commit id
      \item \textbf{message}: the commit message
      \item \textbf{date}: the commit date
      \item \textbf{author name}: name of the commit author
      \item \textbf{author email}: email of the commit author
  \end{itemize}
After some minor preprocessing, the dataset contains 2,249,243 rows.

This dataset has been analyzed to provide insights on commit volumes on repository-basis and developer-basis. This dataset also provides unique information on temporal nature of the commits, which enables us to analyze measures such as commit rates for various time intervals.

\subsection{Project-Level Metadata Information}

\subsubsection{Commit Counts of Projects}
\begin{figure}
    \centering
    \includegraphics[width=10 cm, height= 7 cm]{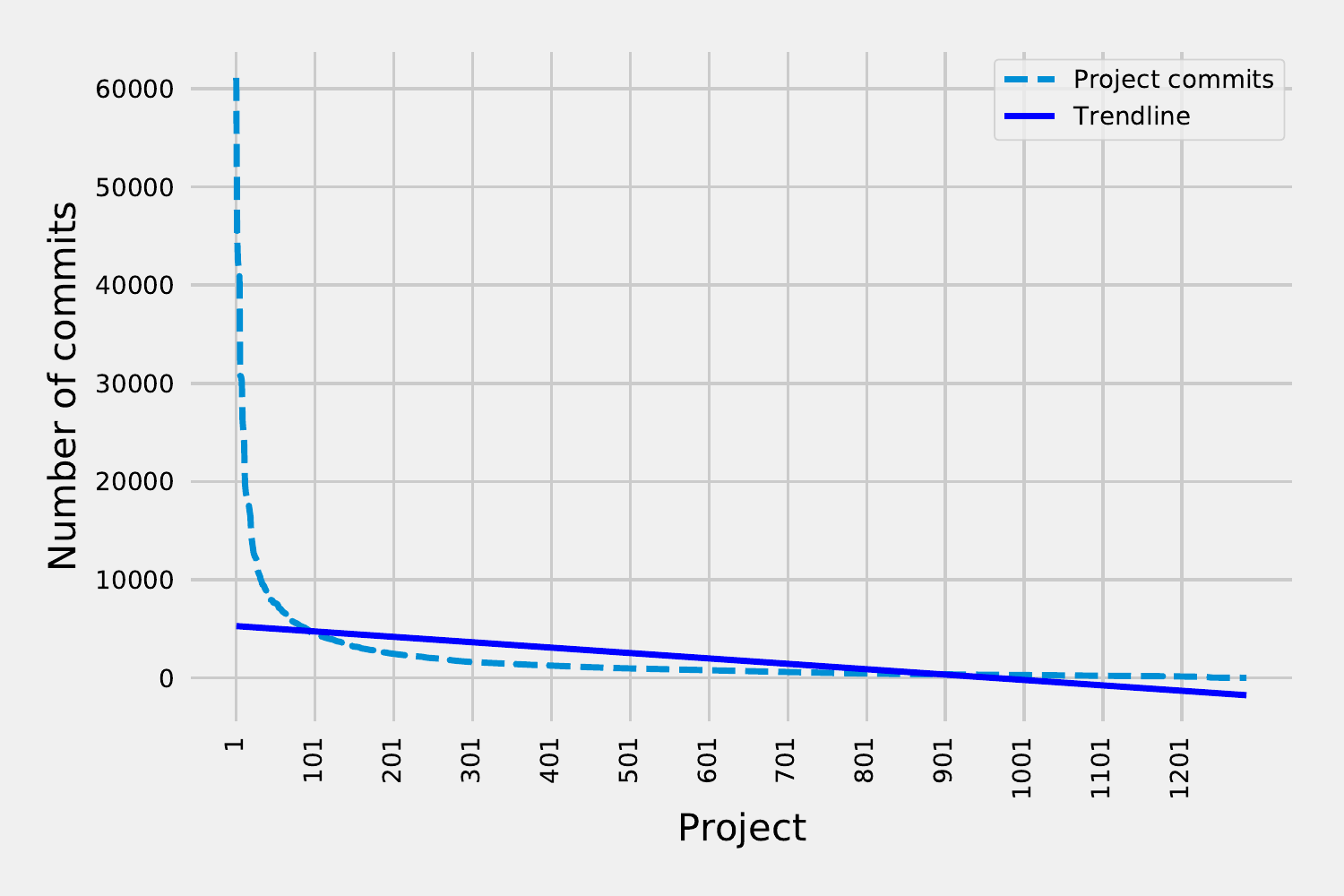}
    \caption{Project commits with trend line}
    \label{fig:project_commit_trend_line}
\end{figure}

Figure~\ref{fig:project_commit_trend_line} shows project-level commits (high to low) with a trend-line. The slope and intercept of the trend-line are -5.49 and 5278.26. 

\subsubsection{Granular View of Commits for Top 100 Projects}

\begin{figure}
    \centering
    \includegraphics[width=10 cm, height= 6.7 cm]{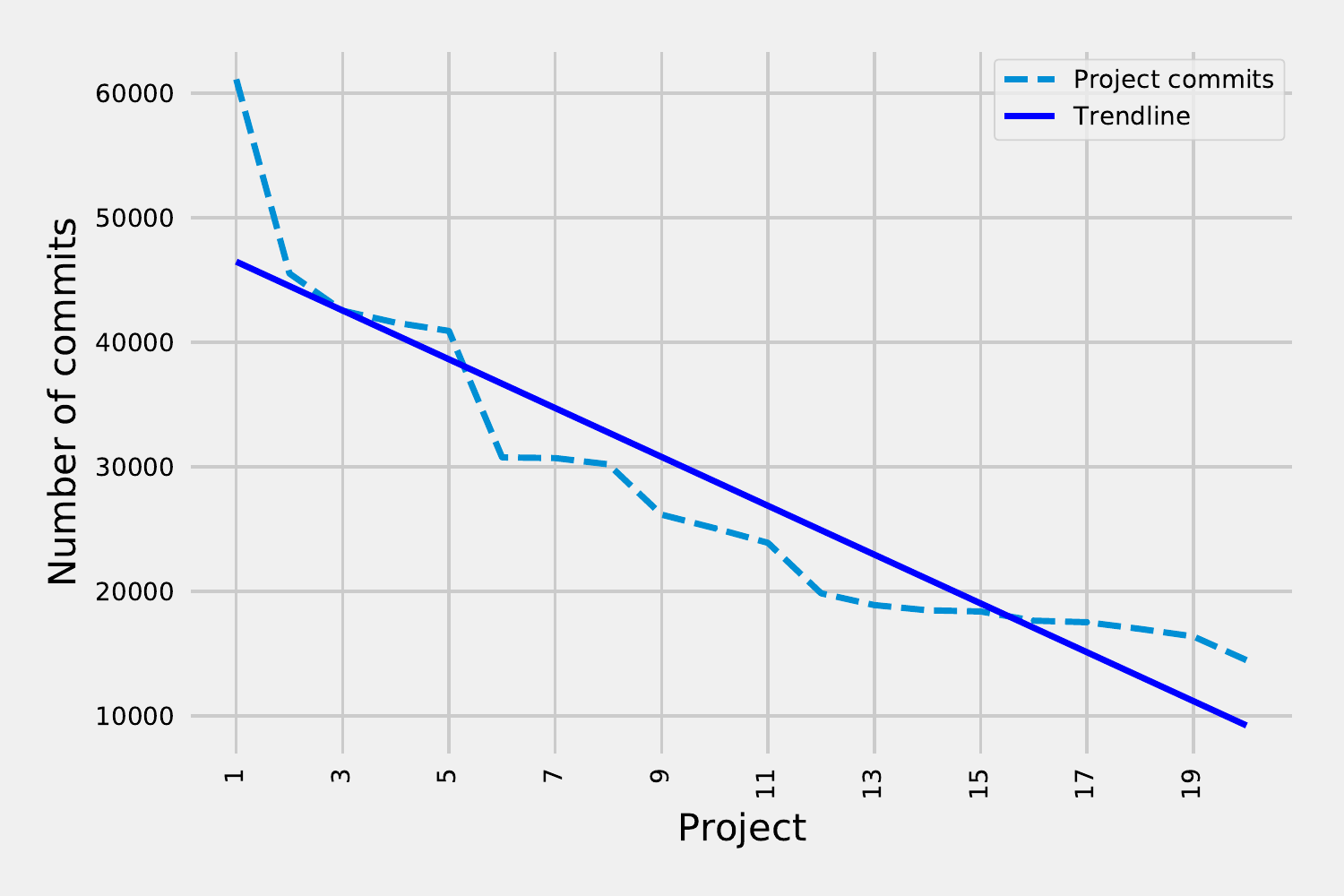}
    \caption{Project commits with trend line for top 20 projects}
    \label{fig:project_commit_trend_line_top_20}
\end{figure}

\begin{figure}
    \centering
    \includegraphics[width=10 cm, height= 8 cm]{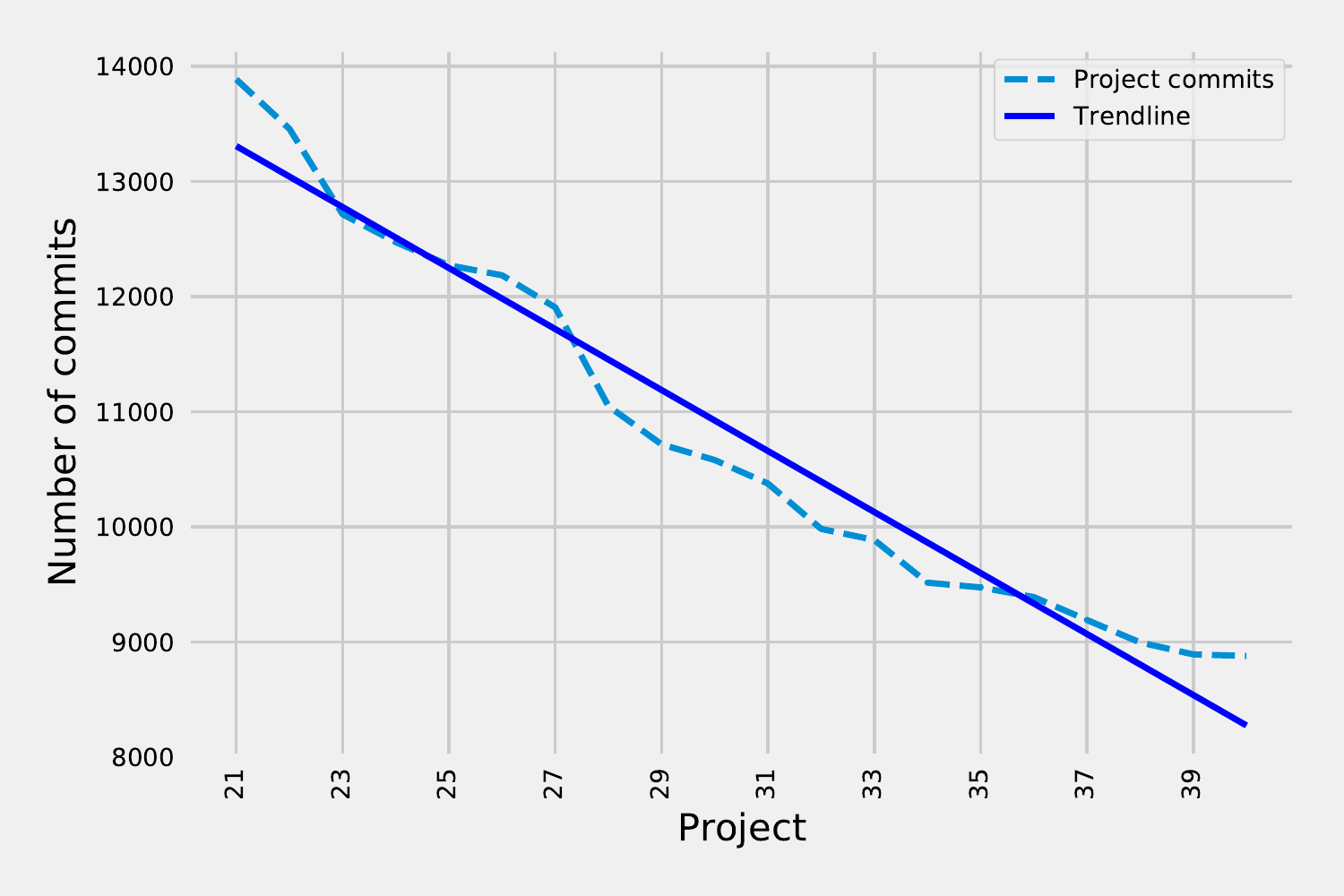}
    \caption{Project commits with trend line for top 21-40 projects}
    \label{fig:project_commit_trend_line_top_21_40}
\end{figure}

\begin{figure}
    \centering
    \includegraphics[width=10 cm, height= 8 cm]{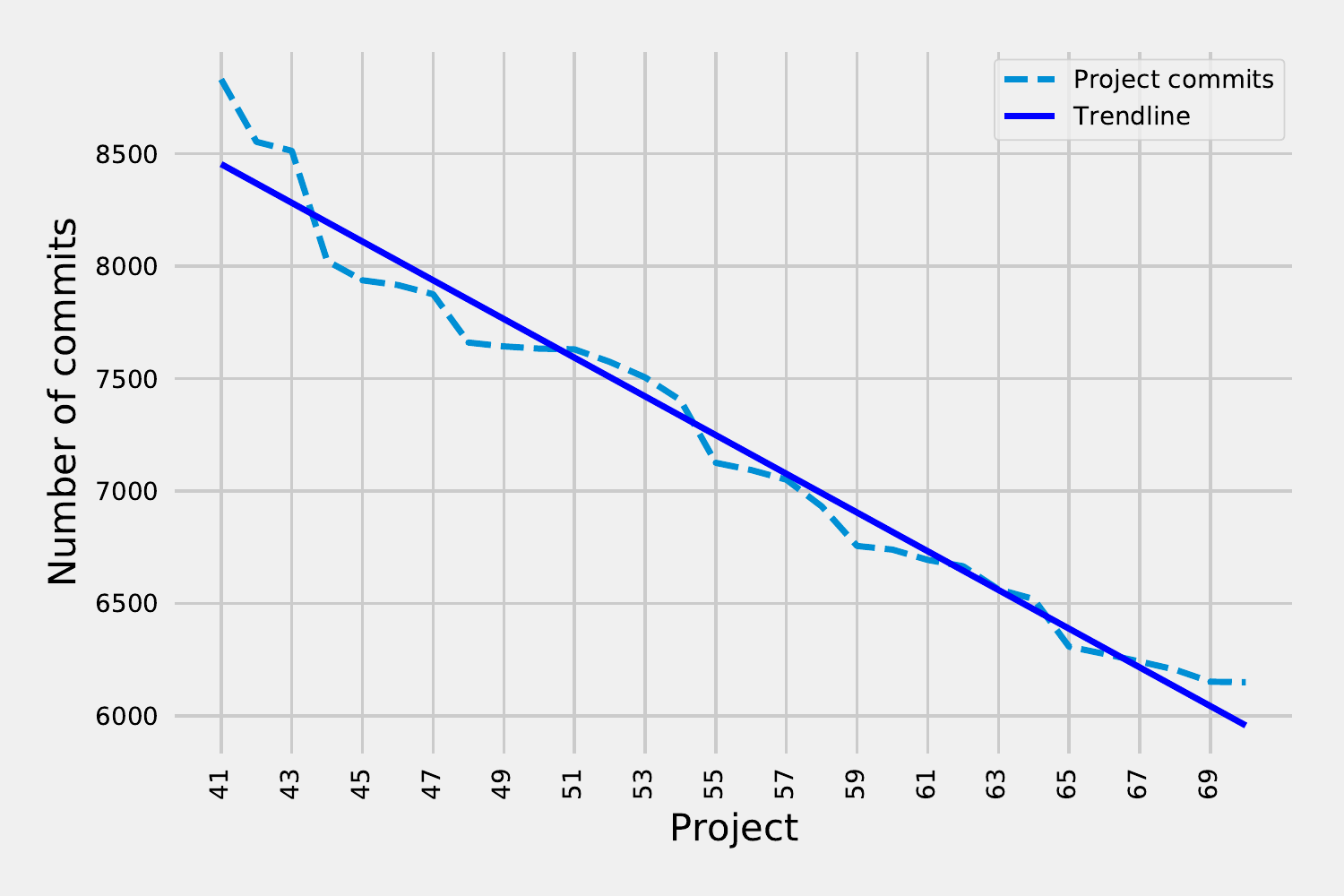}
    \caption{Project commits with trend line for top 41-70 projects}
    \label{fig:project_commit_trend_line_top_41_70}
\end{figure}

\begin{figure}
    \centering
    \includegraphics[width=10 cm, height= 8 cm]{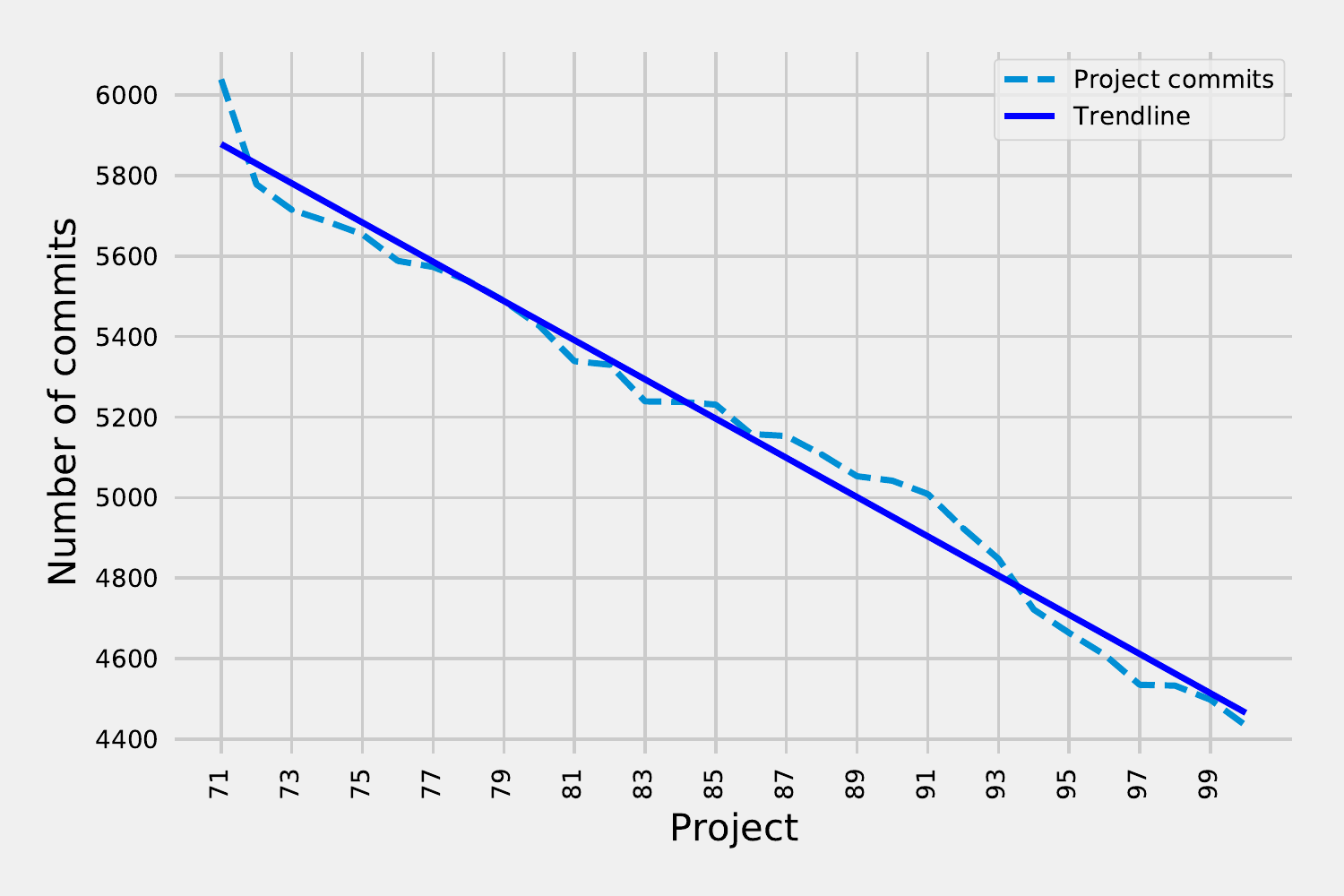}
    \caption{Project commits with trend line for top 71-100 projects}
    \label{fig:project_commit_trend_line_top_71_100}
\end{figure}

Figure~\ref{fig:project_commit_trend_line_top_20},
Figure~\ref{fig:project_commit_trend_line_top_21_40},
Figure~\ref{fig:project_commit_trend_line_top_41_70}, and Figure~\ref{fig:project_commit_trend_line_top_71_100} provides granular view of commits for 1-20, 21-40, 41-70 and 71-100 projects with trend-line. The slopes of the trendline are -1958.95, -264.86, -86.11, and -48.71 respectively. The intercepts of the trend-line are 48426.97, 18869.19, 11984.57, and 9336.82 respectively.

\subsection{Project-Level Commit Rates}

\begin{figure}
    \centering
    \includegraphics[width=10 cm, height= 6 cm]{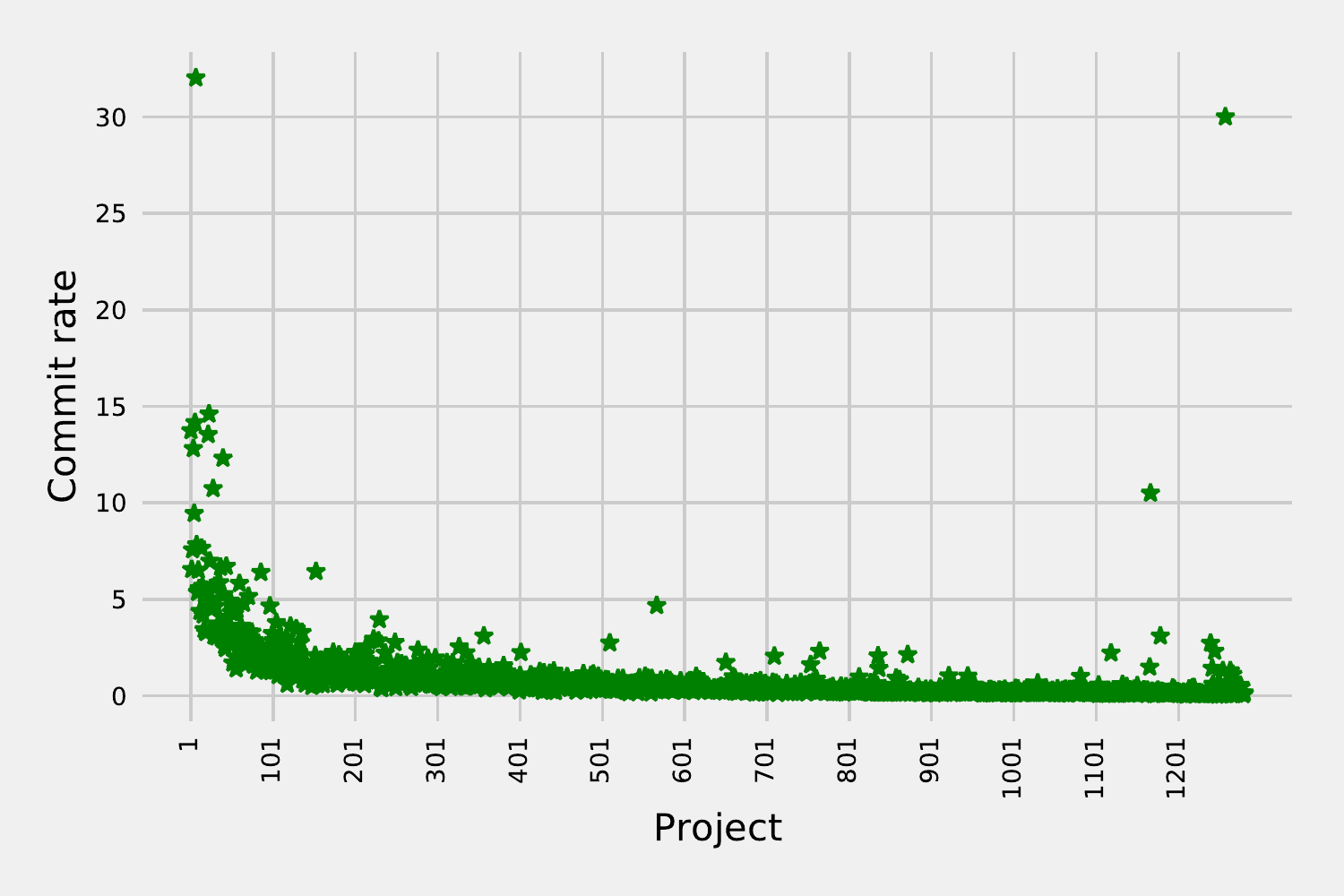}
    \caption{Project-level commit rate for 1281 projects}
    \label{fig:commit_rate_project}
\end{figure}

Figure~\ref{fig:commit_rate_project} shows project-level commit rate for 1281 projects. Although total number of project is 1283, two of the projects had commits on a particular date only. While calculating commit rate, those projects were excluded.

\begin{figure}
    \centering
    \includegraphics[width=10 cm, height= 7 cm]{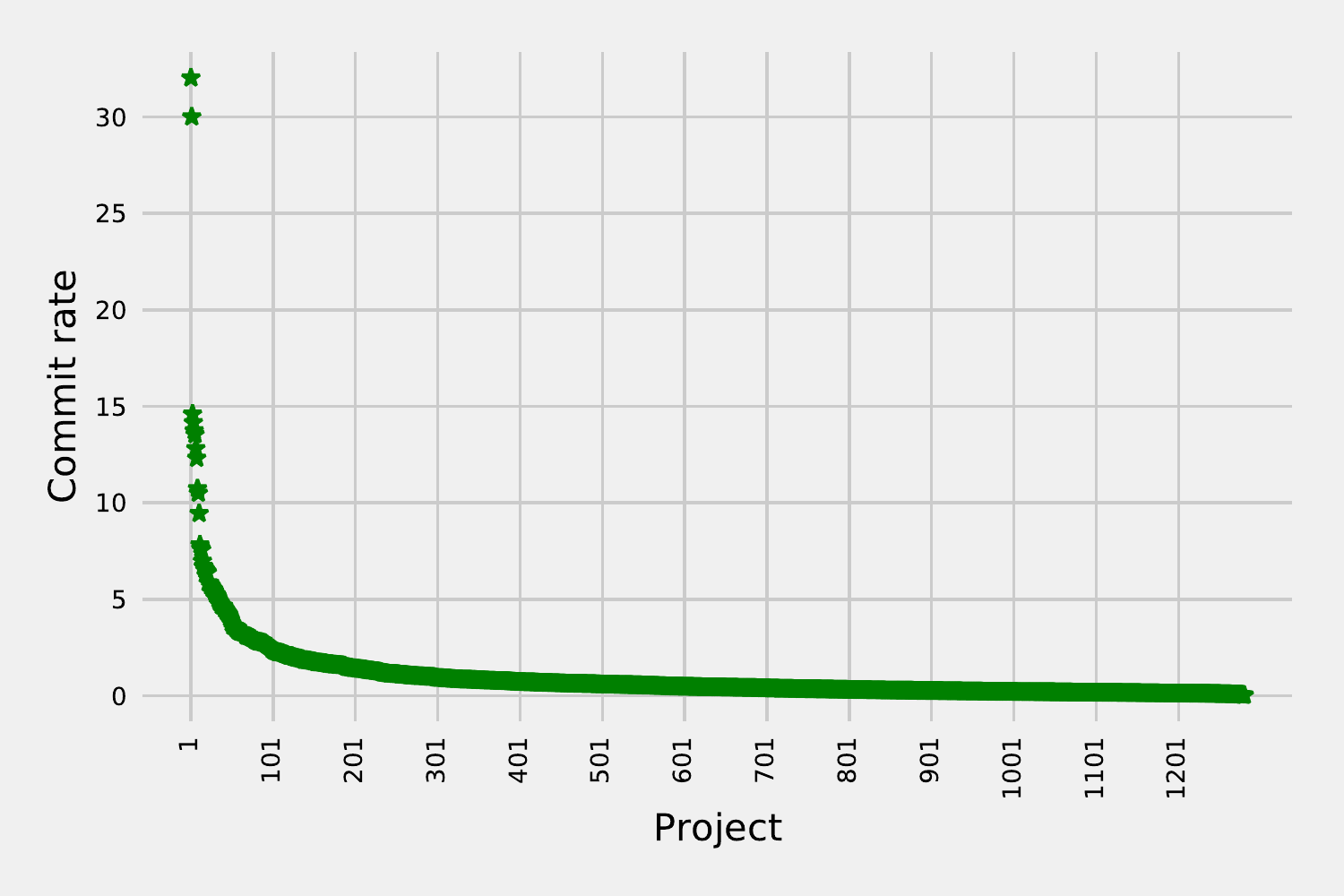}
    \caption{Project-level commit rate (high-low)}
    \label{fig:commit_rate_project_rearranged}
\end{figure}

Figure~\ref{fig:commit_rate_project_rearranged} shows the rearranged (high-low) project-level commit rate for 1281 projects. As we can see from figure~\ref{fig:commit_rate_project_rearranged}, most of the projects had very low (<=2) commit rate. 

\subsection{Author-Level Commit Metadata Information}

\subsubsection{Commit Counts of Authors}

\begin{figure}
    \centering
    \includegraphics[width=10 cm, height= 6.7 cm]{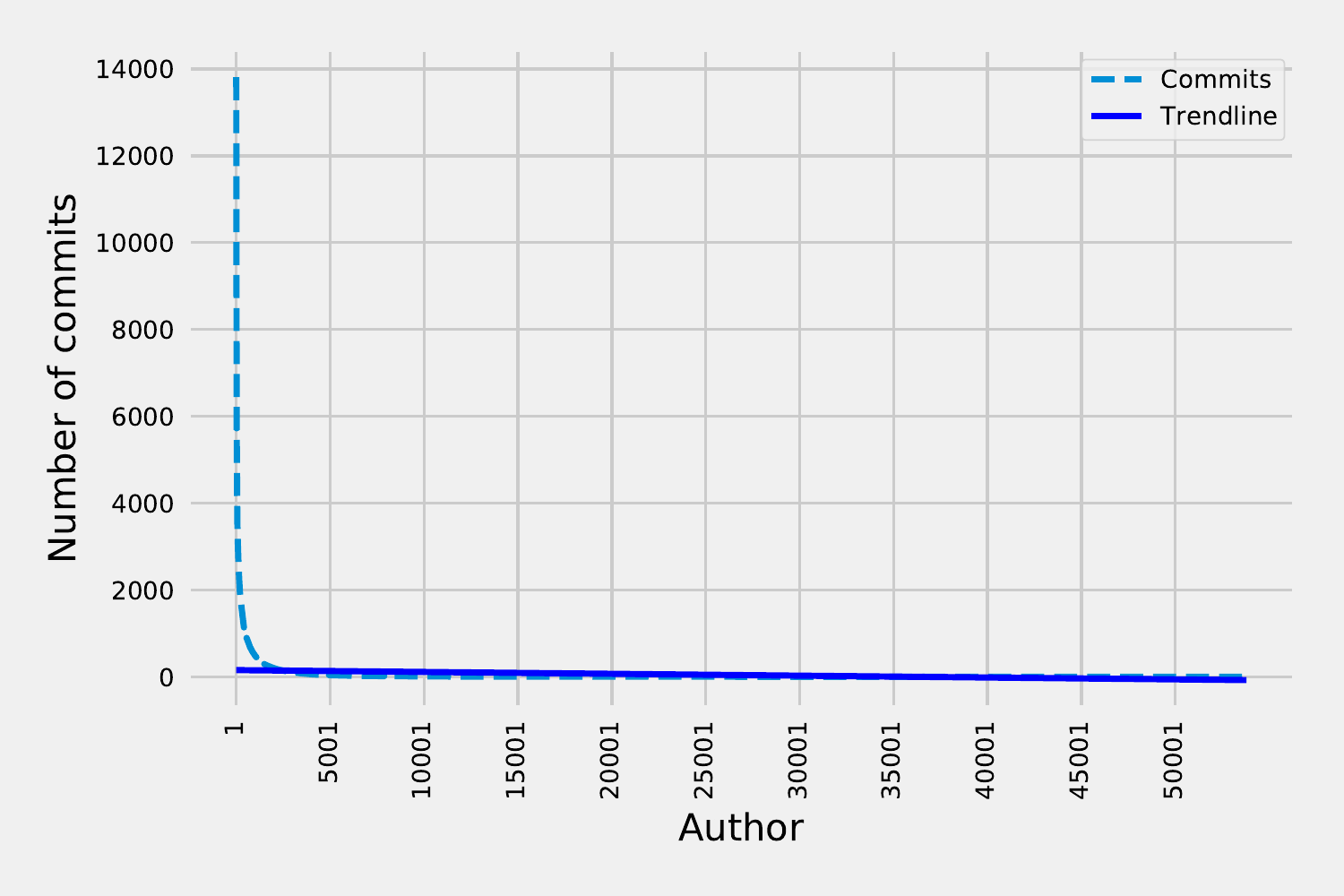}
    \caption{Commit authors with trend line}
    \label{fig:commit_author_trend_line}
\end{figure}

Figure~\ref{fig:commit_author_trend_line} shows author-level commits (high-low) with trend-line. The slope and intercept of the trend-line are -0.00427 and 156.67.

\subsubsection{Granular View of Commits for Top 100 Authors}

\begin{figure}
    \centering
    \includegraphics[width=10 cm, height= 6.6 cm]{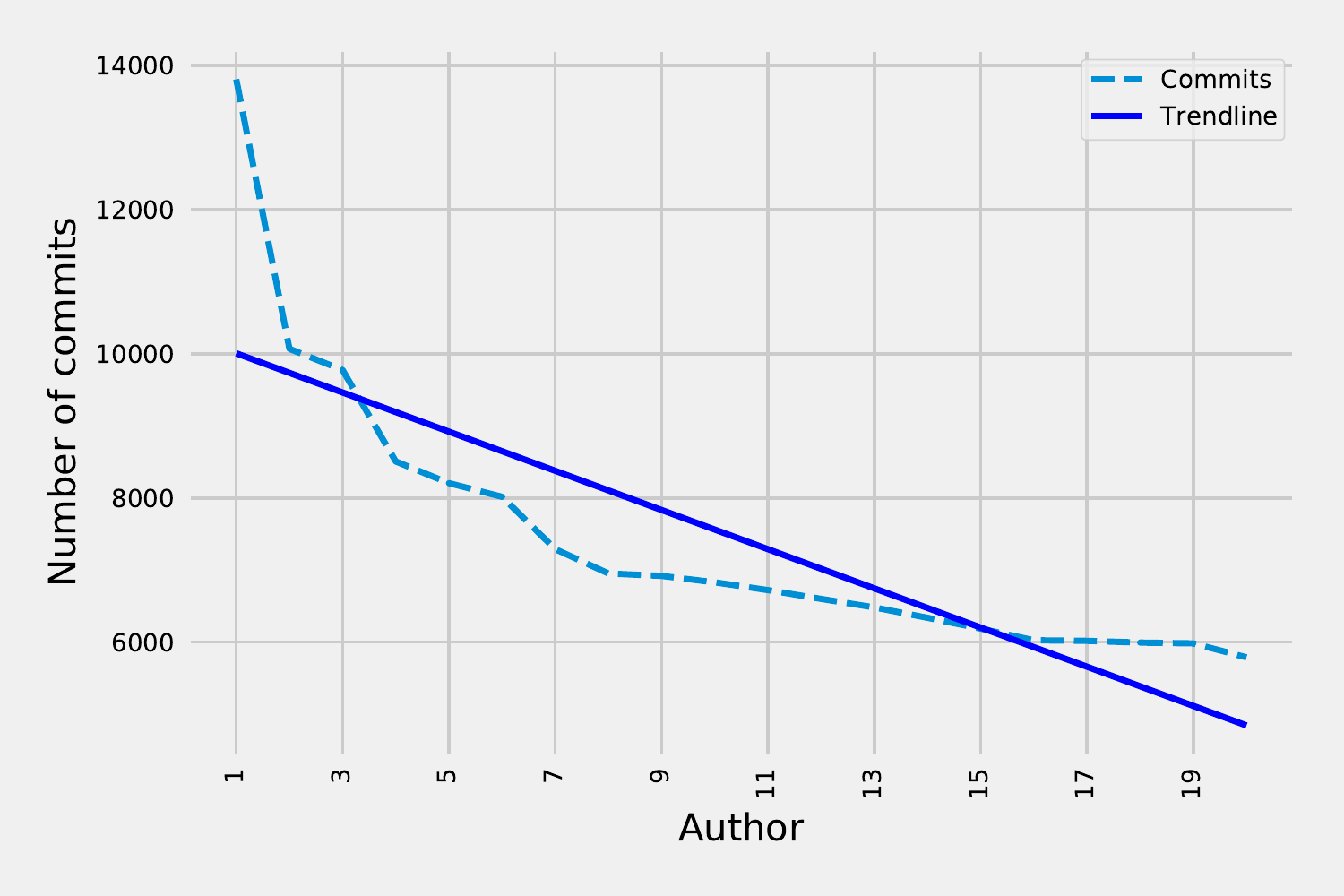}
    \caption{Commit authors with trend line for top 20 authors}
    \label{fig:commit_author_trend_line_top_20}
\end{figure}

\begin{figure}
    \centering
    \includegraphics[width=10 cm, height= 8 cm]{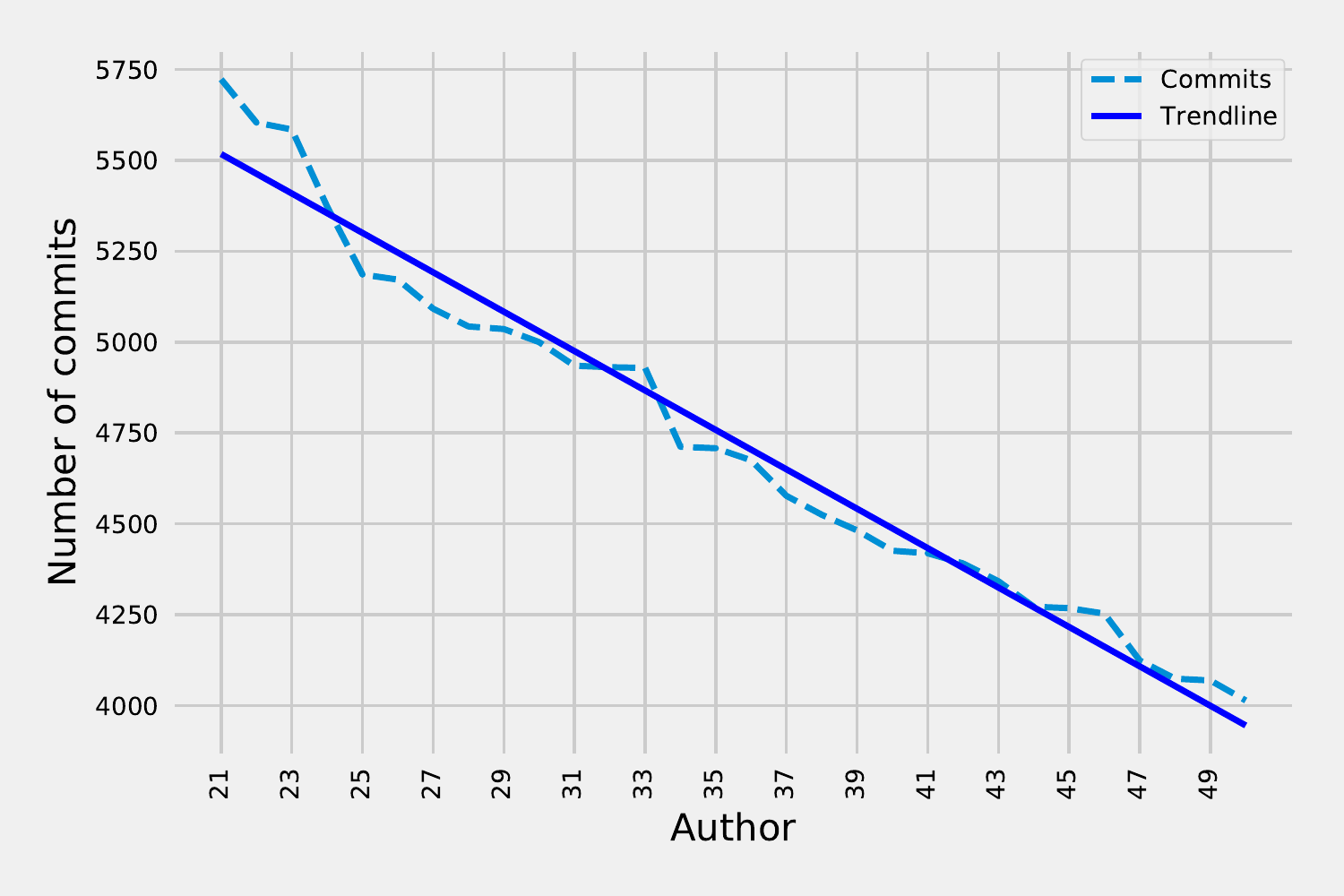}
    \caption{Commit authors with trend line for top 21-50 authors}
    \label{fig:commit_author_trend_line_top_21_50}
\end{figure}

\begin{figure}
    \centering
    \includegraphics[width=10 cm, height= 8 cm]{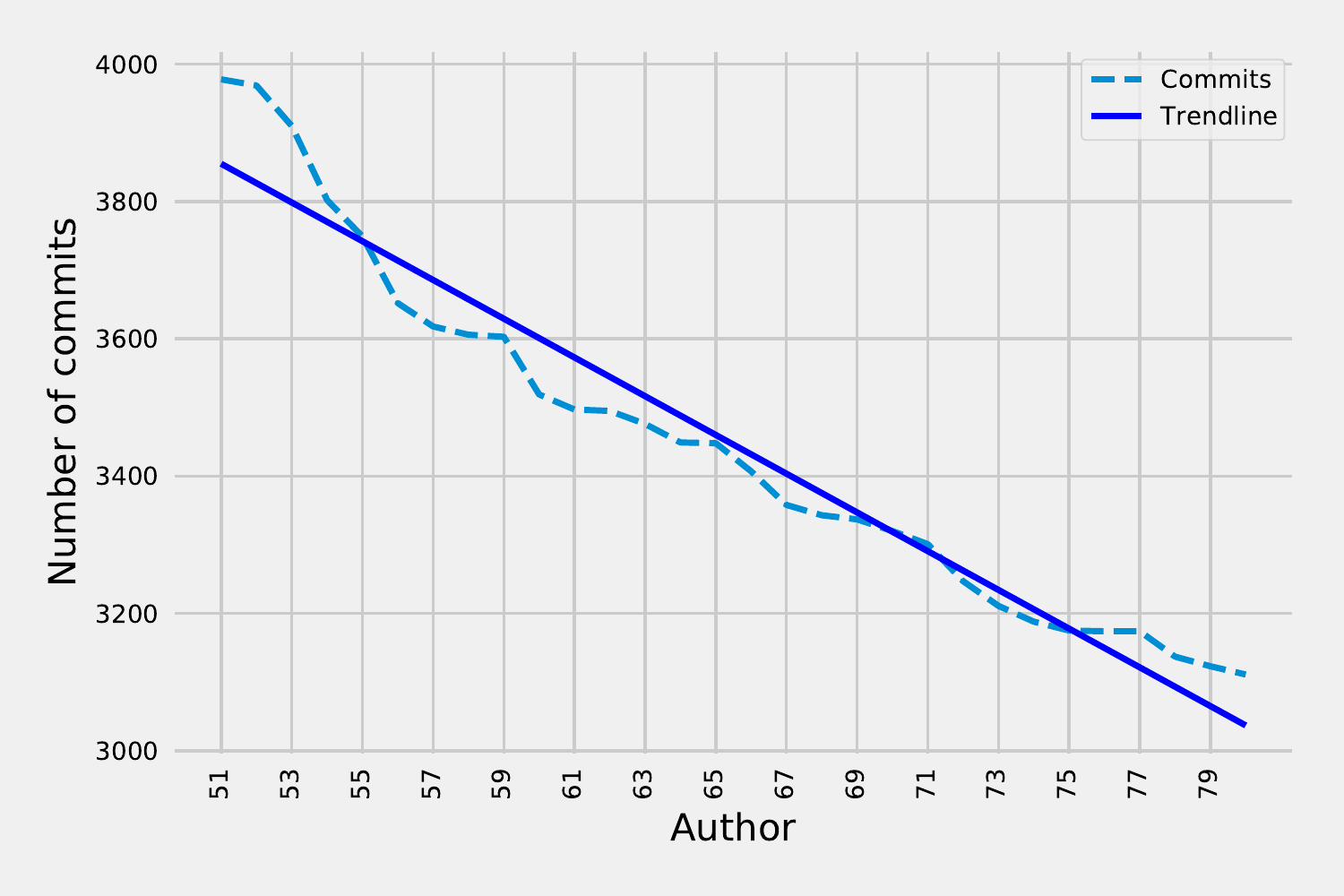}
    \caption{Commit authors with trend line for top 51-80 authors}
    \label{fig:commit_author_trend_line_top_51_80}
\end{figure}

\begin{figure}
    \centering
    \includegraphics[width=10 cm, height= 8 cm]{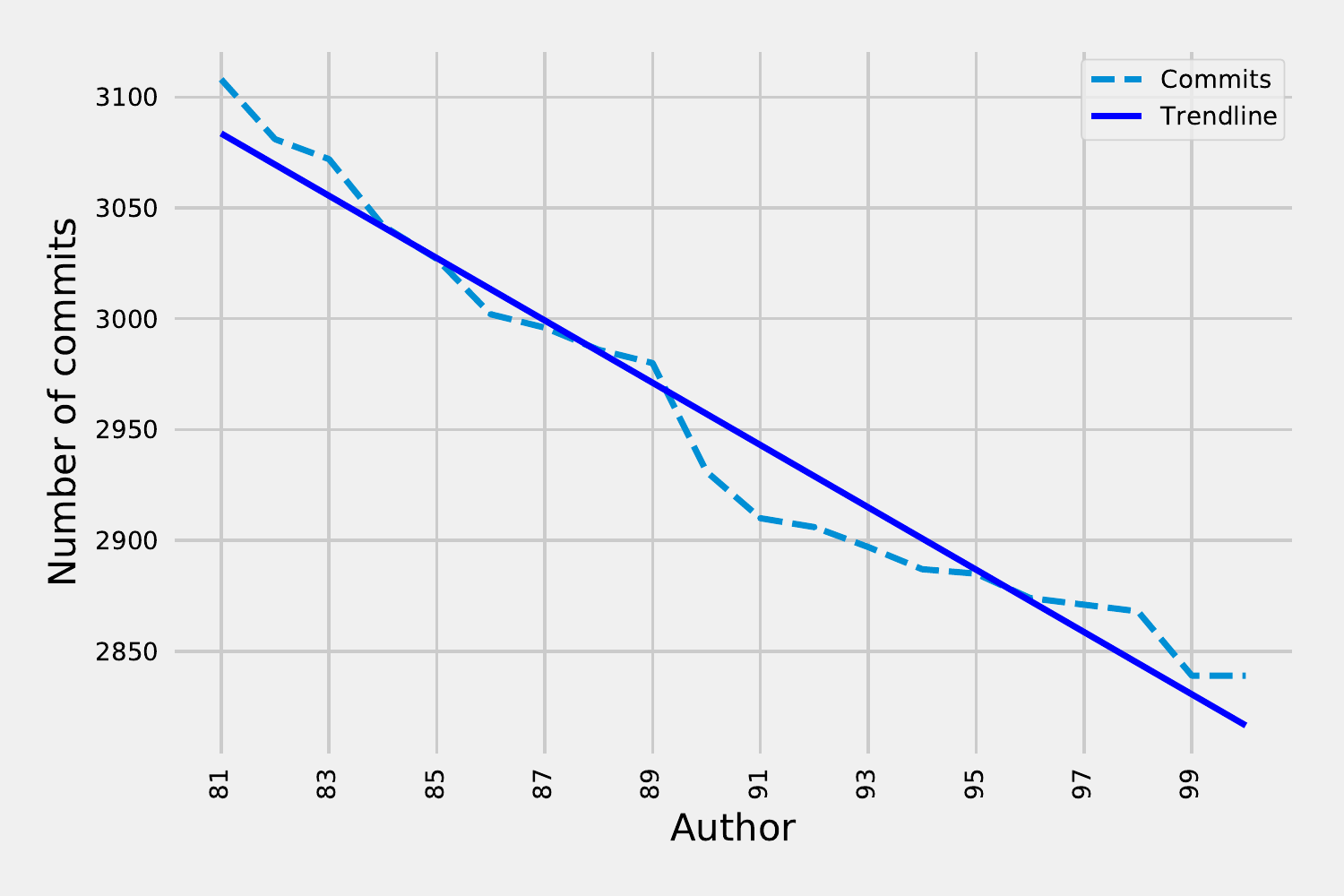}
    \caption{Commit authors with trend line for top 81-100 authors}
    \label{fig:commit_author_trend_line_top_81_100}
\end{figure}

Figure~\ref{fig:commit_author_trend_line_top_20},
Figure~\ref{fig:commit_author_trend_line_top_21_50},
Figure~\ref{fig:commit_author_trend_line_top_51_80}, and Figure~\ref{fig:commit_author_trend_line_top_81_100} provides granular view of commits for 1-20, 21-50, 51-80 and 81-100 authors with trend-line. The slopes of the trendline are -271.74, -54.21, -28.22, and -14.06 respectively. The intercepts of the trend-line are 10278.89, 6655.92, 5294.05, and 4222.15 respectively.

\subsection{Author-Level Commit Rates}

\begin{figure}
    \centering
    \includegraphics[width=10 cm, height= 7.3 cm]{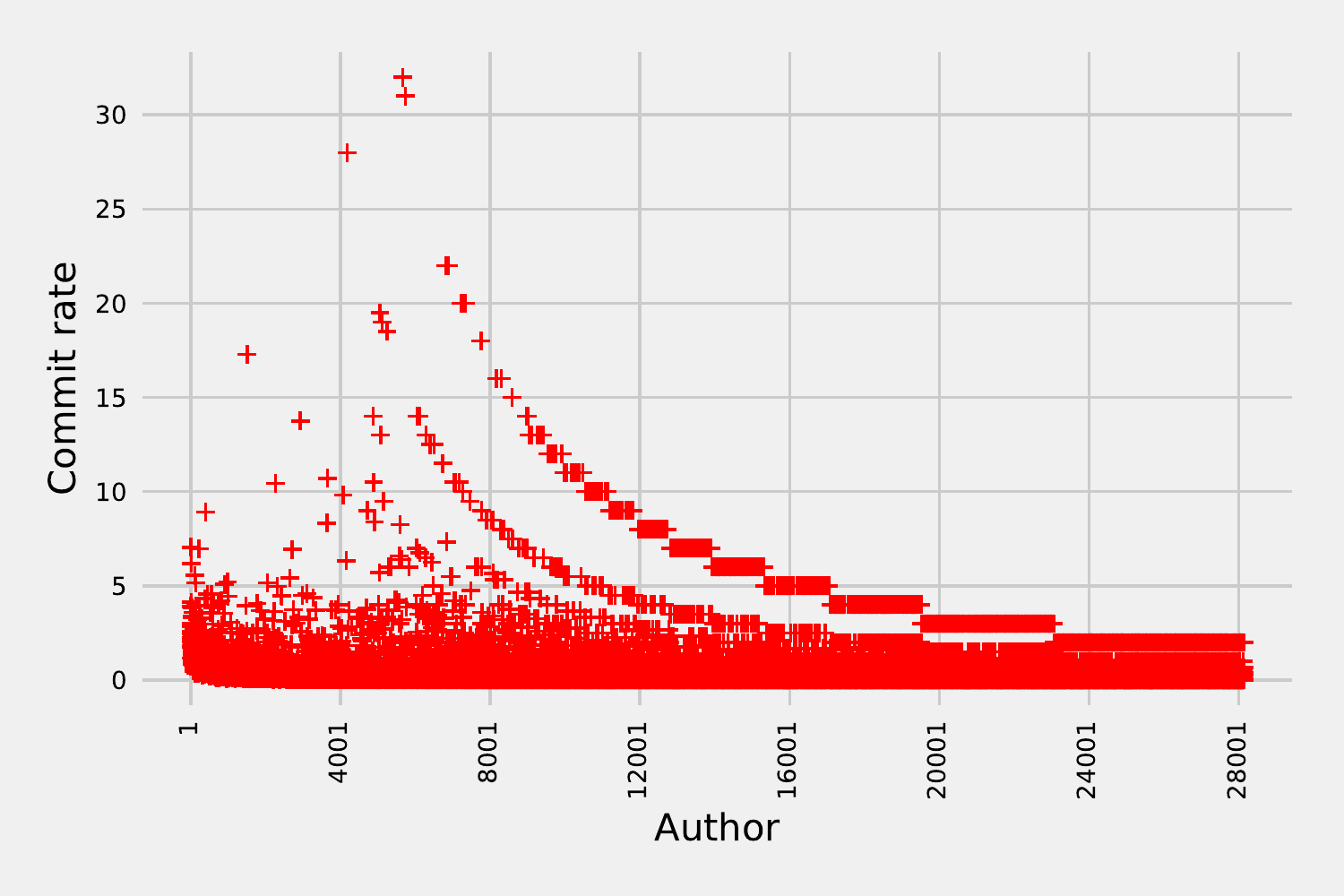}
    \caption{Author-level commit rates}
    \label{fig:commit_rate_author}
\end{figure}

Figure~\ref{fig:commit_rate_author} shows author-level commit rate for over 28k authors. About 50 percent of the authors were excluded because either they committed only once or on a particular date only.

\begin{figure}
    \centering
    \includegraphics[width=10 cm, height= 7.3 cm]{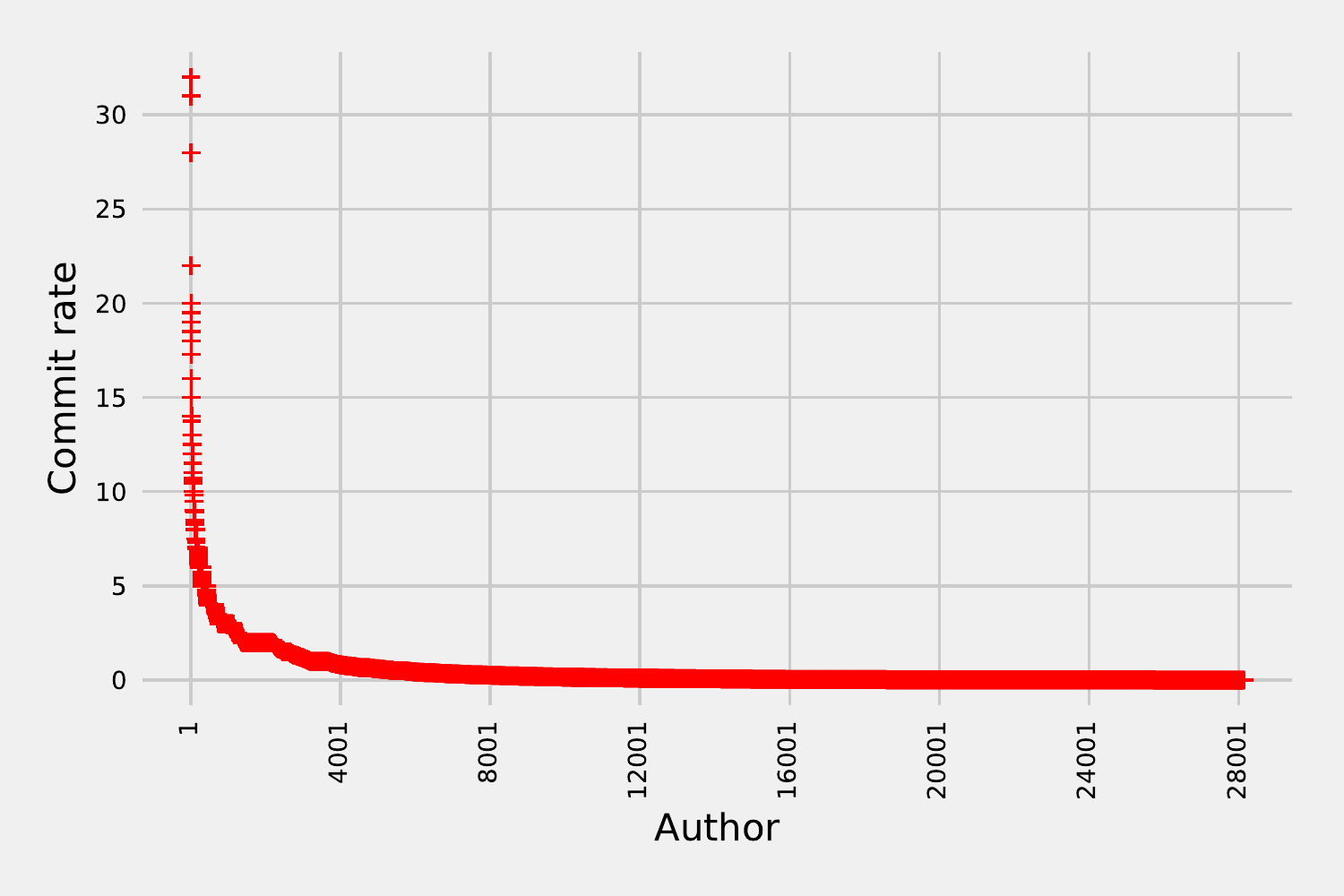}
    \caption{Author-level commit rates (high-low)}
    \label{fig:commit_rate_author_rearranged}
\end{figure}

Figure~\ref{fig:commit_rate_author_rearranged} shows the rearranged (high-low) author-level commit rate for over 28k authors. As we can see from Figure~\ref{fig:commit_rate_author_rearranged}, most of the authors had very low (<=2) commit rate.

\subsection{Weekly Commit Trends}

\begin{figure}
    \centering
    \includegraphics[width=10 cm, height= 7.3 cm]{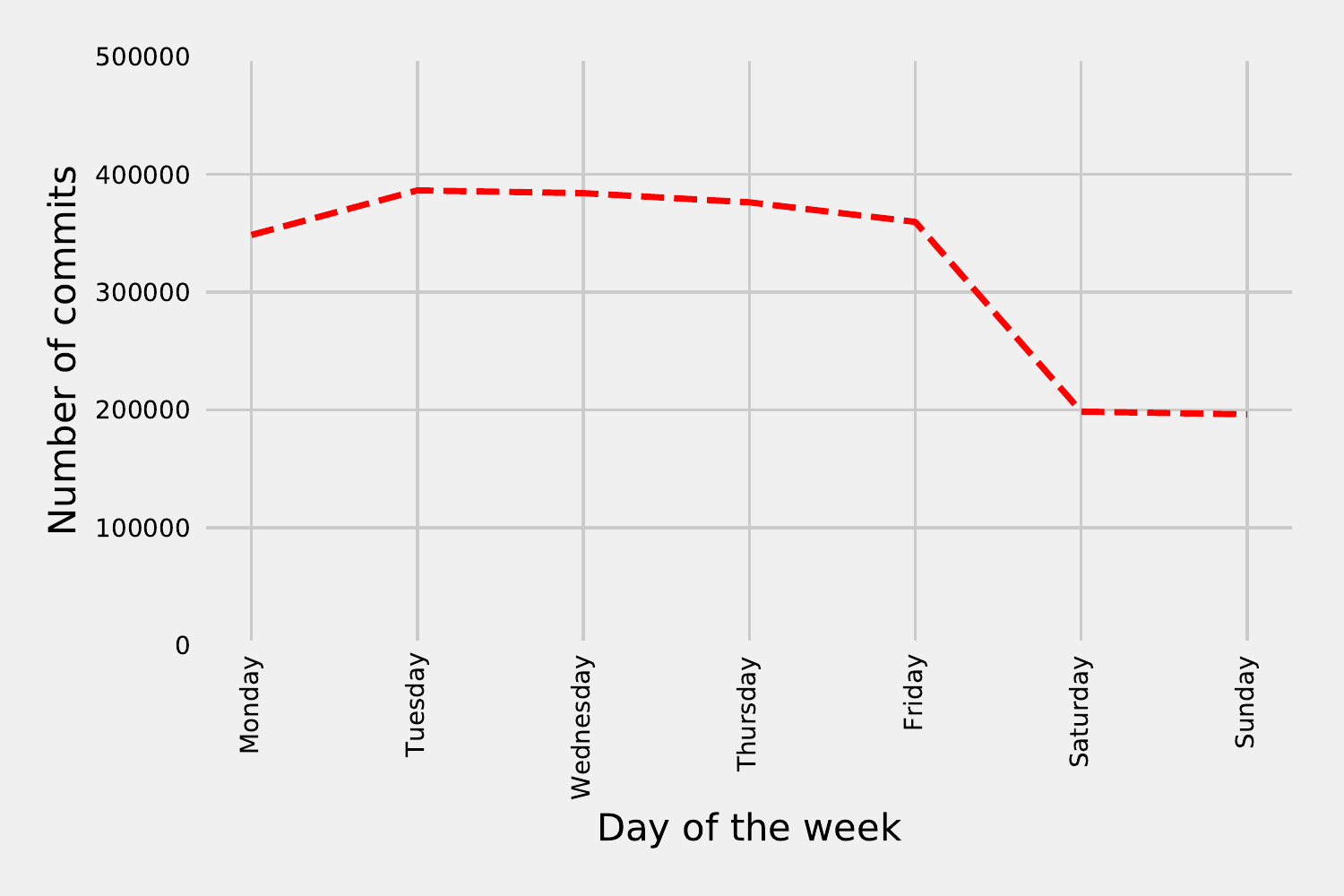}
    \caption{Weekly commit trend for total number of commits}
    \label{fig:weekly_commit_trend}
\end{figure}

Figure~\ref{fig:weekly_commit_trend} shows day-wise commit counts in a typical week for the entire time period (1998-2017) of the data set. 

\subsection{Overall Commit Timelines}

\begin{figure}
    \centering
    \includegraphics[width=10 cm, height= 7 cm]{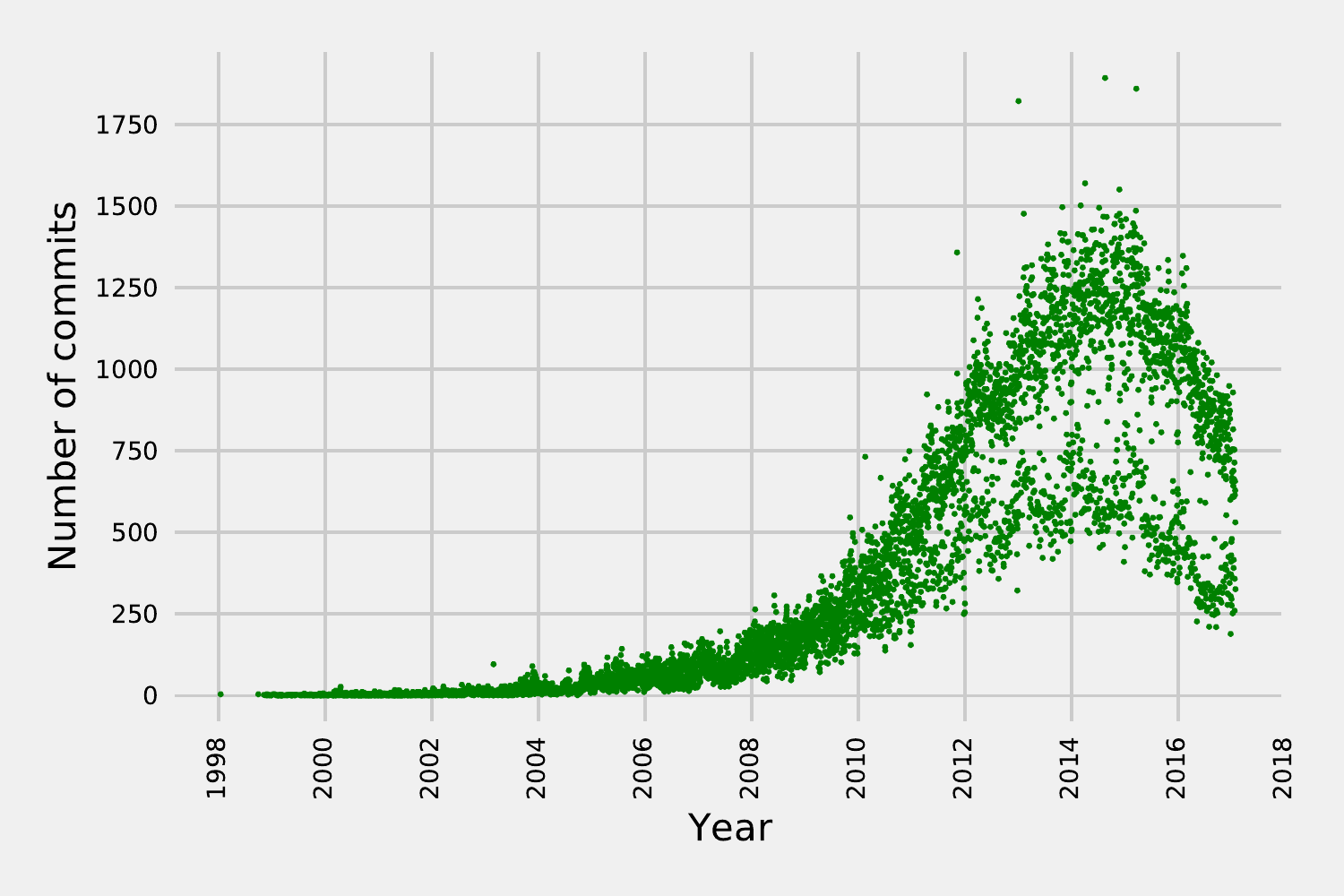}
    \caption{Time series of commit counts for the entire time frame (1998-2017)}
    \label{fig:TravisTorrent timeseries}
\end{figure}

Figure~\ref{fig:TravisTorrent timeseries} shows time series of the commit counts for the entire time frame (1998-2017) of the data set.

\begin{figure}
    \centering
    \includegraphics[width=10 cm, height= 7.3 cm]{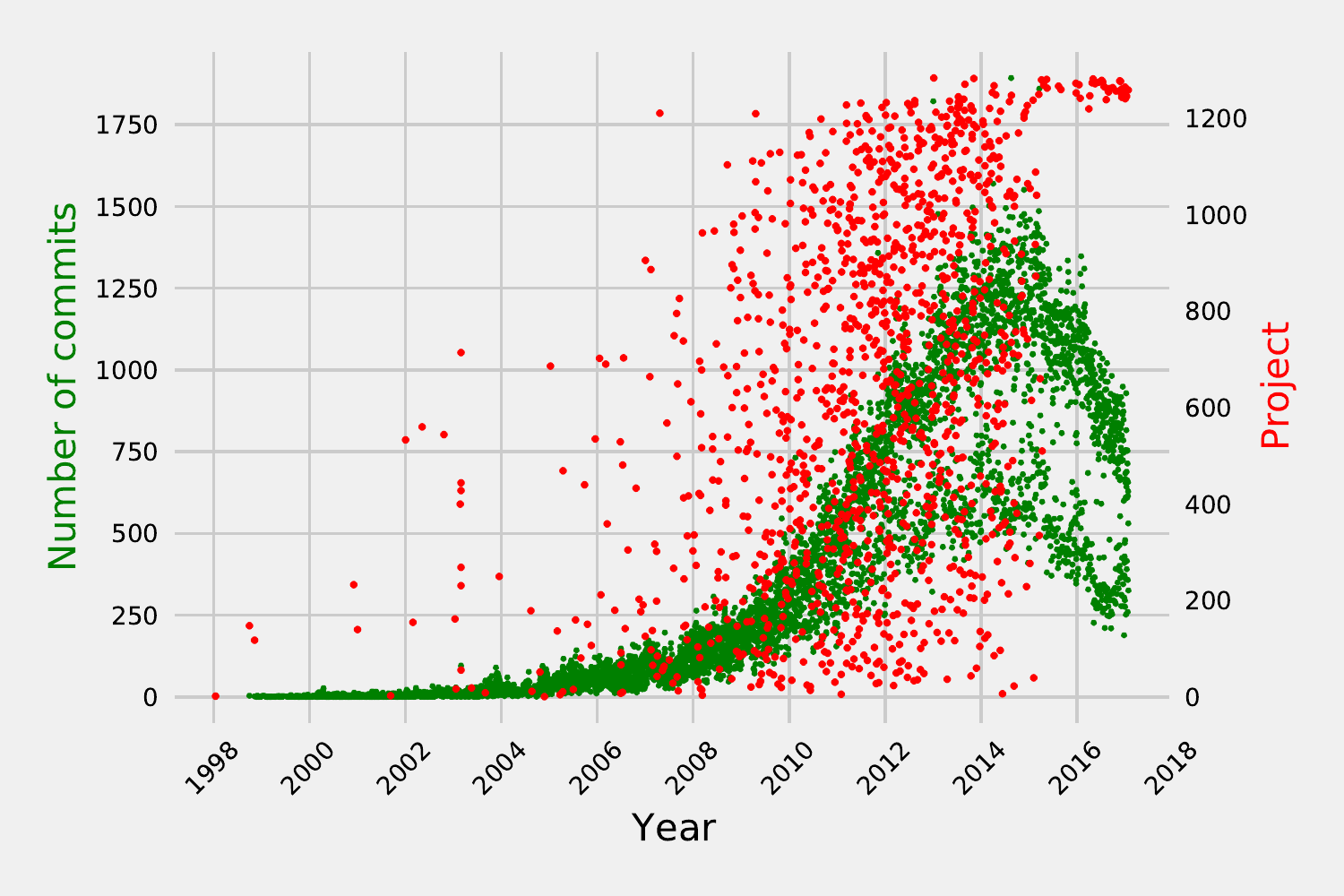}
    \caption{Time series of commit counts for the entire time frame (1998-2017) with project start date overlay}
    \label{fig:TravisTorrent timeseries overlay}
\end{figure}

Figure~\ref{fig:TravisTorrent timeseries overlay} shows the overlay of project start date on the time series of the commit counts for the entire time frame (1998-2017) of the data set. As we can see from the figure~\ref{fig:TravisTorrent timeseries overlay}, most of the projects started just before the year 2014 causing the peak of commit counts in the year 2014.  

\section{Additional Datasets}

\subsection{Bug Datasets}

This is a collection of datasets used in the literature \cite{acoss2019,deeptriage2019,dupbug2014,redis2017,bug_fix2019} and portions of it are available on GitHub (\url{https://github.com/logpai/bugrepo}) for analyzing software bugs.
Table~\ref{tab:bug_dataset} shows the names of the datasets and their number of metadata fields and the number of projects covered in each bug data set.
This collection of datasets is potentially useful in our project to serve as training data for bug patterns in our machine learning methods.

\begin{table}
  \centering
  \caption{Collection of Bug Datasets and their Parameters}
    \begin{tabular}{|l|c|c|}\hline
    \textbf{Dataset} & \textbf{Metadata Fields} & \textbf{Projects Covered} \\\hline\hline
    Bug triage & 8     & 3 \\
    Bugrepo & 11    & 5 \\
    SEOSS & 17    & 33 \\
    Rediscovery & 14    & 3 \\
    10 years bug fixing activity & 53    & 55 \\
    Generating duplicate bug  & 13    & 4 \\\hline
    \end{tabular}%
  \label{tab:bug_dataset}%
\end{table}%

\subsection{Android Application Development Dataset}

The ``Dataset of Commit History of Real-World Android Applications'', also named AndroidTimeMachine, is the first, self-contained, publicly available dataset weaving spread-out data sources about a large number of Android apps \cite{geiger2018graph}.  Covering over 8,431 real-world, open-source Android applications, this dataset contains:
\begin{enumerate}
    \item metadata about the apps’ \texttt{git} projects, with their full commit history, and
    \item metadata from the Google Play store, with app ratings and permissions.
\end{enumerate}

The following is the characterization of file distribution from this dataset available at this \href{https://github.com/AndroidTimeMachine/neo4j_open_source_android_apps/tree/master/data}{link}.

\begin{figure}
    \centering
    \includegraphics[width=0.5\myfigwidth]{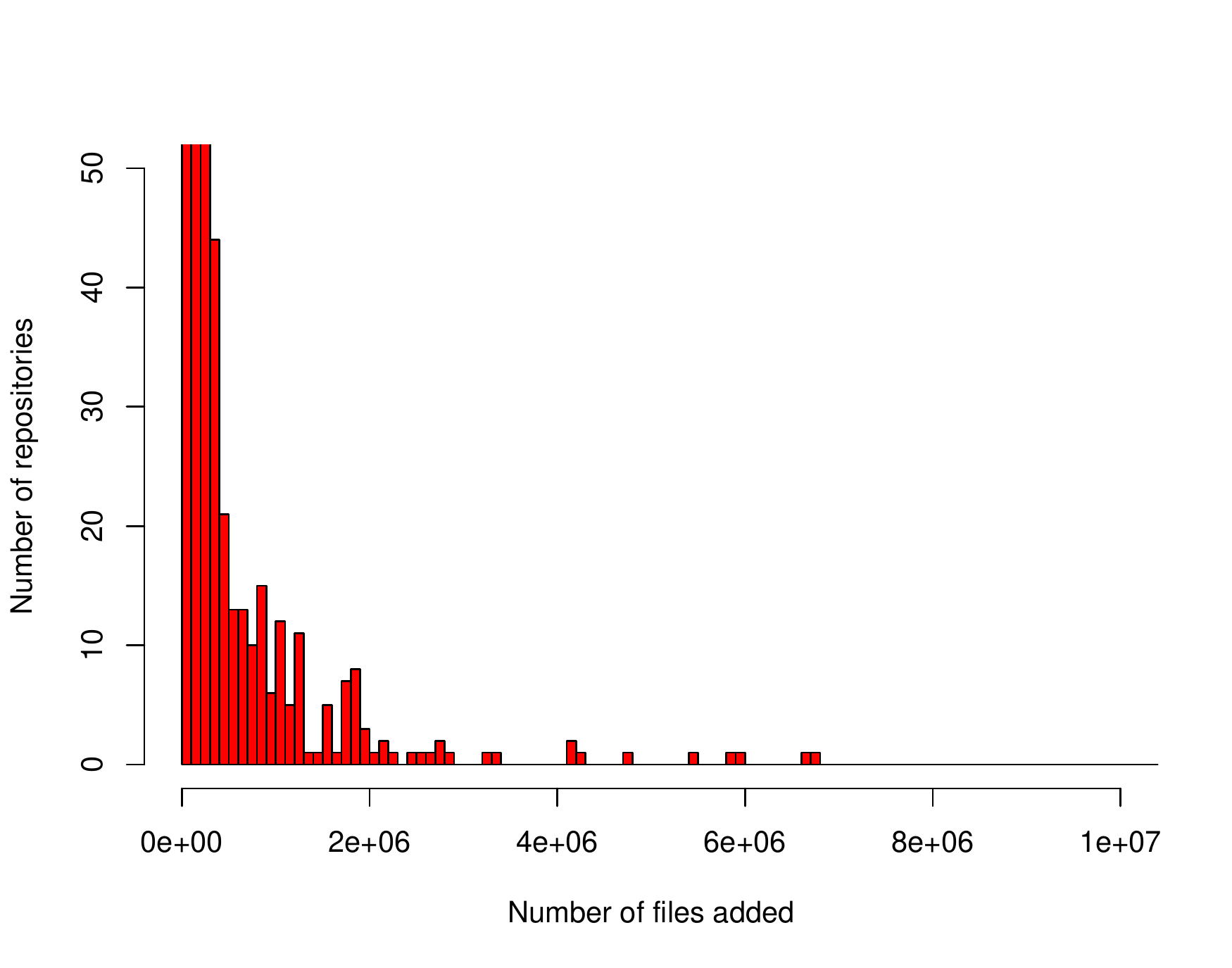}
    \includegraphics[width=0.5\myfigwidth]{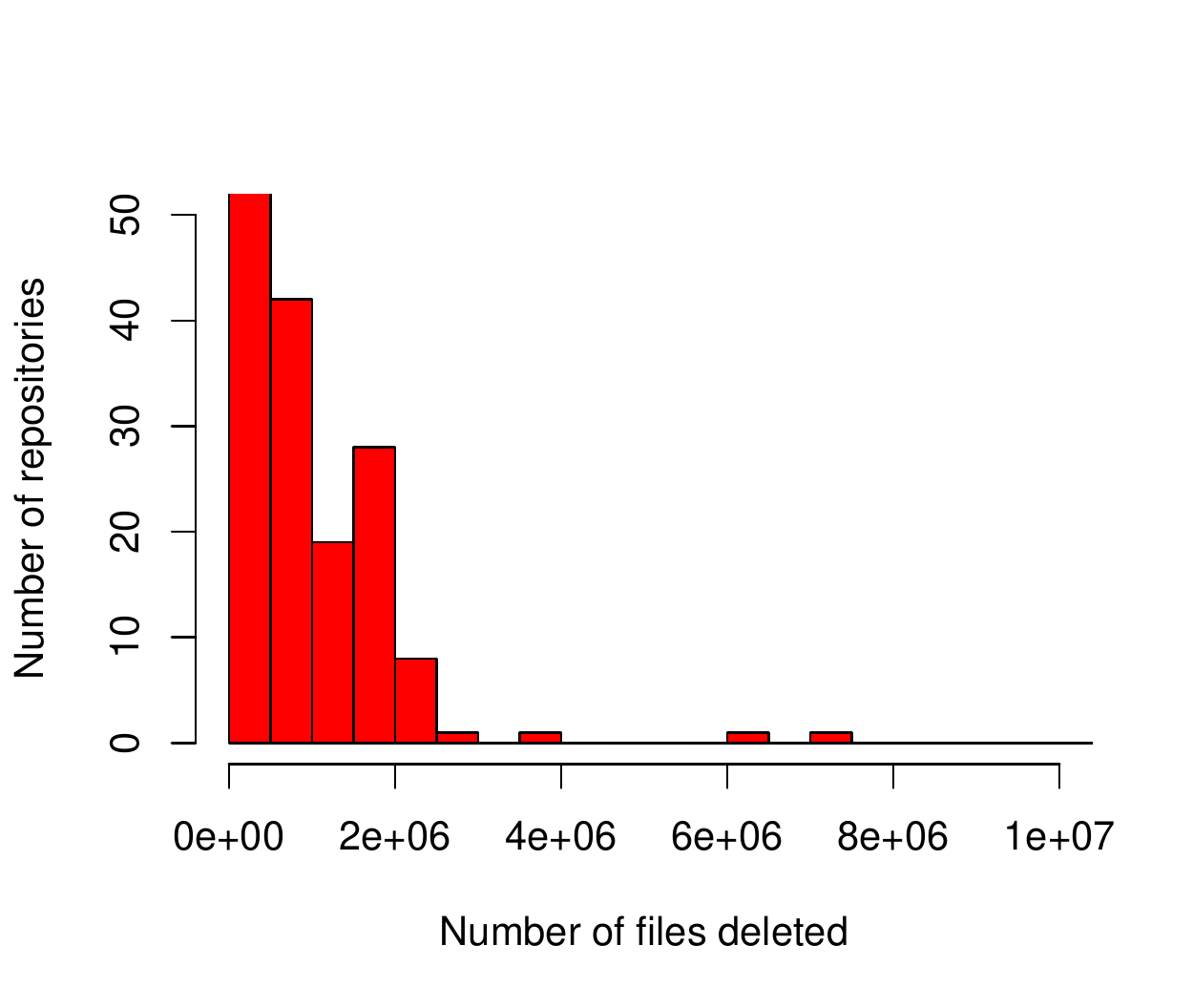}
    \caption{Histograms of additions (left) and deletions (right) of files}
    \label{fig:Android_Files}
\end{figure}

Figure~\ref{fig:Android_Files} shows the histogram of the number of files added to and deleted from in different repositories.

Although our initial analyses were based on the more readily accessible Excel spreadsheet-formatted information from this dataset, this dataset also includes a greater amount of data included inside a container image.  Examining this additional data can potentially provide more training data that can be useful in our machine learning approaches.

\afterpage{\clearpage}

\bibliographystyle{plain}
\bibliography{zeroin.bib}

\end{document}